\documentclass[a4paper,11pt,notitlepage]{article}
\usepackage{jheppub}

\usepackage{amsfonts,amssymb,amsmath,amsthm,mathtools,hyperref,colonequals}
\usepackage{caption}
\usepackage[utf8]{inputenc}
\usepackage{tabularx}
\usepackage{wasysym}
\usepackage{bbm}
\usepackage{mathrsfs}
\usepackage[makeroom]{cancel}
\usepackage{color}
\usepackage{framed}
\definecolor{shadecolor}{rgb}{1,0.9,0.7}

\usepackage{graphicx}
\usepackage{shuffle}
\usepackage{url}
\usepackage{verbatim}

\numberwithin{equation}{section}

\theoremstyle{definition}
\newcommand{\beqa}{\begin{eqnarray}}
\newcommand{\eeqa}{\end{eqnarray}}
\newcommand{\beq}{\begin{equation}}
\newcommand{\eeq}{\end{equation}}

\newcommand{\fP}{\mathsf{P}}
\newcommand{\fp}{\mathsf{p}}
\newcommand{\fs}{\mathsf{s}}
\newcommand{\ft}{\mathsf{t}}

\newcommand{\fTau}{\mathsf{T}}
\newcommand{\fq}{\mathsf{q}}

\newcommand{\calF}{\mathscr{F}}

\newcommand{\calS}{\mathcal{S}}

\newcommand{\calA}{\mathcal{A}}

\newcommand{\calO}{\mathcal{O}}
\newcommand{\calI}{\mathcal{I}}\newcommand{\calZ}{\mathcal{Z}}
\newcommand{\calN}{\mathcal{N}}
\newcommand{\calW}{\mathcal{W}}

\newcommand{\calZt}{\widetilde{\mathcal{Z}}}
\DeclareMathOperator{\Tr}{Tr}

\newcommand{\ver}[1]{\ensuremath{\mathbf{V}(#1)}}
\newcommand{\verop}[3]{\ensuremath{\mathbf{W}_{#1}(#2;#3)}}

\newcommand{\form}[2]{\ensuremath{\left\{ #1 #2\right\}}}
\newcommand{\abra}[2]{\ensuremath{\langle #1 #2\rangle}}
\newcommand{\sbra}[2]{\ensuremath{[ #1 #2]}}
\newcommand{\bradot}[3]{\ensuremath{\langle #1|\, #2|#3]}}

\newcommand{\dd}{\mathrm{d}}
\newcommand{\DD}{\mathrm{D}}
\newcommand{\AAA}{\mathcal{A}}
\newcommand{\WWW}{\mathcal{W}}

\newcommand{\PPP}{\mathbf{F}}
\newcommand{\la}{\lambda}

\newcommand{\n}{\textbf{n}}

\newcommand{\xdot}{\boldsymbol{\cdot}}

\newcommand{\bbA}{\textbf{A}}
\newcommand{\bbB}{\textbf{B}}
\newcommand{\bbC}{\textbf{C}}
\newcommand{\bbD}{\textbf{D}}

\newcommand{\e}{\operatorname{e}}
\newcommand{\ri}{i}

\newcommand{\eqndot}{\, .}
\newcommand{\eqncom}{\, ,}
\newcommand{\eqnsem}{\, ;}

\DeclareMathOperator{\phaneq}{\phantom{{}=}}

\newcommand{\nalpha}{i (4\pi)^2}

\makeatletter
\DeclareRobustCommand*{\bfseries}{%
  \not@math@alphabet\bfseries\mathbf
  \fontseries\bfdefault\selectfont
  \boldmath
}

\begin{document}

\thispagestyle{empty}
\setcounter{page}{0}
\begin{flushright}\footnotesize
\texttt{HU-Mathematik-2016-20}\\
\texttt{HU-EP-16/40}\\
\texttt{MITP/16-123}\\
\vspace{0.5cm}
\end{flushright}
\setcounter{footnote}{0}

\begin{center}
{\huge{
\bf{On Form Factors and \\[4mm] Correlation Functions in  Twistor Space
}
}}
\vspace{10mm}

{\sc 
Laura Koster$^{a}$, Vladimir Mitev$^{b}$,\\ Matthias Staudacher$^a$,
Matthias Wilhelm$^{c}$}\\[5mm]

{\it $^a$Institut f\"ur Mathematik, Institut f\"ur Physik und IRIS Adlershof,\\ Humboldt-Universit\"at zu Berlin,
Zum Gro{\ss}en Windkanal 6,  12489 Berlin, Germany
}\\[2.5mm]
{\it $^b$PRISMA Cluster of Excellence, Institut f\"ur Physik, WA THEP,\\ 
Johannes Gutenberg-Universit\"at Mainz,
Staudingerweg 7, 55128 Mainz, Germany
}\\[2.5mm]
{\it $^c$Niels Bohr Institute, Copenhagen University,\\
Blegdamsvej 17, 2100 Copenhagen \O{}, Denmark
}\\[5mm]

\{\texttt{laurakoster@physik.hu-berlin.de},
\texttt{vmitev@uni-mainz.de},
\texttt{matthias@math.hu-berlin.de},
\texttt{matthias.wilhelm@nbi.ku.dk}\}\\
[15mm]

\textbf{Abstract}\\[2mm]
\end{center}
In this paper, we continue our study of form factors and correlation functions of gauge-invariant local composite operators in the twistor-space formulation of $\calN=4$ super Yang-Mills theory.
Using the vertices for these operators obtained in our recent papers  \cite{Koster:2016ebi,Koster:2016loo}, we show how to calculate the twistor-space diagrams for general N$^k$MHV form factors via the inverse soft limit, in analogy to the amplitude case.
For general operators without $\dot\alpha$ indices, we then reexpress the NMHV form factors from the position-twistor calculation in terms of momentum twistors, deriving and expanding on a relation between the two twistor formalisms previously observed in the case of amplitudes.
Furthermore, we discuss the calculation of generalized form factors and correlation functions as well as the extension to loop level, in particular providing an argument promised in \cite{Koster:2014fva}.

\newpage

\setcounter{tocdepth}{2}
\hrule height 0.75pt
\tableofcontents
\vspace{0.8cm}
\hrule height 0.75pt
\vspace{1cm}

\setcounter{tocdepth}{2}
\setcounter{page}{1}

\section{Introduction}

The maximally supersymmetric Yang-Mills theory in four dimensions ($\mathcal{N}=4$ SYM theory) has many special properties; among others, it has a dual string theory description and is integrable in the planar limit.
In addition, the theory can be formulated in twistor space \cite{Witten:2003nn, Boels:2006ir}. This alternative formulation has proven to be very efficient for describing on-shell quantities such as N$^{k}$MHV amplitudes \cite{Boels:2007qn, ArkaniHamed:2009dn,Mason:2009qx,Bullimore:2010pj, Adamo:2011cb,Adamo:2011pv} and in showing the duality between  planar scattering amplitudes and the planar integrands of supersymmetric Wilson loops in twistor space \cite{Mason:2010yk, Bullimore:2011ni}. 
One advantage of this formulation is the fact that the interactions are reorganized in such a way that the difficulty of a calculation scales with the MHV degree but no longer 
with the number of legs. 
The next step is to start calculating more general quantities that contain also gauge-invariant local composite operators; see \cite{Koster:2014fva, Chicherin:2014uca} for applications of twistor-space techniques to correlation functions. 
This however requires a formulation of general composite operators in twistor space, which cannot be directly obtained by translating the corresponding expressions in space-time \cite{Koster:2016ebi}.

In \cite{Koster:2016ebi,Koster:2016loo}, we have given a formulation of general gauge-invariant local composite operators in twistor space by applying derivatives to a light-like polygonal Wilson loop and taking the limit where this loop is shrunk to a line in twistor space, i.e.\ a point in space-time.
The physical intuition behind this construction stems from recalling three facts. First,  that the field-strength is the curvature of the gauge connection. Second, that the curvature of a connection can be obtained by parallel propagating around a parallelogram, taking derivatives of it and shrinking it to a point.
Third, that all fundamental fields in $\mathcal{N}=4$ SYM theory are related to the field strength via (super)symmetry since they all lie in the singleton representation.%
\footnote{A different formulation of composite operators in the so-called ``Lorentz harmonic chiral'' (LHC) superspace was independently developed in \cite{Chicherin:2016fac,Chicherin:2016fbj,Chicherin:2016qsf} and argued to be related to our twistor-space formulation in \cite{Chicherin:2016soh,Chicherin:2016qsf}.}
As a direct application of the Wilson loop construction, the tree-level MHV form factors of general operators, straightforwardly given by the operator vertices in twistor space, were derived in \cite{Koster:2016loo}. 
Form factors are the simplest objects that contain a local composite operator and are a hybrid between a fully on-shell amplitude and a fully off-shell correlation function. 
Thus, they provide a useful first step for
understanding fully off-shell quantities. 
Moreover, form factors are also interesting on their own; for instance, they are related to the Higgs-to-gluons amplitude in the $m_{\text{top}}\to\infty$ limit and have played an essential role in understanding the infrared divergences of scattering amplitudes \cite{Mueller:1979ih, Collins:1980ih, Sen:1981sd, Magnea:1990zb}.
Although form factors in $\calN = 4$ SYM theory have received increasing attention in recent years, see \cite{vanNeerven:1985ja,Brandhuber:2010ad,Bork:2010wf,Brandhuber:2011tv,
Bork:2011cj,Henn:2011by,Gehrmann:2011xn,Brandhuber:2012vm,Bork:2012tt,
Engelund:2012re,Johansson:2012zv,Boels:2012ew,Penante:2014sza,
Brandhuber:2014ica,Bork:2014eqa,Wilhelm:2014qua,Nandan:2014oga,Loebbert:2015ova,
Bork:2015fla,Frassek:2015rka,Boels:2015yna,Huang:2016bmv,Koster:2016ebi, Koster:2016loo, Chicherin:2016qsf, Brandhuber:2016fni,  Bork:2016hst, Bork:2016xfn, He:2016dol, Caron-Huot:2016cwu, Brandhuber:2016xue, He:2016jdg,Yang:2016ear,Ahmed:2016vgl,Loebbert:2016xkw,Bork:2016egt} at weak coupling and \cite{Alday:2007he,Maldacena:2010kp,Gao:2013dza} at strong coupling, still less is known for form factors than for amplitudes.
For recent reviews, we refer also to \cite{Wilhelm:2016izi, Penante:2016ycx}.

In the present paper, we show how to compute N$^{k}$MHV form factors and correlation functions of $\calN=4$ SYM theory via the twistor formulation and the techniques presented in \cite{Koster:2016ebi, Koster:2016loo}. 
Unlike in \cite{Koster:2016loo}, we now have to face diagrams containing propagators and multiple vertices. In order to efficiently deal with the problem, we extend the inverse soft limit used in \cite{Adamo:2011cb} in the case of amplitudes to our twistor-space vertices for composite operators derived from a Wilson loop \cite{Koster:2016ebi}. A particular stumbling block to overcome is then translating our results from position twistor space to a standard momentum (twistor) space representation, which requires the calculation of a particular type of Fourier integrals.
These integrals distinguish whether or not the operator contains $\dot\alpha$ indices, which correspond to space-time derivatives acting on the Wilson loop.

The paper is organized as follows. In section~\ref{sec2}, we review twistor space, introduce notation, recall the construction of the composite operator from a Wilson loop and derive an inverse soft limit for the corresponding vertices in twistor space. 
In section~\ref{sec3}, we discuss the calculation of general tree-level N$^k$MHV form factors in position twistor space. 
In section~\ref{sec4}, we first show how to translate amplitudes from position twistor space to momentum (twistor) space, deriving a relation between the two twistor formalisms previously observed in \cite{Adamo:2011cb}. We then apply these techniques to the NMHV tree-level form factors of operators without $\dot\alpha$ indices. We conclude section~\ref{sec4} with a computation of the NMHV form factor of an operator with an $\dot\alpha$ index. 
In section~\ref{sec:asketchforloops}, we give a general prescription of how to use the formalism for correlation functions and generalized form factors. We conclude in section~\ref{sec:conclusion} with a summary and a brief overview of open questions.
Three appendices provide further details. We expand on the calculation of the NMHV form factors and the Fourier-type integrals in appendices~\ref{app:NMHVformfactors} and \ref{app:importantFourier}, respectively.
In appendix~\ref{sec:thethreevertexinthetwopointcorrelator}, we revisit the twistor calculation of the one-loop two-point function in the SO$(6)$ sector and provide an argument that was promised in \cite{Koster:2014fva}.

\section{Twistor space, Wilson loops and composite operators}
\label{sec2}

In this section, we set the stage -- introducing all important concepts that will play a role in the sections to come.
These are the twistor action of $\mathcal{N}=4$ SYM theory as introduced for the calculation of scattering amplitudes in \cite{Boels:2006ir, Boels:2007qn} (subsection~\ref{subsec: shortreviewtwistors}), the construction of composite operators in twistor space from Wilson loops as proposed in our recent papers \cite{Koster:2016ebi,Koster:2016loo} (subsection~\ref{subsec: compositeoperatorsfromwilsonloops}), and the inverse soft limit, following \cite{Adamo:2011cb}, which allows to express the $n$-point interaction vertices of the elementary fields and composite operators in terms of $(n-1)$-point vertices (subsection~\ref{subsec: inversesoftlimit}).   

\subsection{The twistor-space formalism}
\label{subsec: shortreviewtwistors}

This section contains a very short introduction to the main ingredients of the twistor-space formulation of $\calN=4$ SYM theory that serves mainly to set up the notation in use throughout the rest of the article. We refer the reader to \cite{Adamo:2013cra, Koster:2014fva} and in particular \cite{Koster:2016ebi,Koster:2016loo} for a more thorough introduction to twistor space, the formalism in use here as well as to our notation. 

Each point $(x,\theta)$ in complexified and conformally compactified super-Minkowski space $\mathbb{M}^{4|8}$ corresponds to a unique projective line $(x,\theta)\cong \mathbb{CP}^1$ in supertwistor space $\mathbb{CP}^{3|4}$.  
Any supertwistor $\calZ$ that is incident with $(x,\theta)$, i.e.\ that lies on the corresponding line, can be written as\footnote{Note that, unlike some texts on the subject, we include an $i$ in the relation between $\chi$ and $\theta$ such that it mirrors the relation between $\mu$ and $x$.
} 
\begin{equation}
\label{eq:definitioncalZonaline}
\calZ=(\la_{\alpha},\mu^{\dot\alpha},\chi^a)=(\la_{\alpha},ix^{\alpha\dot\alpha}\la_{\alpha},i\theta^{a\alpha}\la_{\alpha})\eqncom
\end{equation}
for $\alpha \in\{1,2\}$, $\dot{\alpha}\in \{\dot1,\dot2\}$ and $a\in\{1,2,3,4\}$.
By slight abuse of notation, we will denote lines by $x$ instead of $(x,\theta)$ and will write $\calZ_{x}(\lambda)$ for the supertwistor \eqref{eq:definitioncalZonaline} on the line given by $(x,\theta)$. Furthermore, we shall refer to the bosonic part $(\lambda,\mu)$ of a supertwistor $\calZ$ as $Z$. Contractions of the indices are denoted by corresponding brackets as%
\footnote{Unlike for the angle bracket $\abra{\lambda}{\lambda'}$, where both $\lambda$ and $\lambda'$ live in the same space, $\chi$ and $\eta$ live in dual SU$(4)$ representations. Nevertheless, the shorthand $\form{\chi}{\eta}$ will turn out to be useful.}
\begin{equation}
\label{eq:definitionsproducts}
\abra{\lambda}{\lambda'}\equiv \lambda^{\alpha}\lambda'_{\alpha}\eqncom\qquad \sbra{\mu}{\mu'}\equiv \mu_{\dot{\alpha}}\mu'^{\dot{\alpha}}\eqncom
\qquad \form{\chi}{\eta}\equiv \chi^a\eta_a\eqncom
\end{equation}
with $\epsilon^{\alpha\beta}\lambda_{\alpha}=\lambda^{\beta}$ and $\epsilon^{12}=\epsilon^{\dot{1}\dot{2}}=1=\epsilon_{21}=\epsilon_{\dot 2\dot 1}$. In Minkowski space, we use the mostly-minus metric with the definitions $xy\equiv x_{\alpha\dot \alpha}y^{\alpha\dot\alpha}=2x_\mu y^\mu$ and $x^2\equiv \frac{1}{2}x_{\alpha\dot \alpha}x ^{\alpha\dot \alpha}=x_\mu x^\mu$.

We now define a supertwistor connection $(0,1)$-form $\AAA(\mathcal{Z})$ on supertwistor space $\mathbb{CP}^{3|4}$. When expanded in the Gra{\ss}mann variables $\chi$ as in \cite{Nair:1988bq}, it has the form
\begin{equation}
\label{eq:expansionAAA}
\AAA(\mathcal{Z})=g^+ +\chi^a\bar{\psi}_a+\frac{1}{2}\chi^a\chi^b\phi_{ab}+\frac{1}{3!}\chi^a\chi^b\chi^c\psi^d\epsilon_{abcd}+\chi^1\chi^2\chi^3\chi^4 g^-\eqncom
\end{equation}
i.e.\ it combines all on-shell degrees of freedom of $\calN=4$ SYM theory. 
The action of $\calN=4$ SYM theory  $\calS=\calS_1+\calS_2$ can be split into a self-dual part $\calS_1$ and an interaction part $\calS_2$ that contains the non-self-dual contributions.
Translated to supertwistor space $\mathbb{CP}^{3|4}$, we can write the action in terms of the supertwistor field $\AAA(\mathcal{Z})$ as the sum of a holomorphic Chern-Simons action $\calS_1$, introduced in \cite{Witten:2003nn},
and an interaction part $\calS_2$ \cite{Boels:2006ir}  that is given by an integral over the space of lines in $\mathbb{CP}^{3|4}$, i.e.\ over Minkowski space $\mathbb{M}^{4|8}$. 
From the interaction part $\calS_2$, one can derive the (color-stripped) vertices \cite{Adamo:2011cb}
\begin{equation}
\label{eq:definitionoftheamplitudevertex 2.1}
\ver{\calA_1,\ldots, \calA_{n}}=
\int_{\mathbb{M}^{4|8}} \frac{\dd^4 z\:\dd^8\vartheta}{(2\pi)^4} 
\int_{(\mathbb{CP}^1)^n}\DD\la_1\DD\la_2\cdots \DD\la_n\frac{\AAA_1(\la_1)\AAA_2(\la_2)\cdots\AAA_n(\la_n)}{ \abra{\lambda_1}{\lambda_2} \cdots \abra{\lambda_{n-1}}{ \lambda_{n}} \abra{\lambda_{n}}{\lambda_{1}}}
 \,, 
\end{equation}
where $\DD\lambda\equiv\frac{\abra{\lambda}{\dd\lambda}}{2\pi i }$ and $n\geq2$.
We refer to the action $\calS$ as being the \textit{position twistor space} formulation of $\calN=4$ SYM theory in order to distinguish the twistor space used here from the  \textit{momentum twistor space} introduced in \cite{Hodges:2009hk} for the study of amplitudes in which twistors are introduced for the dual region variables.%
\footnote{Amplitudes are known to be dual to light-like Wilson loops in the space of region (super)momenta, and the latter can be calculated using the twistor action as well \cite{Adamo:2011dq}. However, this is not what we are doing here; we are introducing Wilson loops in twistor space for the operator $\calO(x)$ \eqref{eq:composite operator}.}
For simplicity of notation, we shall from now on assume that $\int \DD\lambda$ is integrated over $\mathbb{CP}^1$ and that $\int \dd^4 z\dd^8\theta $ is integrated over $\mathbb{M}^{4|8}$ and thus omit the space of integration.

Choosing an axial (or light-cone) gauge, 
the action $\calS_1$ becomes free and we obtain the propagator
\begin{equation}
\Delta(\calZ_1,\calZ_2)=\nalpha \bar \delta^{2|4}(\calZ_1,\star,\calZ_2)\eqncom
\end{equation}
where $\star\equiv \calZ_{\star}=(Z_{\star},0)=(0,\zeta^{\dot\alpha},0)$ is a reference twistor. The interested reader is referred to \cite{Adamo:2011cb} for a definition and discussion of the $\bar{\delta}^{i|4}$ functions in supertwistor space. For us, the important properties are that the $\bar{\delta}^{i|4}$ are homogeneous of degree zero in any of their arguments and that they are antisymmetric with respect to the exchange of any two arguments.

In order to obtain amplitudes or form factors, we need external states. An external state in position twistor space is \cite{Adamo:2011cb}
\begin{equation}
\label{eq: external state in position twistor space}
 \calA_{\calZt}(\calZ)=2\pi i \,\bar{\delta}^{3|4}(\calZ,\calZt)\eqncom
\end{equation}
where we denote the external supertwistors as $\calZt$. We are of course interested in obtaining amplitudes and form factors in momentum space, so that we instead need to insert the on-shell momentum representation%
\footnote{Note that the $\lambda_\alpha$ are components of the supertwistor $\mathcal{Z}$ in \eqref{eq:definitioncalZonaline} which are integrated over at each vertex \eqref{eq:definitionoftheamplitudevertex 2.1}. In particular, they are not spinor-helicity variables parametrizing momenta.
When inserting the supermomentum eigenstate \eqref{eq:definitiononshellmomentumeigenstates}, however, the $s$ and $\lambda$ integration in combination with the delta function effectively replace $\lambda_\alpha$ with the spinor-helicity variable $p_\alpha$. 
}
\begin{equation}
 \label{eq:definitiononshellmomentumeigenstates}
\calA_{\fP}(\calZ)=2\pi i \int_{\mathbb{C}}\frac{\dd s}{s}\e^{s(\sbra{\bar{p}}{\mu}+\{\chi\eta\})}
\bar{\delta}^2(s\lambda_{\alpha}-p_{\alpha})\, , \qquad \calZ=(\lambda_{\alpha},\mu^{\dot{\alpha}}, \chi^a)\eqnsem
\end{equation}
see for instance \cite{Adamo:2011cb}.
We denote the on-shell supermomenta in terms of super-spinor-helicity variables as $\fP=(\fp_{\alpha\dot{\alpha}},\eta_a)\equiv (p_{\alpha},\bar{p}_{\dot{\alpha}},\eta_a)$, i.e.\ we write an on-shell momentum as $\fp_{\alpha\dot{\alpha}}=p_{\alpha}\bar{p}_{\dot{\alpha}}$.
Inserting the on-shell states \eqref{eq:definitiononshellmomentumeigenstates} into the interaction vertices \eqref{eq:definitionoftheamplitudevertex 2.1} immediately yields the tree-level MHV super amplitudes:
\begin{equation}
\label{eq:MHVamplitude}
\mathscr{A}^{\text{MHV}}(\fP_1,\ldots, \fP_n)
=\ver{\calA_{\fP_1},\ldots, \calA_{\fP_n}}
= \frac{\delta^{4|8}(\sum_{j=1}^n\fP_j)}{\abra{p_1}{p_2}\cdots \abra{p_n}{p_1}}
\,,
\end{equation}
see \cite{Boels:2007qn} for details. 

\subsection{Composite operators from Wilson loops}
\label{subsec: compositeoperatorsfromwilsonloops}

We now recall the construction of composite operators in twistor space as introduced in \cite{Koster:2016loo}.
The presentation here will be rather brief and we refer the reader to \cite{Koster:2016loo} for details.

Let us start be recalling the classification of composite operators in space-time, see e.g.\ \cite{Beisert:2004ry,Minahan:2010js} for reviews.
In the planar limit, it is sufficient to look at so-called single-trace operators, which are built from traces of products of covariantly transforming fields at a common space-time point $x$:

\begin{equation}
\label{eq:composite operator}
\mathcal{O}(x)=\Tr\left(D^{k_1}A_1(x)D^{k_2}A_2(x)\cdots D^{k_L}A_L(x)\right)\eqndot
\end{equation}
Here, $D$ denotes a covariant derivative and the $A_i$ can be scalars $\phi_{ab}$, fermions $(\psi_{bcd})_{\alpha}=\epsilon_{abcd}\psi^d_{\alpha}$, anti-fermions $\bar{\psi}_{a\dot{\alpha}}$ and the self-dual and anti-self-dual parts of the field strength $F_{\alpha\beta}$ and $\bar{F}_{\dot\alpha\dot\beta}$.
Using the equations of motions for these fields, the definition of the field strength and the Bianchi identity, covariant derivatives of fields with all combinations of spinor indices can be expressed in terms of irreducible fields, which are completely symmetric in all $\alpha$ and $\dot\alpha$ indices.
They form the irreducible singleton representation $\textsf{V}_{\textsf{S}}$ of the symmetry algebra $\mathfrak{psu}(2,2|4)$.
Using superpolarization vectors $\fTau_i=(\tau_i^{\alpha},\bar{\tau}_i^{\dot\alpha}, \xi_i^a)$, we can write the irreducible fields as
\begin{equation}
\label{eq: alphabet of fields with polarizations}
\begin{aligned}
 D^kA\in\{ 
 &+\tau^{\alpha_1}\dots\tau^{\alpha_{k\phantom{+1}}}\bar{\tau}^{\dot\alpha_1}\dots\bar{\tau}^{\dot\alpha_{k+2}}\phantom{\xi^a}\phantom{\xi^a}\phantom{\xi^a}\phantom{\xi^a}
 D_{\alpha_1\dot\alpha_1}\cdots D_{\alpha_k\dot\alpha_k}\bar{F}_{\dot\alpha_{k+1}\dot\alpha_{k+2}},\\
 &+\tau^{\alpha_1}\dots\tau^{\alpha_{k\phantom{+1}}}\bar{\tau}^{\dot\alpha_1}\dots\bar{\tau}^{\dot\alpha_{k+1}}\xi^a\phantom{\xi^a}\phantom{\xi^a}\phantom{\xi^a}
 D_{\alpha_1\dot\alpha_1}\cdots D_{\alpha_k\dot\alpha_k}\bar{\psi}_{\dot\alpha_{k+1}a},\\
 &-\tau^{\alpha_1}\dots\tau^{\alpha_{k\phantom{+1}}}\bar{\tau}^{\dot\alpha_1}\dots\bar{\tau}^{\dot\alpha_{k\phantom{+1}}}\xi^a\xi^b\phantom{\xi^a}\phantom{\xi^a}
 D_{\alpha_1\dot\alpha_1}\cdots D_{\alpha_k\dot\alpha_k}\phi_{ab},\\
 &-\tau^{\alpha_1}\dots\tau^{\alpha_{k+1}}\bar{\tau}^{\dot\alpha_1}\dots\bar{\tau}^{\dot\alpha_{k\phantom{+1}}}\xi^a\xi^b\xi^c\phantom{\xi^a}
 D_{\alpha_1\dot\alpha_1}\cdots D_{\alpha_k\dot\alpha_k}\psi_{\alpha_{k+1}abc},\\
 &+\tau^{\alpha_1}\dots\tau^{\alpha_{k+2}}\bar{\tau}^{\dot\alpha_1}\dots\bar{\tau}^{\dot\alpha_{k\phantom{+1}}}\xi^a\xi^b\xi^c\xi^d
 D_{\alpha_1\dot\alpha_1}\cdots D_{\alpha_k\dot\alpha_k}F_{\alpha_{k+1}\alpha_{k+2}abcd}
 \}\eqncom
\end{aligned}
\end{equation}
where $F_{\alpha_{k+1}\alpha_{k+2}abcd}=\frac{1}{4!}\epsilon_{abcd}F_{\alpha_{k+1}\alpha_{k+2}}$.
The indices are recovered by taking appropriately weighted derivatives with respect to $\tau$, $\bar\tau$ and $\xi$.%
\footnote{See \cite{Koster:2016loo} for a pedagogical example.}

\begin{figure}[tbp]
 \centering
  \includegraphics[height=3.45cm]{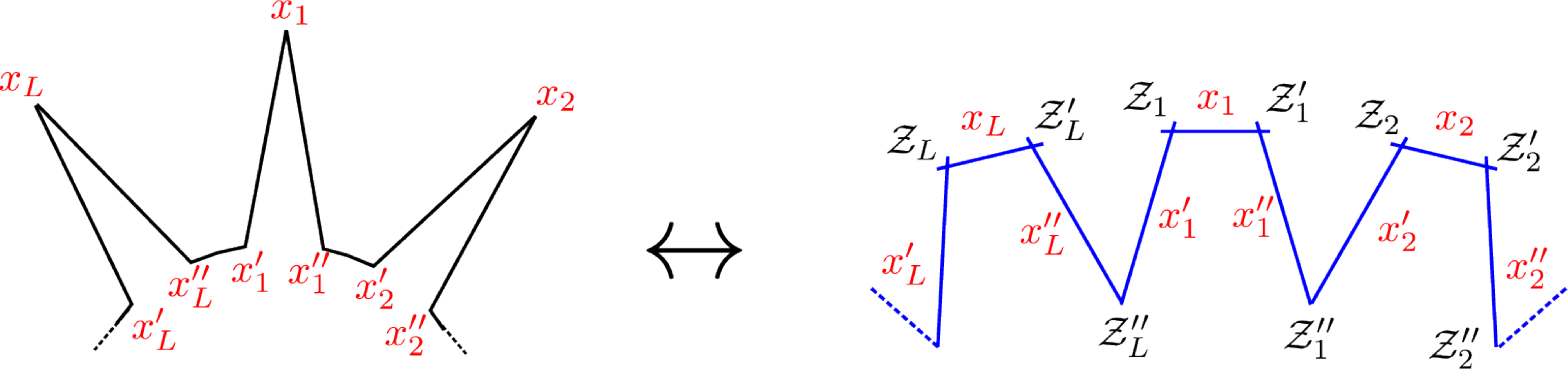}
  \caption{\it This figure illustrates the $3L$-gonal ``cogwheel'' light-like Wilson loop that we use to construct an operator containing $L$ irreducible fields.
    }
  \label{fig:CogwheelZoom}
\end{figure}

We can construct the twistor-space vertex of a composite operator built from $L$ irreducible fields by taking derivatives of a polygonal light-like Wilson loop in twistor space and shrinking the Wilson loop to a line in twistor space, i.e.\ a point in space-time. 
One possible choice for the geometry of the Wilson loop 
has $3L$ vertices in Minkowski space that we label $x_1',x_1,x_1'',\ldots, x_L', x_L,x_L''$.%
\footnote{There is large freedom in the choice of the geometry of the Wilson loop provided that certain minimal requirements are satisfied, see \cite{Koster:2016loo}. Once the Wilson loop is properly shrunk to a line in twistor space, the operator vertex is blind to that choice of geometry.}
It can be drawn in the shape of a cogwheel, see the left-hand side of figure~\ref{fig:CogwheelZoom}. 
In twistor space, each pair of neighboring points intersects and we label the intersection points $\calZ_i=x_i\cap x_{i}'$,  $\calZ_i'=x_i\cap x_{i}''$ and $\calZ_i''=x_{i}''\cap x_{i+1}'$, see the right-hand side of figure~\ref{fig:CogwheelZoom}. Then, the cogwheel Wilson loop is simply obtained by multiplying parallel propagators $U$ connecting adjacent twistors as
\begin{equation}
\label{eq:finaldefinitionWilsonloop}
\begin{aligned}
 \WWW(x_1',x_1,x_1'',\ldots, x_L',x_L,x_L'')= \Tr\Big[&U_{x_1'}(\calZ_L'',\calZ_1) U_{x_1}(\calZ_1,\calZ_1')U_{x_1''}(\calZ_1',\calZ_1'')\\&  U_{x_2'}(\calZ_1'',\calZ_2)U_{x_2}(\calZ_1,\calZ_1')U_{x_2''}(\calZ_1',\calZ_2'') \cdots \\&\cdots U_{x_L'}(\calZ_{L-1}'',\calZ_L) U_{x_L}(\calZ_{L},\calZ_L') U_{x_L''}(\calZ_{L}',\calZ_L'') \Big]\eqncom
\end{aligned}
\end{equation}
where the parallel propagator is \cite{Bullimore:2011ni}
\begin{equation}
\label{eq:frameUdefinitionpart1}
 U_{x}(\calZ_1,\calZ_{2})=1+\sum_{m=1}^\infty \int_{}\frac{\abra{\lambda_1}{\lambda_{2}}\DD\tilde{\la}_1 \cdots  \DD\tilde{\la}_m}{\abra{\lambda_1}{\tilde{\la}_1}\abra{\tilde{\la}_1}{\tilde{\la}_2}\cdots \abra{\tilde{\la}_m}{\lambda_{2}}}\AAA(\calZ_{x}(\tilde{\la}_1)) \cdots \AAA(\calZ_{x}(\tilde{\la}_m))
\eqndot
\end{equation}

In order to obtain vertices for the composite operator $\mathcal{O}$ \eqref{eq:composite operator}, we act on $\mathcal{W}$ with a differential operator, which we call the \emph{forming operator}. It is the product of respective forming operators for the irreducible fields $D^{k_{\ri}}A_{\ri}$ out of which $\mathcal{O}$ is built:
\begin{equation}
\label{formingop}
 \PPP_{\mathcal{O}}=\prod_{\ri=1}^L\PPP_{D^{k_{\ri}}A_{\ri}}\eqndot
\end{equation}
The forming operators of the irreducible fields \eqref{eq: alphabet of fields with polarizations} are
\begin{equation}
 \label{eq:definitionformingfactoronshellstates}
\PPP_{D^{k }A }= -\frac{\la ^{\alpha}\la' {}^{\beta}}{\abra{ \la }{\la' }} 
\big(-i\tau^\gamma \bar{\tau} ^{\dot\gamma}\partial_{ \gamma\dot\gamma}\big)^{k }
\left\{\begin{array}{ll} 
(-i\bar{\tau} ^{\dot\alpha}\partial_{ \alpha\dot\alpha})(-i\bar{\tau} ^{\dot\beta}\partial_{ \beta\dot{\beta}})& 
\text{ for } A =\bar{F}\eqncom\\
(-i\bar{\tau} ^{\dot\alpha}\partial_{ \alpha\dot{\alpha}})(-i\xi ^{a}\partial_{ \beta a})& 
\text{ for } A =\bar{\psi}\eqncom\\
(-i\xi ^{a}\partial_{ \alpha a})(-i\xi ^b\partial_{ \beta b})& 
\text{ for } A =\phi\eqncom\\
(-i\xi ^{a}\partial_{ \alpha a})(-i\xi ^{b}\partial_{ \beta b})(-i\tau ^{\gamma}\xi ^{c}\partial_{ \gamma c})&
\text{ for }A =\psi\eqncom\\
(-i\xi ^{a}\partial_{ \alpha a})(-i\xi ^{b}\partial_{ \beta b})(-i\tau ^{\gamma}\xi ^{c}\partial_{ \gamma c})^2& 
\text{ for } A =F\eqncom
\end{array}\right. 
\end{equation}
where we have used the abbreviations
\begin{equation}
 \partial_{\alpha\dot\alpha}\equiv \frac{\partial}{\partial x^{\alpha\dot\alpha}}\,,\qquad \partial_{\alpha a} \equiv \frac{\partial}{\partial \theta^{\alpha a}}
\end{equation}
and suppressed the index $i$.
The forming operator $\PPP_{D^{k_i }A_i}$ is then simply defined as  \eqref{eq:definitionformingfactoronshellstates} acting on the edge  $(x_i,\theta_i)$ of the Wilson loop of figure~\ref{fig:CogwheelZoom}.

\begin{figure}[tbp]
 \centering
  \includegraphics[height=3.8cm]{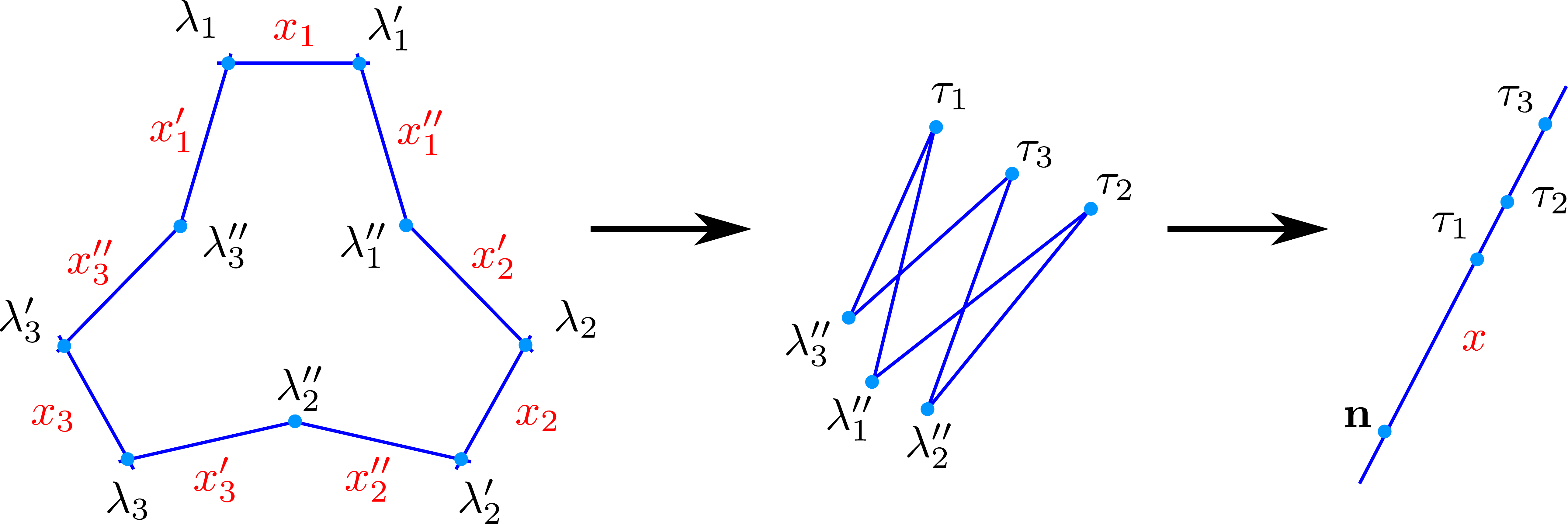}
  \caption{\it This figure illustrates the operator limit for $L=3$. To visualize the process a bit better, as a first step, we set $\lambda_i=\la_i'=\tau_i$ while bringing the $x_i$ closer to each other. The second step then just sends all $x_i$ to $x$ and $\lambda_{i}''\rightarrow \n$. 
  }
  \label{fig:operatorlimitapp}
\end{figure}

Finally, we take the operator limit, which we denote by $\hexagon\rightarrow \xdot$ or by $\hexagon\rightarrow \xdot_x$ if we want to emphasize the point to which the loop shrinks. We refer to appendix $A$ of \cite{Koster:2016loo} for more details on how to take the limit, for now it is only important that the operator limit sets
\begin{equation}
\label{eq:operatorlimit}
x_i,x_i',x_i''\rightarrow x\,,\qquad \lambda_i, \lambda_i'\rightarrow \tau_i\,,\qquad \la_i''\rightarrow \n \,;
\end{equation}
see figure~\ref{fig:operatorlimitapp}. The reference spinor $\n$ has to drop out of any physical quantity.
Moreover, we set $\theta=0$ to obtain only the component operator $\calO$ and not part of its supermultiplet.

Combining everything, we obtain the following expression for the operator vertices:
\begin{equation}
\label{eq: summary operator vertex}
\textbf{W}_{\mathcal{O}(x)}=\lim_{\hexagon\rightarrow \xdot}\PPP_{\mathcal{O}}\, \mathcal{W}\big|_{\theta=0}\,.
\end{equation}
Similar to the case of amplitudes, inserting momentum eigenstates \eqref{eq:definitiononshellmomentumeigenstates} into the operator vertex \eqref{eq: summary operator vertex} immediately yields all MHV form factors, cf.\ \cite{Koster:2016loo}.

\subsection{The inverse soft limit}
\label{subsec: inversesoftlimit}

Via the so-called inverse soft limit, the $n$-point twistor-space vertices of the elementary interactions can be expressed in terms of $(n-1)$-point vertices. This procedure  played a crucial role in the calculation of amplitudes beyond MHV level in position twistor space \cite{Adamo:2011cb}.
Here, we find a similar recursion for the Wilson loop vertices, which will play an equally important role in the calculation of form factors beyond MHV level.
\begin{figure}[htbp]
 \centering
  \includegraphics[height=1.9cm]{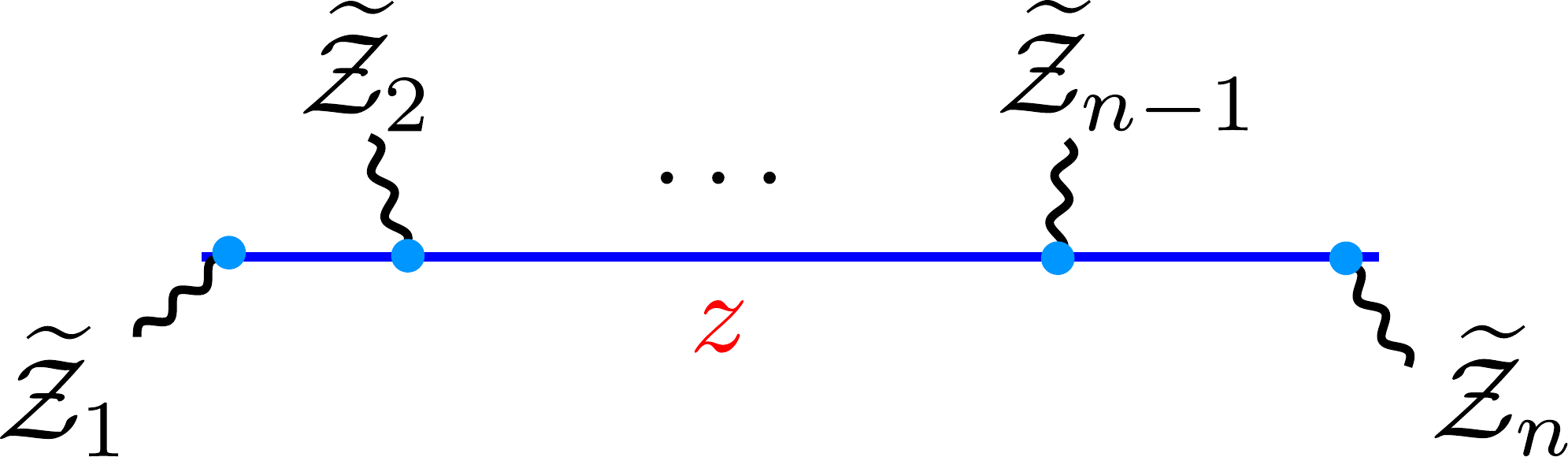}
  \caption{\it The $n$-point vertex \eqref{eq:definitionoftheamplitudevertex 2.1} from the action $\calS_2$. Throughout this paper, labels of the vertices are read off clockwise.}
\label{fig: picture for the vertices}
\end{figure}

We first review the case of the interaction vertices. The $n$-point vertices
\begin{equation}
\label{eq:definitionoftheamplitudevertex}
\ver{\calZt_1,\ldots, \calZt_{n}}\equiv\ver{\calA_{\calZt_1},\ldots, \calA_{\calZt_{n}}}
\end{equation}
of \eqref{eq:definitionoftheamplitudevertex 2.1}, see figure~\ref{fig: picture for the vertices},
can be reduced to $(n-1)$-point vertices via recursion relations \cite{Adamo:2011cb} that allow us to reduce them all the way down to the two-point vertices. Specifically, the inverse soft limit is derived by first parametrizing the $\lambda_{j\alpha}$ as $(1,\sigma_j)_\alpha$ so that $\DD\lambda_j=\tfrac{\dd\sigma_j}{2\pi i }$ and $\abra{\lambda_j}{\lambda_{j+1}}=\sigma_{j+1}-\sigma_{j}$. We now choose a $k\in \{1,\ldots, n\}$ and replace $\sigma_k$ by $s=\tfrac{\sigma_{k}-\sigma_{k+1}}{\sigma_{k-1}-\sigma_{k}}$ such that $\tfrac{\dd s}{s}=-\tfrac{(\sigma_{k+1}-\sigma_{k-1})\dd\sigma_k}{(\sigma_{k+1}-\sigma_k)(\sigma_k-\sigma_{k-1})}$ and $\calZ_z(\sigma_k)=\tfrac{1}{1+s}\left[s\calZ_z(\sigma_{k-1})+\mathcal{Z}_z(\sigma_{k+1})\right]$. Employing  $\int \tfrac{\dd s}{s}\bar{\delta}^{3|4}(s\calZ_z(\sigma_{k-1})+\calZ_z(\sigma_{k+1}),\calZt_k)=-\bar \delta^{2|4}(\calZt_{k-1},\calZt_k,\calZt_{k+1})$, where we have used the $\bar \delta^{3|4}$ to replace the $\mathcal{Z}_z(\sigma_{k\pm 1})$ by $\tilde{\calZ}_{k\pm 1}$, leads to 
\begin{equation}
\label{eq:amplitudeverticesinversesoftlimit}
\ver{\calZt_1,\ldots, \calZt_{n}}=\ver{\calZt_1,\ldots,\calZt_{k-1},\calZt_{k+1},\ldots \calZt_{n}}\bar{\delta}^{2|4}(\calZt_{k-1},\calZt_{k},\calZt_{k+1})\eqndot
\end{equation}
Equation \eqref{eq:amplitudeverticesinversesoftlimit} implies that the vertex can be written as the product
\begin{equation}
\label{eq:inversesoftlimitamplitudes}
\ver{\calZt_1,\ldots, \calZt_{n}}=\ver{\calZt_1,\calZt_2}\prod_{i=3}^n\bar{\delta}^{2|4}(\calZt_1,\calZt_{i-1},\calZt_i)\eqndot
\end{equation}
There are different ways of writing the $n$-point amplitude that are related by identities of the kind 
$\ver{\calZt_1,\calZt_2,\calZt_3}\bar{\delta}^{2|4}(\calZt_1,\calZt_{3},\calZt_4)=\ver{\calZt_2,\calZt_3,\calZt_4}\bar{\delta}^{2|4}(\calZt_2,\calZt_{4},\calZt_1)$, which come from the total antisymmetry of the $\bar\delta^{2|4}$ and from the different choices of points $k$ to remove using \eqref{eq:amplitudeverticesinversesoftlimit}.

\begin{figure}[tbp]
 \centering
  \includegraphics[height=1.6cm]{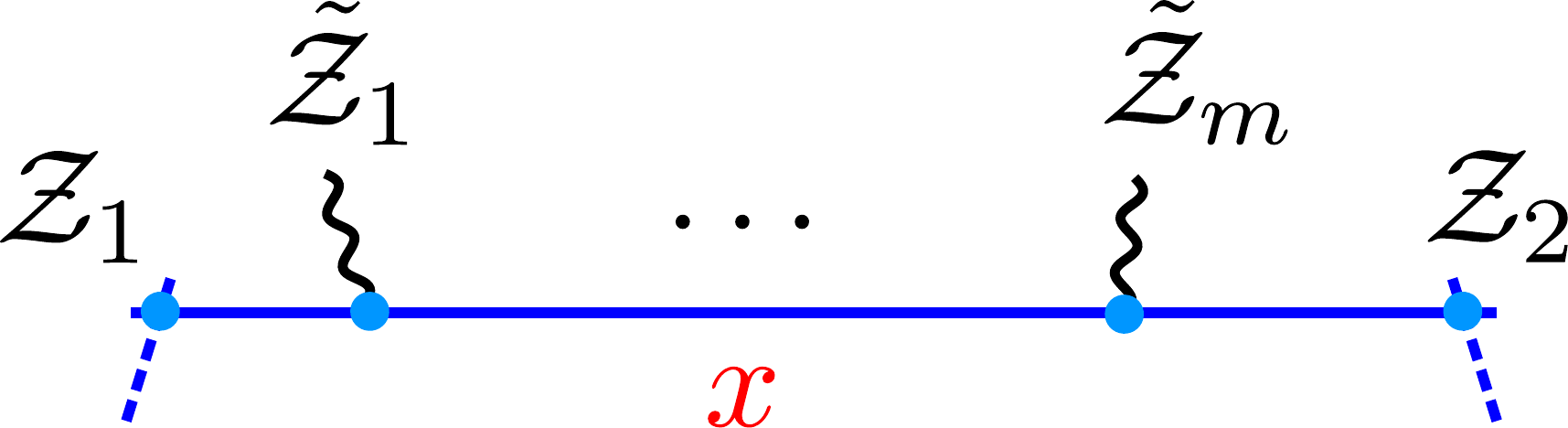}
  \caption{\it The $m$-point vertex \eqref{eq: Wilson line vertex in position space} from the parallel propagator $U_x(\calZ_1,\calZ_2)$. }
\label{fig: picture for the vertices W}
\end{figure}

In space-time, MHV form factors can be constructed via inverse soft limits as well \cite{Nandan:2012rk}. In our twistor-space formulation, this construction is based on the parallel propagator \eqref{eq:frameUdefinitionpart1}.
Using the external states in position twistor space \eqref{eq: external state in position twistor space},
the parallel propagator $U_x(\calZ_1,\calZ_2)$ in \eqref{eq:frameUdefinitionpart1} leads to the following Wilson line vertex: 
\begin{equation}
\label{eq: Wilson line vertex in position space}
 \begin{aligned}
\verop{x}{\calZ_1,\calZ_{2}}{\calZt_1,\dots,\calZt_m}
&\equiv\verop{x}{\calZ_1,\calZ_{2}}{\calA_{\calZt_1},\ldots, \calA_{\calZt_{m}}}\\
&=
\int  \frac{\abra{\lambda_1}{\lambda_{2}}
\prod_{j=1}^m \calA_{\calZt_j}(\calZ_x(\tilde\lambda_j))\DD\tilde\lambda_j}{\abra{\lambda_1}{\tilde\lambda_1} \abra{\tilde\lambda_1}{\tilde\lambda_2}\cdots \abra{\tilde\lambda_{m}}{\lambda_{2}} }
\eqncom
\end{aligned}
\end{equation}
see figure~\ref{fig: picture for the vertices W}.
In particular, for $m=0$ we set $\verop{x}{\calZ_1,\calZ_{2}}{}\equiv1$. We keep in mind the Wilson loop origin of the vertices $\textbf{W}$ and hence denote by $\calZ_1$ and $\calZ_2$ the two twistors at which the line $x$ intersects with the neighboring lines in the Wilson loop, see figure~\ref{fig:CogwheelZoom}. Furthermore, we write $\calZt_i$ for the twistors on the line $x$ where on-shell states or propagators are to be attached.
In particular, the smallest non-trivial vertex is
\begin{equation}
\label{eq: minimal Wilson line vertex}
\begin{aligned}
\verop{x}{\calZ_1,\calZ_{2}}{\calZt}&=\int\DD\tilde{\lambda}\frac{\abra{\lambda_1}{\lambda_2}}{\abra{\lambda_1}{\tilde{\lambda}}\abra{\tilde{\lambda}}{\lambda_2}}2\pi i \,\bar{\delta}^{3|4}(\calZt,\calZ_x(\tilde{\la}))\\&=\int_{\mathbb{C}}\frac{\dd u}{u}\bar{\delta}^{3|4}(\calZ_1+u\calZ_2,\calZt)=\bar{\delta}^{2|4}(\calZ_1,\calZt,\calZ_{2})\eqncom
\end{aligned}
\end{equation}
where in the second line we used the parametrization of the line $x$ given by $\calZ_x(\tilde{\la})=\calZ_1+u\calZ_2$, which implies that 
$\abra{\lambda_1}{\tilde{\la}}=u \abra{\lambda_1}{\lambda_2}$, $\abra{\tilde{\la}}{\lambda_2}=\abra{\lambda_1}{\lambda_2}$ and $\DD\tilde{\la}=\abra{\lambda_1}{\lambda_2}\tfrac{\dd u}{2\pi i }$.

Using now the same method as for the vertices $\textbf{V}$,
we arrive at the identity
\begin{multline}
\label{eq: inverse soft limit frame}
 \verop{x}{\calZ_{1},\calZ_{2}}{\calZt_1,\dots,\calZt_m}
 \\= \verop{x}{\calZ_{1},\calZ_{2}}{\calZt_1,\dots,\calZt_{k-1},\calZt_{k+1},\ldots, \calZt_{m}}
 \bar{\delta}^{2|4}(\calZt_{k-1},\calZt_{k},\calZt_{k+1})\eqncom
\end{multline}
where $k\in \{1,\ldots, m\}$ and it is understood that $\calZt_{0}\equiv\calZ_1$ and $\calZt_{m+1}\equiv\calZ_2$.
Using the  relation \eqref{eq: inverse soft limit frame}, we can reduce every Wilson line vertex to the minimal Wilson line vertex \eqref{eq: minimal Wilson line vertex}:
\begin{equation}
 \label{eq: total inverse soft limit frame}
\begin{aligned}
  \verop{x}{\calZ_{1},\calZ_{2}}{\calZt_1,\dots,\calZt_m}&=\prod_{j=1}^m\bar{\delta}^{2|4}(\calZt_{j-1},\calZt_{j},\calZ_{2}) \eqndot
\end{aligned}
\end{equation}
The expression \eqref{eq: total inverse soft limit frame} has a very geometric interpretation, which makes its origin obvious and goes as follows.
The vertices $\verop{x}{\calZ_1,\calZ_{2}}{\calZt_1,\dots,\calZt_m}$ in \eqref{eq: Wilson line vertex in position space} describe a Wilson line in position twistor space. Hence, all position twistors $\calZ_1,\calZ_{2},\calZt_1,\dots,\calZt_m$ have to lie on one line.
This is expressed in \eqref{eq: total inverse soft limit frame} by enforcing each of the first $m$ pairs of neighboring position twistors to be collinear with the last position twistor. However, a lot of equivalent ways exist to express the collinearity of all $m+2$ position twistors by forcing $m$ triplets of them to be collinear.

\section{\texorpdfstring{N$^k$MHV}{NkMHV} form factors in position twistor space}
\label{sec3}

Using the building blocks set up in the previous section, we can now calculate the diagrams for N$^k$MHV tree-level form factors of general operators in position twistor space.
We start discussing the case of NMHV form factors and then generalize to arbitrary $k$.
This calculation is similar to the one for amplitudes \cite{Adamo:2011cb}, and we will frequently refer to this case for intuition. 

\subsection{NMHV amplitudes and form factors}
\label{NMHVamplitudesandformfactorsinpositiontwistorspace}

In the twistor-space formalism, the MHV degree $k$ directly translates to twistor-space diagrams involving $k$ propagators at tree level.
At the NMHV level, we therefore have exactly $k=1$ propagator.

\begin{figure}[tbp]
 \centering
  \includegraphics[height=3.5cm]{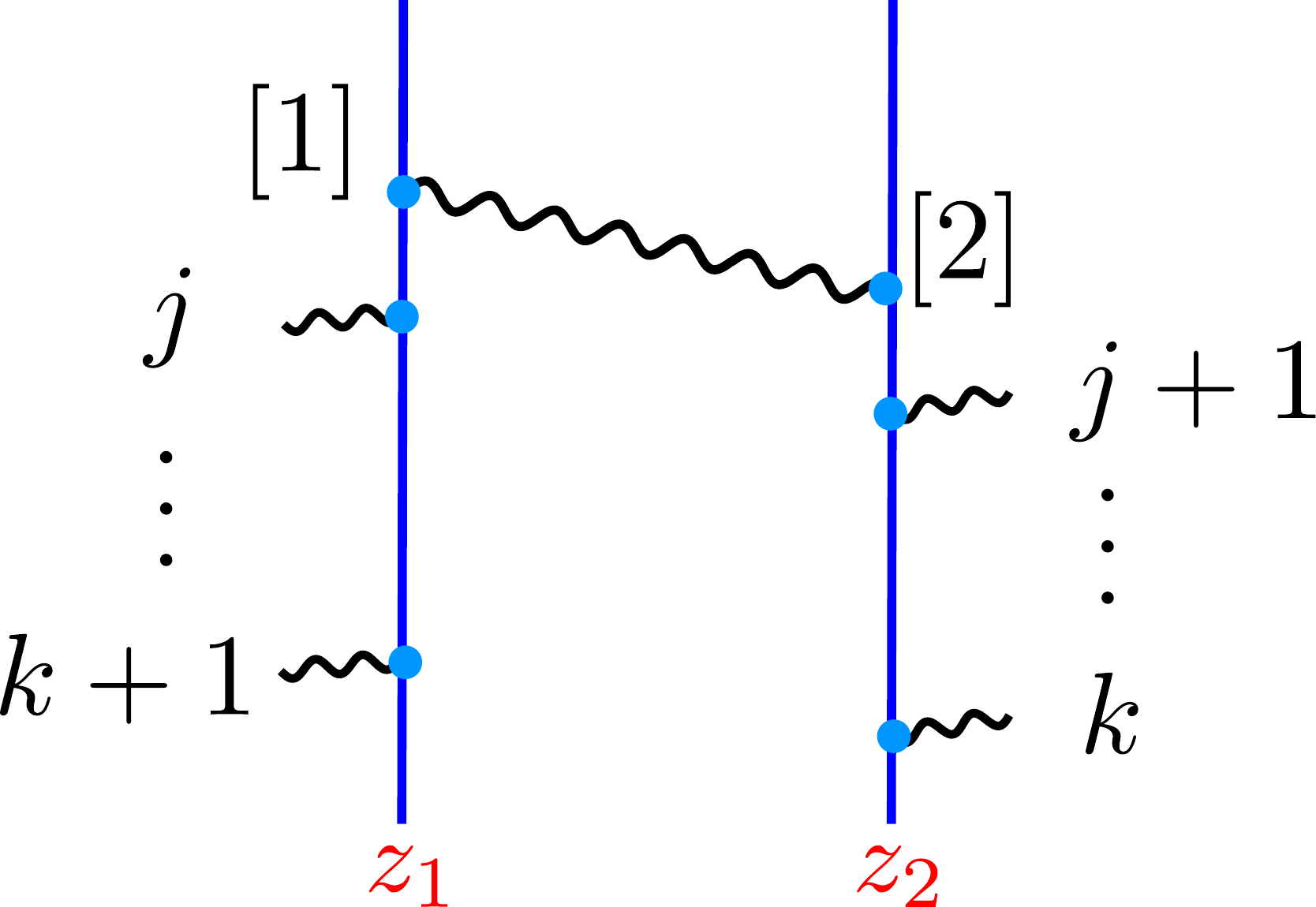}
  \caption{
  \it 
  The most general diagram for amplitudes at NMHV level in position twistor space. The points $z_1$ and $z_2$ are integrated over as required by the definition of the vertex $\textbf{V}$.}
\label{fig: general NMHV amplitude diagram}
\end{figure}

\paragraph{Amplitudes}
Let us first review the case of amplitudes \cite{Adamo:2011cb}. 
In this case, the single propagator has to connect two interaction vertices \eqref{eq:definitionoftheamplitudevertex}.
The most general such diagram is shown in figure~\ref{fig: general NMHV amplitude diagram}.%
\footnote{From now on, in figures we shall refrain from writing $\calZt_{j}$ for the supertwistors of the external on-shell states and just label them by their index $j$.} 
The interaction vertex $\mathbf{V}$ can involve two or more fields.  
However, the two-point vertex yields zero when it is directly connected to an external field, see \cite{Adamo:2011cb}. 
Hence, we can restrict ourselves to the case of at least three fields in $\mathbf{V}$, which allows us to use the inverse soft limit.
Using \eqref{eq:amplitudeverticesinversesoftlimit}, we find
\begin{equation}
\label{eq: NMHV amplitude in position twistor space}
\begin{aligned}
&\int \DD^{3|4}\calZ_{[1]}\DD^{3|4}\calZ_{[2]}\ver{\ldots, \calZt_j,\calZ_{[1]},\calZt_{k+1},\ldots}\Delta(\calZ_{[1]},\calZ_{[2]})\ver{\ldots, \calZt_{k},\calZ_{[2]},\calZt_{j+1},\ldots}\\
&=\nalpha\ver{\ldots, \calZt_{j},\calZt_{k+1},\ldots}\ver{\ldots, \calZt_k,\calZt_{j+1},\ldots}\\
&\phaneq\times \int \DD^{3|4}\calZ_{[1]}\DD^{3|4}\calZ_{[2]}
\bar{\delta}^{2|4}(\calZt_{j},\calZ_{[1]},\calZt_{k+1})
\bar{\delta}^{2|4}(\calZ_{[1]},\star,\calZ_{[2]})
\bar{\delta}^{2|4}(\calZt_{k},\calZ_{[2]},\calZt_{j+1})\\
&=\nalpha \ver{\ldots, \calZt_{j},\calZt_{k+1},\ldots}\ver{\ldots, \calZt_k,\calZt_{j+1},\ldots}[\calZt_{k+1},\calZt_j,\star,\calZt_{j+1},\calZt_{k}]\eqncom 
\end{aligned}
\end{equation}
where the five-bracket is defined as
\begin{equation}
\begin{aligned}
\label{eq:fivebracket}
[\calZ_{1},\calZ_{2},\calZ_{3},\calZ_{4},\calZ_{5}]&\equiv
\bar{\delta}^{0|4}(\calZ_{1},\calZ_{2},\calZ_{3},\calZ_{4},\calZ_{5})=
\frac{\prod_{a=1}^4(\chi_{1}(2 3 4 5)+\mathrm{cyclic})^a}{(1234)(2 3 4 5)(3451)(4512)(5123)}
\end{aligned}
\end{equation} 
with the four-bracket $(ijkl)\equiv \det(Z_i,Z_j,Z_k,Z_l)$ given by the determinant of four bosonic twistors. We recall that the five-bracket, like $\bar\delta^{i|4}$, is totally antisymmetric in its arguments.
Summing over all diagrams of this type yields the NMHV amplitude 
\begin{equation}
\label{eq: NMHV amplitude in position twistor space final}
\begin{aligned}
\mathbb{A}^{\text{NMHV}}= -\nalpha \sum_{1\leq j<k\leq n}&\ver{\ldots, \calZt_{j},\calZt_{k+1},\ldots}\\& \times [\calZt_{j},\calZt_{j+1},\star,\calZt_{k},\calZt_{k+1}]\ver{\ldots, \calZt_k,\calZt_{j+1},\ldots}\eqncom
\end{aligned}
\end{equation}
where we have used said antisymmetry.
In section~\ref{eq:NMHV amplitudes in momentum twistor space}, we shall derive the usual expression for the NMHV amplitudes in momentum twistor space from \eqref{eq: NMHV amplitude in position twistor space final} by inserting the momentum eigenstates and computing the integrals that are left implicit in $\mathbf{V}$.

\paragraph{Form factors:}
Let us now return to form factors and calculate the NMHV form factors of Wilson loops as well as of the corresponding general composite operators.
The propagator in this case has to connect a Wilson line vertex \eqref{eq: Wilson line vertex in position space} corresponding to an edge of the Wilson loop \eqref{eq:finaldefinitionWilsonloop} with an interaction vertex \eqref{eq:definitionoftheamplitudevertex}.
In the end, we have to sum over all such edges and over all possible combinations to distribute the external fields on the interaction vertex $\mathbf{V}$ and the Wilson line vertices $\mathbf{W}$ of the different edges.

To simplify the notation and keep the presentation transparent, we will consider a particular edge $X$ with vertex $\verop{X}{\calZ_{X1},\calZ_{X2}}{\calZt_{a+1},\dots,\calZt_b}$, where $\calZ_{X1}$ ($\calZ_{X2}$) denotes the twistor corresponding to the intersection with the previous (next) edge.
Concretely, $(X,\calZ_{X1},\calZ_{X2})$ is an element of the union $
\{(x_i,\calZ_i,\calZ_i')|i=1,\dots,L\}\cup
\{(x_i'',\calZ_i',\calZ_i'')|i=1,\dots,L\}\cup
\{(x_i',\calZ_{i-1}'',\calZ_i)|i=1,\dots,L\}$.
Moreover, we let $\calZt_{a+1}$ ($\calZt_b$) denote the first (last) external field on that edge.
\begin{figure}[tbp]
 \centering
  \includegraphics[height=2.3cm]{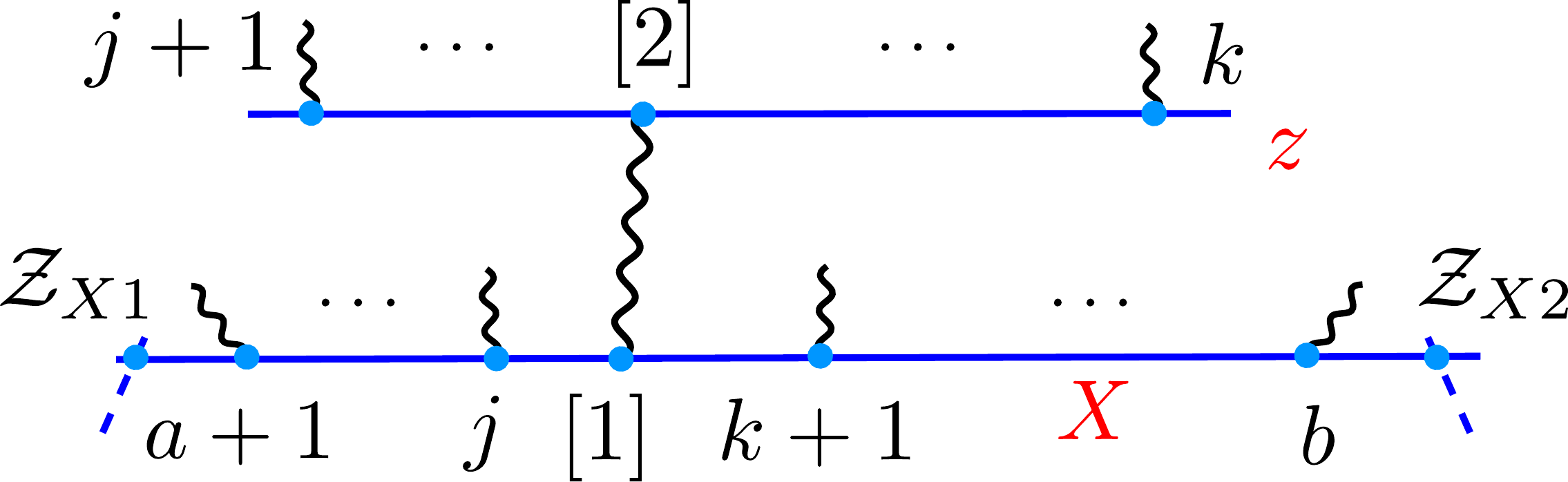}
   \caption{\it A generic NMHV diagram in which an interaction vertex $\textbf{V}$ is connected to a $\textbf{W}_X$ one. The line $X$ in twistor space contains the two twistors $\calZ_{X1}$ and $\calZ_{X2}$ at which it intersects the lines of the cogwheel Wilson loop adjacent to it.} 
  \label{fig:AllTermsNMHVGeneric}
\end{figure}
For a generic distribution of the external fields, shown in figure~\ref{fig:AllTermsNMHVGeneric}, we then find 
\begin{equation}
\label{eq: NMHV diagram for form factors}
 \begin{aligned}
&\int \DD^{3|4}\calZ_{[1]}\DD^{3|4}\calZ_{[2]}\verop{X}{\calZ_{X1},\calZ_{X2}}{\calZt_{a+1},\ldots, \calZt_j,\calZ_{[1]},\calZt_{k+1},\ldots,\calZt_b}\Delta(\calZ_{[1]},\calZ_{[2]})\\
&\phaneq\times\ver{\ldots, \calZt_{k},\calZ_{[2]},\calZt_{j+1},\ldots}\\
&=\nalpha\verop{X}{\calZ_{X1},\calZ_{X2}}{\calZt_{a+1},\ldots, \calZt_{j},\calZt_{k+1},\ldots,\calZt_b}\ver{\ldots, \calZt_k,\calZt_{j+1},\ldots}\\
&\phaneq\times \int \DD^{3|4}\calZ_{[1]}\DD^{3|4}\calZ_{[2]}\bar{\delta}^{2|4}(\calZt_{j},\calZ_{[1]},\calZt_{k+1})
\bar{\delta}^{2|4}(\calZ_{[1]},\star,\calZ_{[2]})
\bar{\delta}^{2|4}(\calZt_{k},\calZ_{[2]},\calZt_{j+1})\\
&=\nalpha\verop{X}{\calZ_{X1},\calZ_{X2}}{\calZt_{a+1},\ldots, \calZt_{j},\calZt_{k+1},\ldots,\calZt_b}\ver{\ldots, \calZt_k,\calZt_{j+1},\ldots}
\\&\phaneq\times[\calZt_{k+1},\calZt_j,\star,\calZt_{j+1},\calZt_k]\eqncom
\end{aligned}
\end{equation}
which is completely analogous to \eqref{eq: NMHV amplitude in position twistor space} except that we have also used the inverse soft limit \eqref{eq: inverse soft limit frame} for the Wilson line vertex.
We have to consider two special distributions.
If $a=j$, the external supertwistor $\calZt_j$ is not on the edge $X$ and we have to use $\calZ_{X1}$ for the inverse soft limit. We then need to replace $\calZt_j\rightarrow \calZ_{X1}$ in the five-bracket of the last line of \eqref{eq: NMHV diagram for form factors}.
Similarly, if $b=k$, the external supertwistor $\calZt_{k+1}$ is not on the edge $X$, which leads to a replacement $\calZt_{k+1}\rightarrow \calZ_{X2}$ in the five-bracket. In order to write the result in a condensed form, we shall denote by 
 \begin{equation} 
 \label{eq:definitionreplacementrule}
 \calZt_j\stackrel{j=a}{\longrightarrow} \calZ_{X1}
 \end{equation} 
  the supertwistor that is $\calZt_j$ if $j\neq a$ and becomes equal to $\calZ_{X1}$ if $j=a$.

It follows from the preceding discussion that the total position twistor space form factor is
\begin{equation}
\label{eq:NMHVWilsonloopformfactor}
\begin{aligned}
\mathbb{F}^{\text{NMHV}} &=-\nalpha
\sum_{X}\Big\{\sum_{a\leq j<k\leq b}
\verop{X}{\calZ_{X1},\calZ_{X2}}{\calZt_{a+1},\ldots, \calZt_{j},\calZt_{k+1},\ldots,\calZt_b}\\
&\phaneq\times \ver{\ldots, \calZt_k,\calZt_{j+1},\ldots}\big[\calZt_j\stackrel{j=a}{\longrightarrow} \calZ_{X1},\calZt_{j+1},\star,\calZt_k,\calZt_{k+1}\stackrel{k=b}{\longrightarrow} \calZ_{X2}\big]\Big\}\\
&\phaneq \times 
\prod_{X'\neq X}\verop{X'}{\calZ_{X1'},\calZ_{X2'}}{\ldots}\,,
\end{aligned}
\end{equation}
where we have kept the sum over all distributions of the remaining $n-b+a$ external fields on the $3L-1$ remaining edges implicit.

To compute the NMHV form factor $\mathbb{F}^{\text{NMHV}}_{\calO}$ of a specific operator $\calO$ \eqref{eq:composite operator}, we must now act with the forming factor, do the integral and take the operator limit as in \eqref{eq: summary operator vertex}. 
We will do this and rephrase the result in momentum twistor space in several cases in section~\ref{sec4}.

\subsection{\texorpdfstring{N$^k$MHV}{NkMHV} form factors}
\label{sec:NkMHVtreelevelformfactors}

Having shown how to compute the NMHV form factors in position twistor space, let us now consider the case of arbitrary high MHV degree. The discussion parallels the one for amplitudes in \cite{Adamo:2011cb}, and hence we shall mostly concern ourselves with highlighting the differences. There are three different kinds of twistor-space diagrams for amplitudes, namely {\it generic, boundary} and {\it boundary-boundary}, and the same applies to form factors. The types of diagrams differ in the relative positions of the propagators in the (interaction) vertices.

\paragraph{Generic:} For these diagrams, no adjacent propagators occur at any vertex such that they can be calculated in complete analogy to the NMHV diagrams. The corresponding contribution to the amplitude/form factor is given by products of twistor-space vertices and of R-invariants of the kind $[\calZt_{a_i},\calZt_{a_j},\star,\calZt_{a_k},\calZt_{a_l}]$ with the $\calZt_{a_n}$ indicating some external supertwistors. For some diagrams, there are no external particles to the left (right) of a propagator on the Wilson line vertex $\textbf{W}_x$.
In this case, we need to replace the external twistors $\calZt_j$ ($\calZt_{k+1}$) by the appropriate twistors fixing the line $x$, cf.~\eqref{eq:definitionreplacementrule}.

\begin{figure}[tbp]
 \centering
  \includegraphics[height=3.9cm]{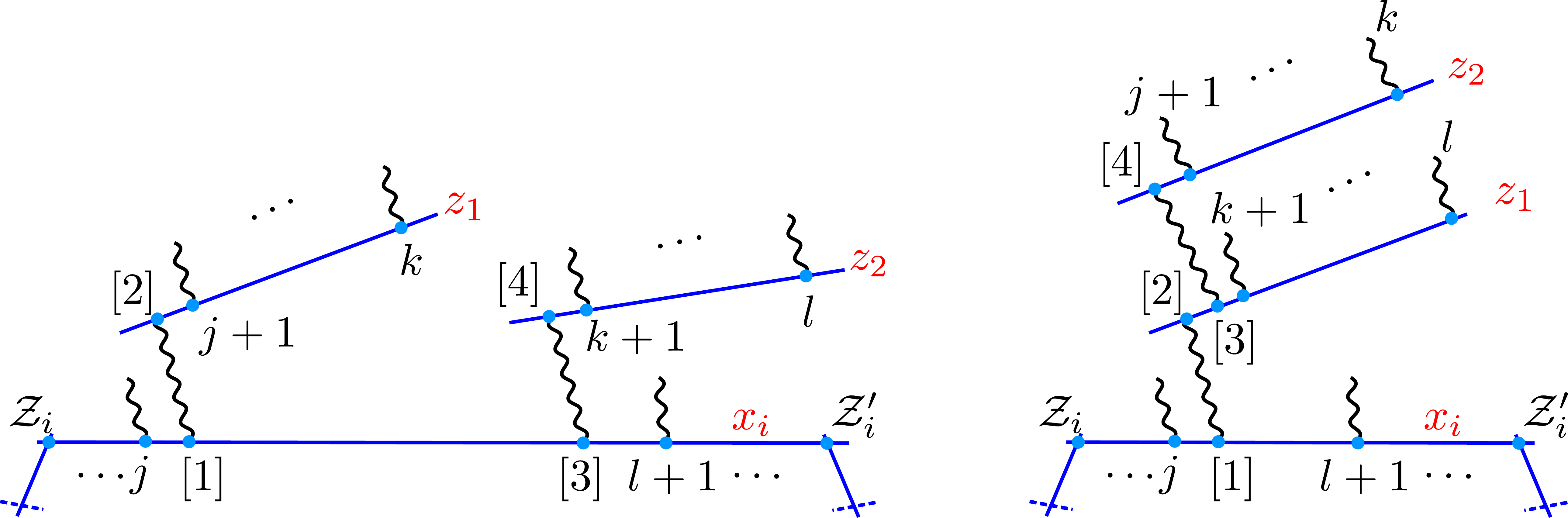}
  \caption{\it This figure depicts two boundary cases. On the left, the propagator insertions at the twistors $[1]$ and $[3]$ are adjacent, but they are on the line $x_i$ corresponding to an edge of the Wilson loop. On the right, the  twistors $[2]$ and $[3]$ are adjacent but there are at least two external particles on the line $z_1$ if $l>k+1$. 
  We denote the two interaction lines by $z_1$ and $z_2$, respectively. 
  }
  \label{fig:BoundaryTerm}
\end{figure}

\begin{figure}[tbp]
 \centering
  \includegraphics[height=4cm]{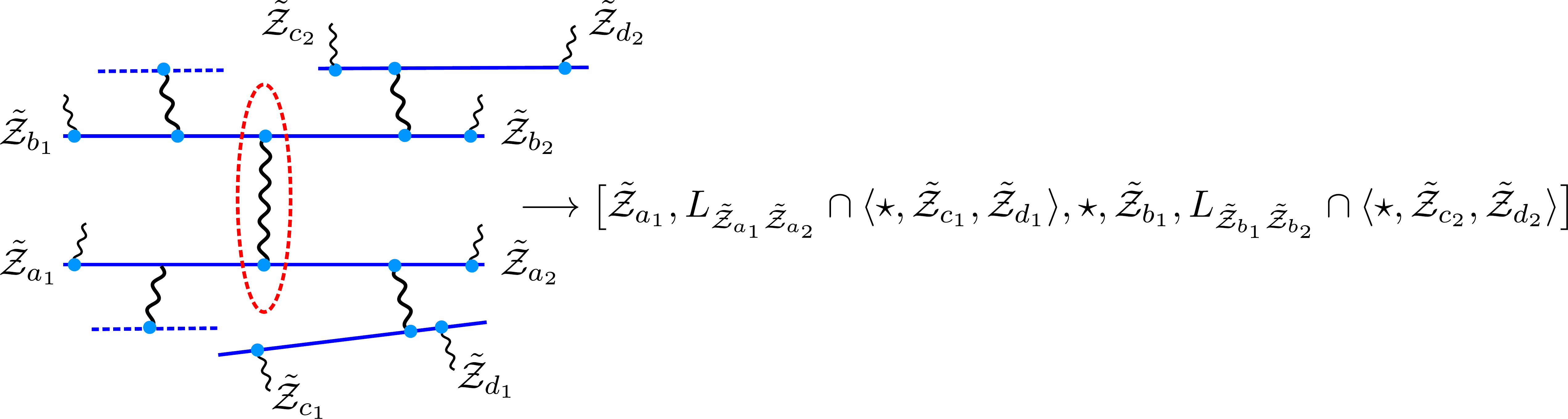}
  \caption{\it 
  The propagator encircled in red is replaced by the R-invariant \eqref{eq:ruleforboundaryRinvariant} with two twistors replaced by intersections \eqref{eq:intersectionlineplane} as indicated.
  }
\label{fig:BoundaryTermExample}
\end{figure}

\paragraph{Boundary:} In this case, some propagators are inserted next to each other, but each vertex for which that happens is either a Wilson line vertex $\mathbf{W}$ or it is an interaction vertex $\mathbf{V}$ but has at least two external particles, see figure~\ref{fig:BoundaryTerm} for one example of each. 
This allows us to use the inverse soft limit and the resulting delta functions to do all twistor integrals. 
The diagram in position twistor space thus evaluates to a product of R-invariants and twistor-space vertices $\textbf{V}$ and $\textbf{W}$. However, unlike in the generic case, the supertwistors entering the R-invariants are not the external ones but rather are obtained by simple geometric means, namely intersecting a line and a plane, as explained in the following.
Assuming that there is a propagator between the line given by the twistors 
$\calZt_{a_1}$ and $\calZt_{a_2}$ and another line given by $\calZt_{b_1}$ and $\calZt_{b_2}$, the contribution is given by the R-invariant
\begin{equation}
\label{eq:ruleforboundaryRinvariant}
[\calZt_{a_1},\skew{5}{\widehat}{\calZt}_{a_2},\star,\calZt_{b_1},\skew{5}{\widehat}{\calZt}_{b_2}]\,.
\end{equation}
Here, $\skew{5}{\widehat}{\calZt}_{a_2}=\calZt_{a_2}$ if there is no propagator to the right of the insertion twistor on the line $L_{\calZt_{a_1}\calZt_{a_2}}$ given by the twistors $\calZt_{a_1}$ and $\calZt_{a_2}$. Otherwise, $\skew{5}{\widehat}{\calZt}_{a_2}$ is the intersection of the line $L_{\calZt_{a_1}\calZt_{a_2}}$ and the plane $\langle \star, \calZt_{c_1},\calZt_{d_1}\rangle$ spanned by the twistors $\calZt_{c_1}$, $\calZt_{d_1}$ and the reference twistor $\star$. Here, $\calZt_{c_1}$ and $\calZt_{d_1}$ define the line that is connected with a propagator to the twistor on the right of the insertion twistor on the line $L_{\calZt_{a_1}\calZt_{a_2}}$:
\begin{equation}
\label{eq:intersectionlineplane}
\skew{5}{\widehat}{\calZt}_{a_2}=L_{\calZt_{a_1}\calZt_{a_2}}\cap\langle \star,\calZt_{c_1},\calZt_{d_1}\rangle=(\star,\calZt_{c_1},\calZt_{d_1},\calZt_{a_1})\calZt_{a_2}-(\star,\calZt_{c_1},\calZt_{d_1},\calZt_{a_2})\calZt_{a_1}\eqncom
\end{equation}
see figure~\ref{fig:BoundaryTermExample}. We further remark that in any computation in which some of the external twistors have to be replaced by the twistors on the corners of the Wilson loop as in \eqref{eq:definitionreplacementrule}, the replacement rule just carries through the computation, i.e.\ we simply have to replace the respective twistors in \eqref{eq:ruleforboundaryRinvariant} and \eqref{eq:intersectionlineplane} via \eqref{eq:definitionreplacementrule}.


\paragraph{Boundary-boundary:} This case is similar to the boundary one, but differs in that there are now less than two external particles at a vertex $\mathbf{V}$. Thus, we cannot use the inverse soft limit to do all twistor integrals, as the inverse soft limit cannot be applied to the two-point vertex.
It seems difficult to obtain an expression built only out of R-invariants and MHV vertices for this case, but, as mentioned in \cite{Adamo:2011cb}, the boundary-boundary case is still fully described by the twistor formalism.
As no boundary-boundary case can occur at the Wilson line vertices $\mathbf{W}$,%
\footnote{Recall that we can always use the inverse soft limit for Wilson line vertices.}
the boundary-boundary diagrams for form factors are however no more difficult than for amplitudes.

\section{From position twistors to momentum twistors} 
\label{sec4}

In the previous section, we have shown how to calculate general tree-level N$^k$MHV form factors in position twistor space.
For comparisons with the literature as well as for applications e.g.\ in unitarity calculations, it is however more useful to write the form factors in momentum space, in terms of spinor-helicity variables or even momentum twistors. 
In particular, this requires the explicit integration over the positions of all vertices.
In subsection~\ref{eq:NMHV amplitudes in momentum twistor space}, we will perform this for amplitudes, and then move on to form factors, restricting ourselves to the NMHV case for simplicity.
This is easiest for forming operators that contain only fermionic derivatives, which correspond to operators without $\dot\alpha$ indices; they trivially commute with the integrations.
We treat this case in subsection~\ref{subsec:NMHV form factors in momentum twistor space}.
In subsection~\ref{subsec: non-chiral operators}, we demonstrate using a simple example how our formalism continues to work for general operators, where the space-time derivatives that occur for operators with $\dot\alpha$ do not commute with the integration.
The evaluation of the occurring Fourier-type integrals is relegated to appendix~\ref{app:importantFourier}.

\subsection{NMHV amplitudes}
\label{eq:NMHV amplitudes in momentum twistor space}
\newcommand{\rmom}{y}
\newcommand{\rsmom}{\Gamma}
\newcommand{\mtwistor}{\mathcal{W}}

In \cite{Adamo:2011cb}, it was remarked that expression \eqref{eq: NMHV amplitude in position twistor space final} for the NMHV amplitude $\mathbb{A}^{\text{NMHV}}$ in position twistor space closely resembles the corresponding expression in momentum twistor space:
\begin{equation}
\label{eq: NMHV amplitude in momentum twistors}
\mathscr{A}^{\text{NMHV}}(1,\dots,n)= \mathscr{A}^{\text{MHV}}(1,\dots,n)\sum_{1\leq j<k\leq n}[\mtwistor_{j},\mtwistor_{j+1},\star,\mtwistor_{k},\mtwistor_{k+1}]\eqncom
\end{equation}
where $\mathscr{A}^{\text{MHV}}$ was given in \eqref{eq:MHVamplitude} and the definition of the momentum twistors $\mtwistor$ is given below.
We will now derive this remark starting from formula \eqref{eq: NMHV amplitude in position twistor space final} for the NMHV amplitudes in position twistor space. 

\begin{figure}[tbp]
 \centering
  \includegraphics[height=3.5cm]{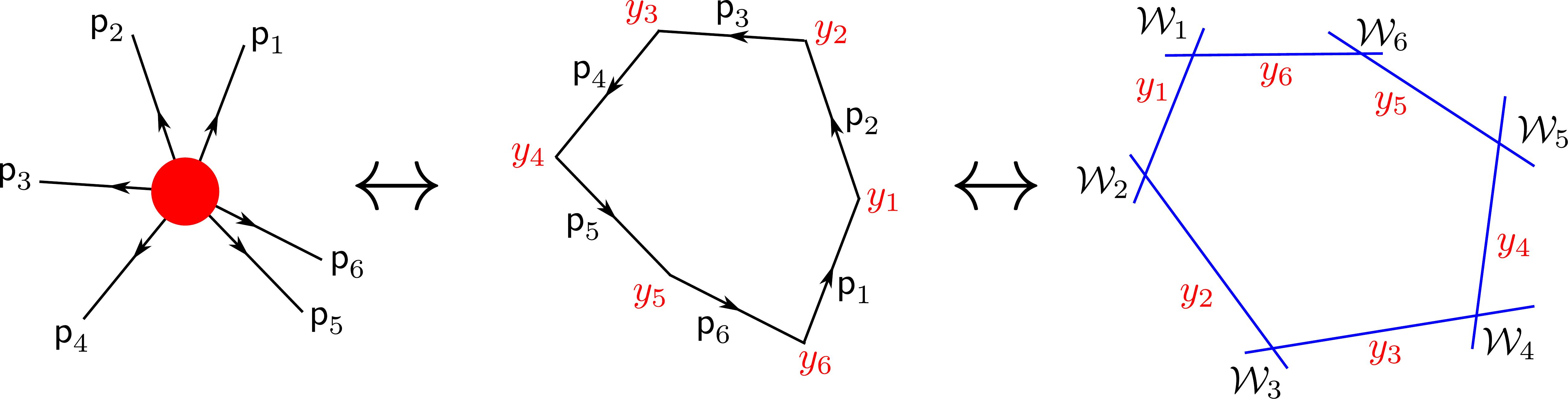}
  \caption{\it This figure illustrates the relationship between (super)momenta, region momenta and momentum twistors for $n=6$. We suppress the fermionic part.
  }
  \label{fig:MomentumTwistors}
\end{figure}

Let us first briefly review the definition of momentum twistors \cite{Hodges:2009hk}.
Starting with $n$ supermomenta $\fP_i=(\fp_i,\eta_i)$, we define dual, or region, momenta and supermomenta by
\begin{equation}
\label{eq: regional momenta and supermomenta}
 \begin{aligned}
  \fp_i^{\alpha \dot\alpha}=(\rmom_{i}-\rmom_{i-1})^{\alpha\dot\alpha}\eqncom \qquad  p_i^{\alpha}\eta_{i,a}=(\rsmom_{i}-\rsmom_{i-1})^{\alpha}_{\phantom{\alpha}a}\eqncom
 \end{aligned}
\end{equation}
see figure~\ref{fig:MomentumTwistors} for an illustration in the case $n=6$. 
The origin in dual (super)momentum space is arbitrary.
For example, we can choose
\begin{equation}
\label{eq:definitionofdualregionvariables}
 \rmom_{i}^{\alpha\dot\alpha}=\sum_{j=1}^i p_j^\alpha\bar{p}_j^{\dot\alpha}\eqncom \qquad (\rsmom_{i})^{\alpha}{}_a=\sum_{j=1}^i p_j^\alpha\eta_{j,a}
 \eqndot
\end{equation}
The momentum twistors $\mtwistor_j$ are then defined as the intersections of the lines $(\rmom_{j-1},\rsmom_{j-1})$ and $(\rmom_{j},\rsmom_{j})$ in twistor space:
\begin{equation}
\label{eq: definition momentum twistors}
 \mtwistor_j\equiv (\rmom_{j-1},\rsmom_{j-1})\cap(\rmom_{j},\rsmom_{j})=(p_{j\alpha},i \rmom_j^{\alpha\dot\alpha}p_{j \alpha},i (\rsmom_j)^{\alpha}_{\phantom{\alpha}a}p_{j\alpha})\equiv\calZ_{(\rmom_{j},\rsmom_{j})}(p_j)\eqndot
\end{equation}
As in the case of position twistors, we will frequently abbreviate $\calZ_{\rmom_{j}}(p_j)\equiv\calZ_{(\rmom_{j},\rsmom_{j})}(p_j)$.

Having introduced our notation for the momentum twistors, we can now compute the NMHV amplitudes and rephrase them in momentum twistor space. 
Inserting the momentum eigenstates \eqref{eq:definitiononshellmomentumeigenstates} into \eqref{eq: NMHV amplitude in position twistor space} 
is effectively accomplished by the following integration
\begin{equation}
 \mathscr{A}^{\text{NMHV}}(1,\dots,n)=\int \Big[\prod_{s=1}^n\DD\calZ_s\frac{\calA_{\fP_s}(\calZ_s)}{2\pi i}\Big]\mathbb{A}^{\text{NMHV}}(1,\dots,n)\eqncom
\end{equation}
cf.\ \eqref{eq: external state in position twistor space} and \eqref{eq:definitiononshellmomentumeigenstates}.
Now using the methods of for example \cite{Adamo:2011cb},
we find 
\begin{align}
\label{eq: amplitude}
\mathscr{A}^{\text{NMHV}}(1,\dots,n)
&=\nalpha\sum_{1\leq j<k\leq n}\int \frac{\dd^{4}z_1\,\dd^8\vartheta_1}{(2\pi)^4}\frac{ \dd^{4}z_2\,\dd^8\vartheta_2}{(2\pi)^4} \frac{1}{\prod_{i=1}^n\abra{p_i}{p_{i+1}}}\frac{\abra{p_j}{p_{j+1}}\abra{p_k}{p_{k+1}}}{\abra{p_{k+1}}{p_j}\abra{p_{j+1}}{p_k}}
\nonumber\\&\phaneq \times [\calZ_{(z_1,\vartheta_1)}(p_{k+1}),\calZ_{(z_1,\vartheta_1)}(p_j),\star,\calZ_{(z_2,\vartheta_2)}(p_{j+1}),\calZ_{(z_2,\vartheta_2)}(p_k)]\\&\phaneq
\times \e^{i\sum_{s={k+1}}^j(z_1\fp_s+\vartheta_1 p_s\eta_s)}\e^{i\sum_{s={j+1}}^k(z_2\fp_s+\vartheta_2 p_s\eta_s)}\eqndot \nonumber
\end{align}
Using \eqref{eq:Rinvariantforxandxprime},
we see that the five-bracket only depends on the differences $z_2-z_1$ and $\vartheta_2-\vartheta_1$.
Thus, we shift the integration variables as $z_2\to z_1+z_2$ and $\vartheta_2\to \vartheta_1+\vartheta_2$.
As a result, the five-bracket no longer depends on $z_1$ and $\vartheta_1$.
Moreover, considering the explicit form \eqref{eq:fivebracket}
of the five-bracket, the replacement allows us to use a Fourier-type integration identity derived in appendix~\ref{app:importantFourier} for $(z_2,\vartheta_2)$.
In the five-bracket, the formula \eqref{eq:superFourierRinvariant} for the $(z_2,\vartheta_2)$ integration effectively replaces 
\begin{equation}
\label{eq:replacementzvarthetawithmomentumtwistors}
 z_2\to\sum_{s=j+1}^{k}p_s\bar{p}_s = \rmom_k-\rmom_j\,,\qquad  \vartheta_2\to\sum_{s=j+1}^{k}p_s\eta_s = \rsmom_k-\rsmom_j 
\end{equation}
and multiplies by an overall factor of $\tfrac{1}{\nalpha}$.
Hence, \eqref{eq: amplitude} becomes
\begin{equation}
\label{eq: amplitude after integration}
\begin{aligned}
\mathscr{A}^{\text{NMHV}}(1,\dots,n)&= \sum_{1\leq j<k\leq n}\int \frac{\dd^{4}z_1\dd^8\vartheta_1}{(2\pi)^4} 
\frac{\e^{i(z_1\sum_{s=1}^n\fp_s+\vartheta_1\sum_{s=1}^n p_s\eta_s)}}{\prod_{i=1}^n\abra{p_i}{p_{i+1}}}\frac{\abra{p_j}{p_{j+1}}\abra{p_k}{p_{k+1}}}{\abra{p_{k+1}}{p_j}\abra{p_{j+1}}{p_k}}\\
&\phaneq \times \big[\calZ_{0}(p_{k+1}),\calZ_{0}(p_j),\star,\calZ_{\rmom_k-\rmom_j}(p_{j+1}),\calZ_{\rmom_k-\rmom_j}(p_k)\big]\,, 
\end{aligned}
\end{equation}
where we keep in mind that $\star=(0,\zeta,0)$.
The only remaining dependence on the integration variables $z_1$ and $\vartheta_1$ is in the exponential factor, which reduces to the momentum- and supermomentum-conserving delta functions.
Hence, we find 
\begin{align}
\label{eq: amplitude simplified}
&\mathscr{A}^{\text{NMHV}}(1,\dots,n)\nonumber\\&=\sum_{1\leq j<k\leq n}\frac{\delta^{4|8}(\sum_{i=1}^n\fP_i)}{\prod_{i=1}^n\abra{p_i}{p_{i+1}}}
\frac{\abra{p_j}{p_{j+1}}\abra{p_k}{p_{k+1}}}{\abra{p_{k+1}}{p_j}\abra{p_{j+1}}{p_k}}[\calZ_{0}(p_{k+1}),\calZ_{0}(p_j),\star,\calZ_{\rmom_k-\rmom_j}(p_{j+1}),\calZ_{\rmom_k-\rmom_j}(p_{k})]\nonumber\\
&=\mathscr{A}^{\text{MHV}}(1,\dots,n)
\sum_{1\leq j<k\leq n}[\calZ_{0}(p_j),\calZ_{0}(p_{j+1}),\star,\calZ_{\rmom_k-\rmom_j}(p_k),\calZ_{\rmom_k-\rmom_j}(p_{k+1})]
\eqncom
\end{align}
where in the last step we made use of \eqref{eq:Rinvariantforxandxprime} and the total antisymmetry of the five-bracket. 
The prefactor in the last step of \eqref{eq: amplitude simplified} is exactly the MHV amplitude \eqref{eq:MHVamplitude}, while the five-bracket equals 
\begin{equation}
 \label{eq: amplitude five bracket simplified}
\begin{aligned}
&[\calZ_{0}(p_j),\calZ_{0}(p_{j+1}),\star,\calZ_{\rmom_k-\rmom_j}(p_k),\calZ_{\rmom_k-\rmom_j}(p_{k+1})]\\
&=[\calZ_{\rmom_j}(p_j),\calZ_{\rmom_j}(p_{j+1}),\star,\calZ_{\rmom_k}(p_k),\calZ_{\rmom_k}(p_{k+1})]\\
&=[\calZ_{\rmom_j}(p_j),\calZ_{\rmom_{j+1}}(p_{j+1}),\star,\calZ_{\rmom_k}(p_k),\calZ_{\rmom_{k+1}}(p_{k+1})]\\
&\equiv[\mtwistor_{j},\mtwistor_{j+1},\star,\mtwistor_{k},\mtwistor_{k+1}]\,.
\end{aligned}
\end{equation}
In the first step of \eqref{eq: amplitude five bracket simplified}, we have used the invariance of the five-bracket under shifts \eqref{eq:Rinvariantforxandxprime} and in the second step the fact that, due to the definition \eqref{eq:definitionofdualregionvariables}, we have the relations
\begin{equation}
 p_{i+1}\rmom_{i+1}=p_{i+1}\rmom_{i}\eqncom \qquad p_{i+1}\rsmom_{i+1}=p_{i+1}\rsmom_{i}\eqndot
\end{equation}
Thus, we indeed obtain the desired amplitude in momentum twistor space \eqref{eq: NMHV amplitude in momentum twistors}.

\subsection{NMHV form factors for operators without \texorpdfstring{$\dot\alpha$}{dotted} indices}
\label{subsec:NMHV form factors in momentum twistor space}

Having understood how to transition from position to momentum twistor space for amplitudes in the previous subsection, we now want to do the same for the NMHV form factors of subsection~\ref{NMHVamplitudesandformfactorsinpositiontwistorspace}. 
In the case of form factors, the forming operator \eqref{eq:definitionformingfactoronshellstates} occurs inside of the (super) Fourier-type integrals over the positions of the vertices and the operator. While the fermionic $\theta_i$ derivatives in the forming operator trivially commute with the fermionic integrals, this is in general not the case for the bosonic $x_i$ derivatives and the space-time integrals -- as we will demonstrate explicitly in appendix~\ref{app:importantFourier}.
In the second half of this subsection, we restrict ourselves to operators without $\dot\alpha$ indices, i.e.\ to forming operators without $x_i$ derivatives.

\begin{figure}[tbp]
 \centering
  \includegraphics[height=3.1cm]{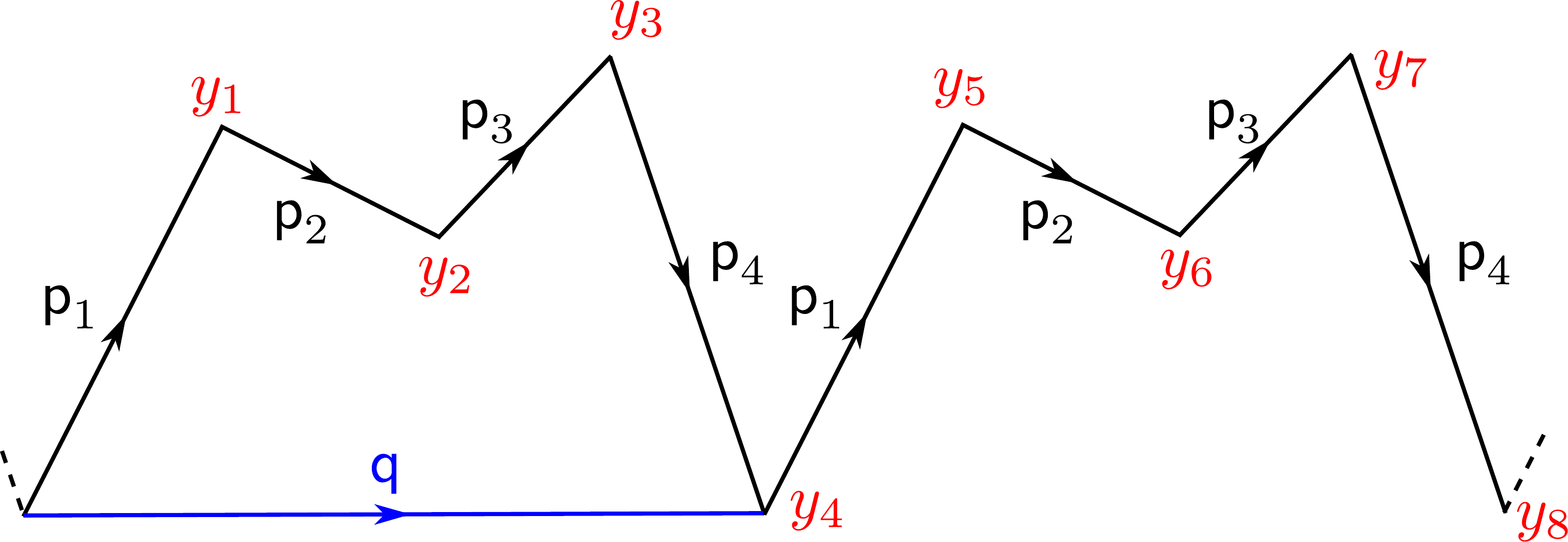}
  \caption{\it An illustration of the region momenta used for form factors. Here, we consider a case with $n=4$. }
  \label{fig:NonperiodicMomentumtwistors}
\end{figure}

One difference from the amplitude case lies in the definition of the momentum twistors for form factors. In contrast to the amplitude case, for form factors the on-shell momenta $\fp_i$ of the $n$ external on-shell states do not add up to zero but to the off-shell momentum $\fq$ of the composite operator. Hence, the contour they form in the space of region momenta is periodic instead of closed \cite{Alday:2007he}:
\begin{equation}
 \rmom_{i+n}=\rmom_i +\fq \eqnsem
\end{equation}
see figure~\ref{fig:NonperiodicMomentumtwistors}.
One way to define momentum twistors in this case is to use two periods of the contour, which results in $2n$ momentum twistors \cite{Brandhuber:2011tv,Bork:2014eqa} still defined via \eqref{eq: definition momentum twistors} but with $i=1,\dots,2n$.%
\footnote{Alternatively, the momentum twistors can be defined by closing a single period of the contour via two auxiliary on-shell momenta as done in \cite{Frassek:2015rka}. This results in $n+2$ momentum twistors but requires to consider the different possible ways to close the contour.}

The computation of the NMHV form factor parallels the calculations done for amplitudes in the preceding subsection. Inserting momentum eigenstates in $\mathbb{F}^{\text{NMHV}}$ of  \eqref{eq:NMHVWilsonloopformfactor}, including the forming operator \eqref{formingop} and taking the operator limit, we find   
\begin{equation}
\label{eq:NMHVWilsonloopformfactor momentumspace}
\begin{aligned}
\mathscr{F}^{\text{NMHV}}_{\calO}(1,\dots,n;\fq) &=\nalpha
\sum_{X}
\sum_{a\leq j<k\leq b}
\int \frac{\dd^{4}x}{(2\pi)^4}\e^{-ix\fq}\int\frac{ \dd^{4}z\,\dd^8\vartheta}{(2\pi)^4} \lim_{\hexagon\rightarrow \xdot}\bigg\{\PPP_{\mathcal{O}}
\\
&\phaneq\times \frac{1}{\prod_{i=a+1}^{b}\abra{p_i}{p_{i+1}}}
\frac{\abra{(p_j\stackrel{j=a}{\longrightarrow }\lambda_{X1})}{p_{j+1}}\abra{p_k}{(p_{k+1}\stackrel{k=b}{\longrightarrow }\lambda_{X2})}}{\abra{(p_j\stackrel{j=a}{\longrightarrow }\lambda_{X1})}{(p_{k+1}\stackrel{k=b}{\longrightarrow }\lambda_{X2})}\abra{p_k}{p_{j+1}}}
\\&\phaneq \times 
[\calZ_{X}(p_{k+1}\stackrel{k=b}{\longrightarrow }\lambda_{X2}),\calZ_{X}(p_j\stackrel{j=a}{\longrightarrow }\lambda_{X1}),\star,\calZ_{(z,\vartheta)}(p_{j+1}),\calZ_{(z,\vartheta)}(p_k)]
\\&\phaneq \times \e^{i\sum_{s={a+1}}^j(X\fp_s+\theta p_s\eta_s)}\e^{i\sum_{s={j+1}}^k(z\fp_s+\vartheta p_s\eta_s)}\e^{i\sum_{s={k+1}}^b(X\fp_s+\theta p_s\eta_s)}\\
&\phaneq\times \prod_{X'\neq X}\text{contributions from edge }X'\bigg\}
\bigg{|}_{\theta=0}\,,
\end{aligned}
\end{equation}
where we have suppressed the contributions from the other edges.
In general, we have to act with the forming operator on the last four lines before doing the integration.
In the special case that the forming operator does not contain $x_i$ derivatives, it does however commute with the integration; we can then do the integral as in the amplitude case.
We will focus on this special case for the rest of this subsection and treat a simple example of the general case in the next subsection. The fact that the integral and the $x_i$ derivatives do in general not commute is exemplified in appendix~\ref{app:importantFourier}.

Let us restrict ourselves to forming operators without $x_i$ derivatives, i.e.\ to composite operators built of irreducible fields whose forming operators \eqref{eq:definitionformingfactoronshellstates} contain only $\theta$ derivatives.
These are the scalars $\phi$, the fermions $\psi$ and the self-dual part of the field strength $F$; they require $N=n_\theta=2,3,4$ $\theta$ derivatives, respectively.
We can then commute the $(z,\vartheta)$ integral past the forming operator and evaluate this integral in complete analogy to the amplitude case treated in the previous subsection.
The full calculation involves the sum over all distributions of the external fields on the different edges, which replace $a$ and $b$ in \eqref{eq:NMHVWilsonloopformfactor momentumspace}.
We parametrize it via  $1\leq m_1\leq m_1'\leq m_1''< m_2\leq m_2'\leq m_2''<m_3\leq \cdots <m_L\leq m_L'\leq m_L''< m_1+n\leq 2n$ and abbreviate this set as $\{m_i,m_i',m_i''\}$.
The details of this calculation are quite technical, and we relegate them to appendix~\ref{app:NMHVformfactors}.
Here, we just present the result:
\begin{equation}
\label{eq:NMHFformfactorMomentumPart2}
\begin{aligned}
\mathscr{F}^{\text{NMHV}}_{\calO}(1,\dots,n;\fq)\
= {}&{}\frac{\delta^{4}\big(\sum_{r=1}^n\fp_r-\fq\big)}{\prod_{r=1}^n\abra{p_r}{p_{r+1}}}\sum_{\{m_i,m_i',m_i''\}}
\widetilde{\mathscr{F}}_{\calO}(\{m_i,m_i',m_i''\})
\\
\phaneq\times \sum_{i=1}^L \Biggl\{
\sum_{m_{i-1}''\leq j<k\leq m_i-1}&\Big[\calW_j\stackrel{j=m_{i-1}''}{\longrightarrow }\calZ_{\rmom_j}(\textbf{n}),\calW_{j+1},\star, \calW_{k},\calW_{k+1}\stackrel{k=m_i-1}{\longrightarrow }\calZ_{\rmom_k}(\tau_i)\Big]\\
+ \sum_{m_i-1\leq j<k\leq m_i'} &\Big[\calW_j\stackrel{j=m_i-1}{\longrightarrow }\calZ_{\rmom_j}(\tau_i),\calW_{j+1},\star, \calW_{k},\calW_{k+1}\stackrel{k=m_i'}{\longrightarrow }\calZ_{\rmom_k}(\tau_i)\Big]\\
+ \sum_{m_{i}'\leq j<k\leq m_i''}&\Big[\calW_j\stackrel{j=m_{i}'}{\longrightarrow }\calZ_{\rmom_j}(\tau_i),\calW_{j+1},\star, \calW_{k},\calW_{k+1}\stackrel{k=m_i''}{\longrightarrow }\calZ_{\rmom_k}(\textbf{n})\Big]
\Biggr\}\,,
\end{aligned}
\end{equation}
where we have defined
\begin{align}
\widetilde{\mathscr{F}}_{\calO}(\{m_i,m_i',m_i''\})&= \prod_{r=1}^L 
\left\{\delta_{m_{r-1}''+1,m_r}+(1-\delta_{m_{r-1}''+1,m_r})\frac{\abra{ \n}{\tau_r}}{\abra{\n}{p_{m_{r-1}''+1}}\abra{p_{m_r-1}}{\tau_r}} \right\}
\nonumber\\&\quad \times \abra{p_{m_r-1}}{p_{m_r}}
\frac{\left(\sum_{s=m_r}^{m_r'}\abra{\tau_r}{p_s}\{\xi_r\eta_s\}\right)^{N_r}
}{\abra{p_{m_r}}{\tau_r}\abra{p_{m_r'}}{\tau_r}}\abra{p_{m_r'}}{p_{m_r'+1}}
\label{eq: F tilde}
\\&\quad \times\left\{\delta_{m_r',m_r''}+(1-\delta_{m_r',m_r''})
\frac{\abra{\tau_r}{\n}}{\abra{\tau_r}{p_{m_r'+1}}\abra{p_{m_r''}}{\n}}
\right\}\abra{p_{m_r''}}{p_{m_r''+1}}\nonumber\,.
\end{align}

Two remarks are in order. First, the result \eqref{eq:NMHFformfactorMomentumPart2} still contains the auxiliary spinor  $\textbf{n}$ from the operator limit \eqref{eq:operatorlimit}.
For the MHV form factor 
\begin{equation}
\begin{aligned}
\label{eq:MHVformfactorMomentum}
\mathscr{F}^{\text{MHV}}_{\calO}(1,\ldots, n;\fq)
&=\frac{\delta^{4}\big(\sum_{r=1}^n\fp_r-\fq\big)}{\prod_{r=1}^n\abra{p_r}{p_{r+1}}}\sum_{\{m_i,m_i',m_i''\}}
\widetilde{\mathscr{F}}_{\calO}(\{m_i,m_i',m_i''\})\,,
\end{aligned}
\end{equation}
we have explicitly shown in \cite{Koster:2016loo} how $\textbf{n}$ drops out thanks to a repeated application of the Schouten identity. 
In the $n=(L+1)$-point case,  the spinor $\textbf{n}$ trivially drops out of the NMHV form factor due to the summation ranges and the Kronecker deltas in \eqref{eq:NMHFformfactorMomentumPart2} and \eqref{eq: F tilde}.
We leave a proof of the independence of $\mathscr{F}^{\text{NMHV}}_{\calO}$ on $\textbf{n}$ in the general case for future work.
Moreover, \eqref{eq:NMHFformfactorMomentumPart2} contains the reference twistor $\calZ_\star$. As in the case of general amplitudes, it is not immediate to see how $\calZ_\star$ drops out.

Second, we have numerically checked in several cases that \eqref{eq:NMHFformfactorMomentumPart2} reproduces the known result \cite{Brandhuber:2011tv,Bork:2014eqa} for the chiral half of the stress-tensor supermultiplet $T$:
\begin{equation}
\label{eq: T NMHV form factor}
 \mathscr{F}^{\text{NMHV}}_{T}(1,\ldots,n;\fq)= \mathscr{F}^{\text{MHV}}_{T}(1,\ldots,n;\fq) \sum_{j=1}^{n}\sum_{k=j+2}^{n+j-1} \big[\calW_j,\calW_{j+1},\star,\calW_k,\calW_{k+1}\big]\eqncom
\end{equation}
which factorizes similarly to the amplitude \eqref{eq: NMHV amplitude in momentum twistors}. 
It would be desirable to find a factorized form of \eqref{eq:NMHFformfactorMomentumPart2} also for general operators.%
\footnote{For more general operators, the immediate generalization of \eqref{eq: T NMHV form factor} is \emph{not} true.
}

\subsection{NMHV form factors for general operators}
\label{subsec: non-chiral operators}

We now turn to form factors of general operators, for which also $x_i$ derivatives occur in the forming operator \eqref{eq:definitionformingfactoronshellstates}.
We show in appendix~\ref{app:importantFourier} that we cannot commute these derivatives out of the integral, which complicates the calculation.
To demonstrate that our formalism nevertheless continues to work in this case, we consider a simple example.

We study the next-to-minimal (i.e.\ $(L+1)$-point) NMHV form factor of the operator $\Tr(\bar\psi_2(\phi_{13})^{L-1})$ and consider the external state $1^{\phi_{12}}2^{\psi_{234}}3^{\phi_{13}}\dots(L+1)^{\phi_{13}}$.
The advantage of this setup is that only one diagram contributes, in which $\bar\psi_2\rightarrow \phi_{12}\psi_{234}$ and which in particular must be gauge invariant.
This diagram is the counterpart of \eqref{eq:NMHVWilsonloopformfactor momentumspace} with $a=j=n=L+1$, $j+1=1$ and $k=b=2$.
Picking up the calculation at this point, we find
\begin{align}
\label{eq: start of calculation}
&\calF_{\Tr(\bar\psi_2(\phi_{13})^{L-1})}(1^{\phi_{12}}2^{\psi_{234}}3^{\phi_{13}}\dots(L+1)^{\phi_{13}};x)\nonumber\\
&=\nalpha\lim_{\hexagon\rightarrow \xdot}\Bigg\{\frac{\partial^5}{\partial \eta_1^1\partial \eta_1^2\partial \eta_2^2\partial \eta_2^3\partial\eta_2^4}\int \frac{\dd^{4}z\dd^8\vartheta}{(2\pi)^4} \frac{\e^{iz(\fp_1+\fp_2)+i\vartheta(p_1\eta_1+p_2\eta_2)}}{\abra{p_1}{p_2}\abra{p_2}{p_1}}\nonumber\\
&\phaneq\times
\left(-\frac{\lambda_{X1}^\alpha \lambda_{X2}^{\beta}}{\abra{\lambda_{X1}}{\lambda_{X2}}}\right)
\left(-i\bar{\tau}^{\dot\alpha}\frac{\partial}{\partial x^{\alpha\dot{\alpha}}}\right)
\left(-i\xi^a\frac{\partial}{\partial \theta^{\beta a }}\right)\\
&\phaneq\times
 \big[\calZ_{(x,\theta)}(\la_{X2}),\calZ_{(x,\theta)}(\la_{X1}),\star,\calZ_{(z,\vartheta)}(p_1),\calZ_{(z,\vartheta)}(p_2)\big] \Bigg\}\Bigg|_{\xi^a=\delta^{a2},\theta=0} \e^{ix\sum_{i=3}^{L+1}\fp_i}\nonumber\\
&=- \nalpha \int\frac{\dd^{4}z}{(2\pi)^4} \e^{iz(\fp_1+\fp_2)}\left[\left(-i\tau^{\alpha}\bar{\tau}^{\dot\alpha}\frac{\partial}{\partial x^{\alpha\dot\alpha}}\right)\frac{1}{(x-z)^2}\frac{\bradot{p_2}{x-z}{\zeta}}{\bradot{\tau}{x-z}{\zeta}}\right]\e^{ix\sum_{i=3}^{L+1}\fp_i}\,,\nonumber
\end{align}
where the $\eta$ derivatives serve to select the $\phi_{12}\psi_{234}$ component of the super form factor and the only effective contribution of the other edges is the phase $\e^{ix\sum_{i=3}^{L+1}\fp_i}$. 
The polarization vectors $\tau$ and $\bar\tau$ correspond to the field $\bar\psi_2$ and we have dropped the index as the polarization vectors of all other fields have already dropped out. We remind the reader that in the operator limit $\lambda_{X1}\rightarrow \tau$  and $\lambda_{X2}\rightarrow \tau$.
Thanks to the following identity,
\begin{equation}
\label{eq:cuteidentity}
\bradot{A}{x}{C}\bradot{B}{x}{D}-\bradot{A}{x}{D}\bradot{B}{x}{C}=-x^2\abra{A}{B}\sbra{C}{D}\,,
\end{equation}
we obtain after evaluating the derivative inside of the $z$ integral
\begin{align}
\label{eq:fpsi1}
&\calF_{\Tr(\bar\psi_2(\phi_{13})^{L-1})}(1^{\phi_{12}}2^{\psi_{234}}3^{\phi_{13}}\dots(L+1)^{\phi_{13}};x)\nonumber\\
&=-(4\pi)^2\int \frac{\dd^4z}{(2\pi)^4}\frac{\e^{iz(\fp_1+\fp_2)}}{(x-z)^4}\frac{(x-z)^2\abra{\tau}{p_2}\sbra{\zeta}{\bar\tau}-\bradot{\tau}{x-z}{\bar\tau}\bradot{p_2}{x-z}{\zeta}}{\bradot{\tau}{x-z}{\zeta}}\e^{ix\sum_{i=3}^{L+1}\fp_i}\nonumber\\
&=-(4\pi)^2\int \frac{\dd^4z}{(2\pi)^4} \e^{iz(\fp_1+\fp_2)}\frac{\bradot{p_2}{z-x}{\bar\tau}}{(z-x)^4}\e^{ix\sum_{i=3}^{L+1}\fp_i}\nonumber\\
&=-\frac{\e^{ix\sum_{i=1}^{L+1}\fp_i}}{(\fp_1+\fp_2)^2} \bradot{p_2}{\fp_1+\fp_2}{\bar\tau}\,,
\end{align}
where we have used \eqref{eq:IxMinkowskiFinalResult}. 
Fourier transforming in $x$ and using $\bradot{p_2}{\fp_1+\fp_2}{\bar\tau}=\abra{p_1}{p_2}\sbra{\bar \tau}{\bar p_1}$ as well as  $(\fp_1+\fp_2)^2=-\abra{p_1}{p_2}\sbra{\bar p_1}{\bar p_2}$, we find
\begin{equation}
\calF_{\Tr(\bar\psi_2(\phi_{13})^{L-1})}(1^{\phi_{12}}2^{\psi_{234}}3^{\phi_{13}}\dots(L+1)^{\phi_{13}};\fq)=\delta^4\left(\fq-\sum_{i=1}^{L+1}\fp_i\right)\frac{\sbra{\bar\tau}{\bar{p}_1}}{\sbra{\bar{p}_1}{\bar{p}_2}}\, .
\end{equation}
This is indeed the expected gauge-invariant expression; it can be obtained e.g.\ from the MHV form factor $\calF_{\Tr(\psi_{134}(\phi_{24})^{L-1})}(1^{\phi_{34}}2^{\bar\psi_{1}}3^{\phi_{24}}\dots(L+1)^{\phi_{24}};\fq)$ by conjugation.

In the above computation, it was essential to first act with the space-time derivative in the forming operator before integrating out the interaction, so that we may simplify the expression and in particular get rid of the reference twistor through \eqref{eq:cuteidentity}. In appendix~\ref{subsec:non-chiralFourierAppendix}, we show that reversing the order of integration and derivative does not work. 
It would be desirable to be able to evaluate the integrals occurring in \eqref{eq:NMHVWilsonloopformfactor momentumspace} for general operators or to find a general form for the commutator of the forming operator and the integrals that allows for a more efficient evaluation.
We leave this for future work.

\section{Generalized form factors and correlation functions}
\label{sec:asketchforloops}

In this section, we will look at quantities that contain more than one composite operator, namely generalized form factors and correlation functions.

Generalized form factors contain $m$ composite operators and $n$ external on-shell fields, see \cite{Engelund:2012re}. 
For the case $n=0$, they reduce to correlation functions.
Similar to form factors, we can calculate generalized form factors and correlation functions of general local composite operators in twistor space using the operator vertices \eqref{eq: summary operator vertex}. 
Equally, we can calculate generalized form factors and correlation functions of Wilson loops and extract the corresponding expressions for composite operators by applying the forming operators \eqref{formingop} and taking the operator limit $\hexagon\rightarrow \xdot$ for every space-time point -- provided we are careful in treating the situations where integrals and forming operators do not commute as discussed in the last section.

\begin{figure}[htbp]
 \centering
  \includegraphics[height=3cm]{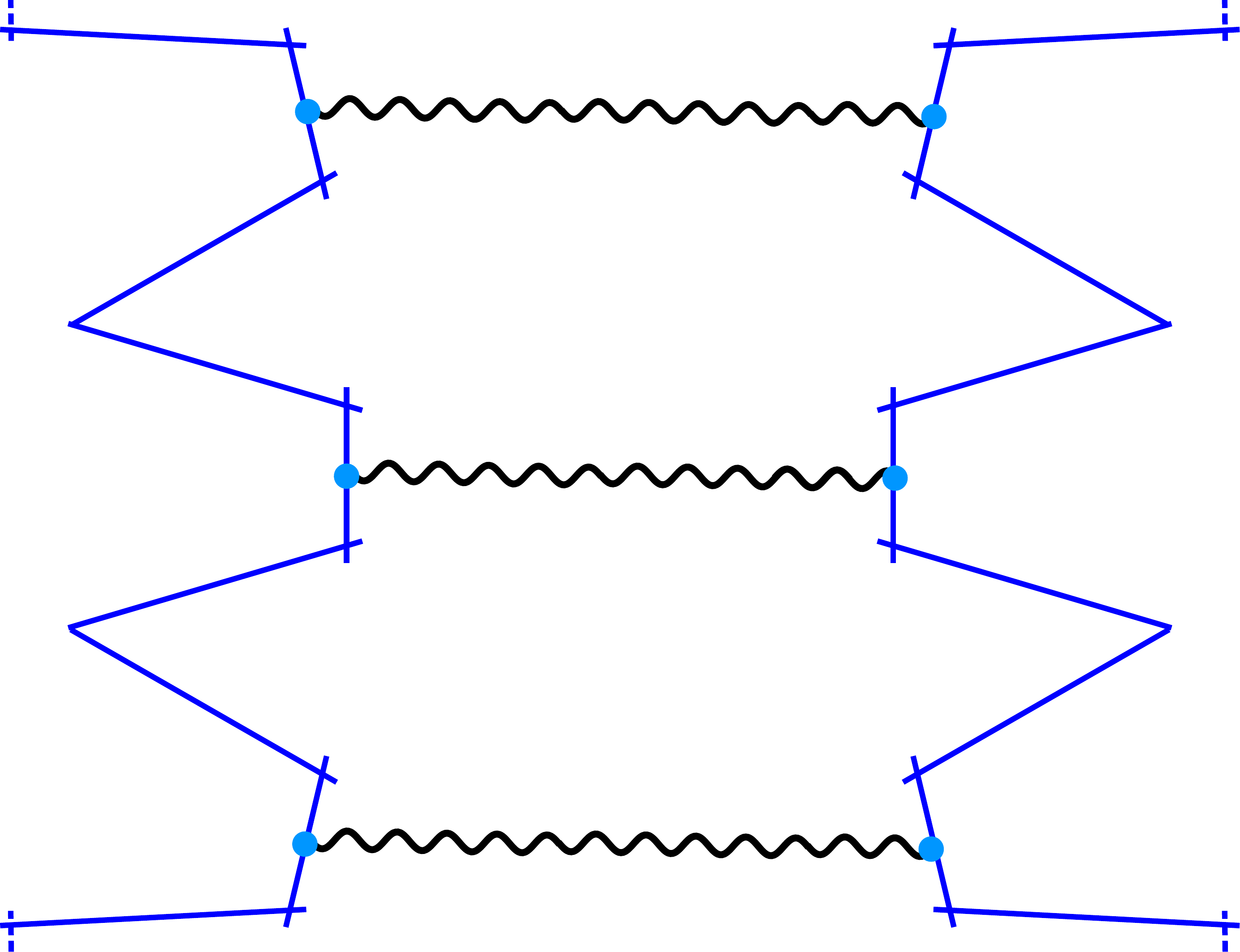}
  \caption{\it A diagram for the tree-level two-point correlation function.}
  \label{fig:TreeLevelCorrelation}
\end{figure}

Let us consider correlation functions in some generality. 
At tree level, the correlation function of two local composite operators before taking the operator limit is shown in figure~\ref{fig:TreeLevelCorrelation}.
By a trivial counting of Gra\ss{}mann numbers, one sees that at tree level only the minimal operator vertices  contribute.%
\footnote{These are given in (4.1) of \cite{Koster:2016loo}.}
Operators of different lengths have vanishing correlation functions at tree level.
For two operators of the same length $L$, the tree-level correlation function can be calculated using the inverse soft limit. Using underlined symbols for the second operator, we find
\begin{equation}
 \langle \calO(x) \underline{\calO}(y)\rangle_{\text{tree-level}} = \lim_{\hexagon\rightarrow \xdot_x,\underline{\hexagon}\rightarrow \xdot_y}\PPP_{\mathcal{O}}\underline{\PPP}_{\underline{\mathcal{O}}}
 \sum_{\sigma}
 \prod_{i=1}^L\big[\calZ_i,\calZ_i',\star,\underline{\calZ}_{\sigma(i)},\underline{\calZ}_{\sigma(i)}'\big]
 {\big{|}}_{\theta,\underline{\theta}=0}\eqncom
\end{equation}
where the permutation $\sigma$ is cyclic in the planar limit. We immediately see that this is zero unless complementary fermionic derivatives act on all pairs $i$ and $\sigma(i)$. 

\begin{figure}[htbp]
 \centering
  \includegraphics[height=3cm]{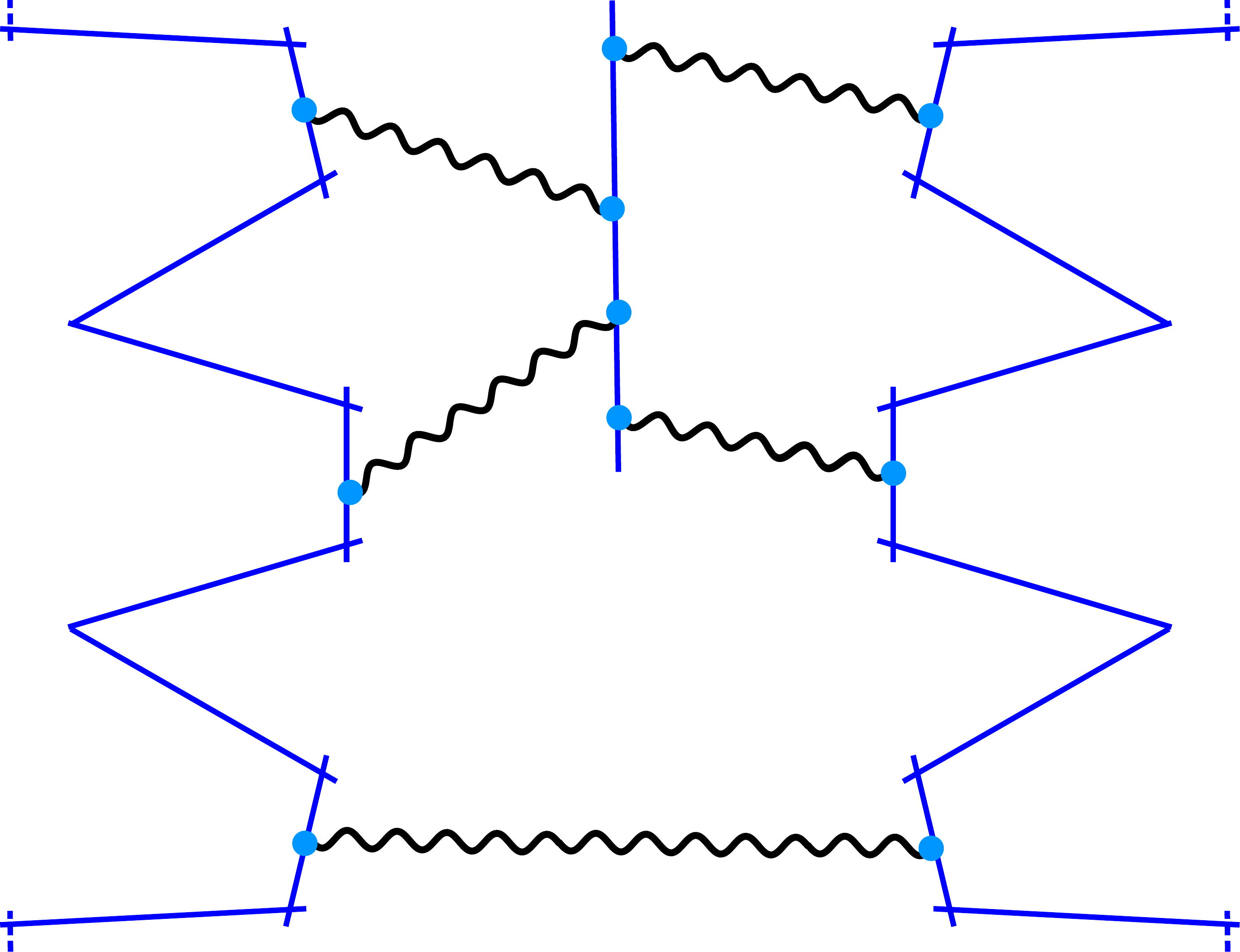}
  \caption{\it A diagram with a four-point vertex contributing to the one-loop two-point correlation function.}
  \label{fig:oneloopCorrelation4}
\end{figure}

\begin{figure}[htbp]
 \centering
  \includegraphics[height=2.4cm]{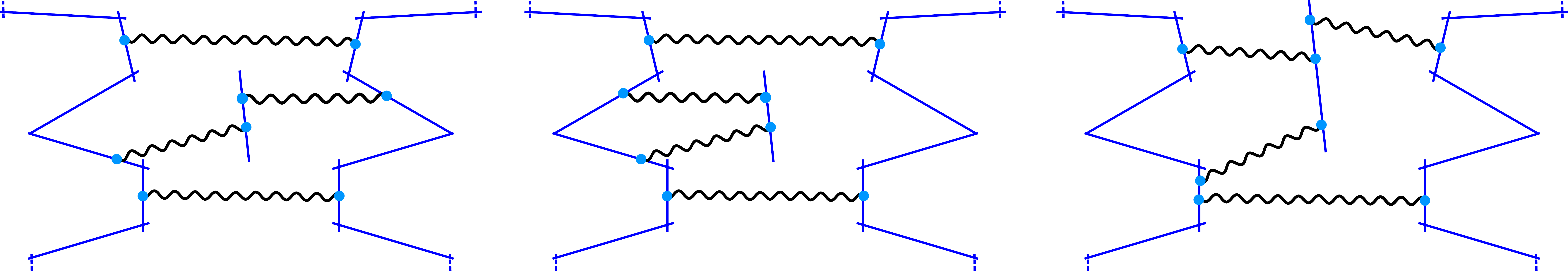}
  \caption{\it Two diagrams with a two-point vertex and one with a three-point vertex contributing to the one-loop two-point correlation function.}
  \label{fig:oneloopCorrelation3}
\end{figure}

At one-loop order, we have one interaction vertex in addition to the Wilson loops.
Counting the Gra\ss{}mann degree, we find that several diagrams contribute.
The first one consists of a four-point interaction vertex connected to minimal operator vertices, an example of which is shown in figure~\ref{fig:oneloopCorrelation4}.
The second possibility consists of a three-point interaction vertex connected to a minimal operator vertex and a next-to-minimal operator vertex, an example of which is shown on the right-hand side of figure~\ref{fig:oneloopCorrelation3}.
The third possibility consists of a two-point interaction vertex connected to two next-to-minimal operator vertices, an example of which is shown in on the left-hand side of figure~\ref{fig:oneloopCorrelation3}. The fourth possibility is a two-point vertex connected to a next-to-next-to-minimal operator vertex, an example of which is shown in the middle of figure~\ref{fig:oneloopCorrelation3}.
Note that the figures show the correlation functions before taking the operator limit, such that the contraction with the minimal operator vertex receives contributions from the additional field being connected to several different edges of the Wilson loop, of which only one is shown.


In \cite{Koster:2014fva}, some of the present authors calculated the one-loop dilatation operator in the SO$(6)$ sector via the UV divergence of the one-loop two-point correlation functions in twistor space. 
It was stated that the contributions to the dilatation operator stem only from the diagram with the four-point vertex shown in figure~\ref{fig:oneloopCorrelation4}, see \cite{Koster:2014fva} for details of this calculations.
In appendix~\ref{sec:thethreevertexinthetwopointcorrelator}, we provide the argument promised in \cite{Koster:2014fva} that the contributions of the three-vertices and two-vertices shown in figure~\ref{fig:oneloopCorrelation3} vanish.

\section{Conclusion and outlook}
\label{sec:conclusion}

In this article, we have shown
how to apply the general formalism for composite operators in position twistor space developed in \cite{Koster:2016ebi, Koster:2016loo} to N$^k$MHV form factors as well as to correlation functions.

In \textit{position twistor space}, the form factors of general composite operators can be efficiently calculated using the twistor action in close analogy to the amplitude case treated in \cite{Adamo:2011cb}. 
This calculation is based on the inverse soft limit for the position-twistor-space vertices, which we have extended to composite operators in section~\ref{subsec: inversesoftlimit}. In section~\ref{sec3}, we wrote the NMHV form factors in position twistor space explicitly and described how to generalize them to the N$^k$MHV case. In order to obtain explicit results for form factors in \textit{momentum (twistor) space}, the implicit Fourier-type integrals over the positions of the interactions have to be calculated.
After revisiting the case of NMHV amplitudes, we have shown how this can be efficiently done for NMHV form factors of operators whose forming operators contain no space-time derivatives, i.e.\ for operators without $\dot\alpha$ indices.
In this case, we can commute the forming operator past the aforementioned integrals, and our final result in this case is given in \eqref{eq:NMHFformfactorMomentumPart2}.
For composite operators $\calO$ whose forming operators contain space-time derivatives, our formalism works as well but requires to act with the forming operator before the integration,
as we illustrate in section~\ref{subsec: non-chiral operators}. 
Finally, in section~\ref{sec:asketchforloops} we looked at the computation of generalized form factors and correlation functions of composite operators using twistor-space techniques. 
The last part on one-loop correlation functions in particular closes a gap in an argument in \cite{Koster:2014fva} of three of the present authors, as previously promised.

Let us now conclude with an overview of possible directions of future research. We begin by emphasizing that the points raised in the outlook of our last article \cite{Koster:2016loo}, namely the connection to integrability, the Yangian symmetry of Wilson loops, multi-loop correlation functions in twistor space and the application of twistor-space techniques to theories with less supersymmetry, remain as pertinent as before. 
Furthermore, the results of our paper allude to several new open questions worth investigating.  

First, in section~\ref{sec:NkMHVtreelevelformfactors}, we discussed the computation of form factors in position twistor space. The method relies on the computation of amplitudes and there are some points, concerning specifically the boundary-boundary case, in the position twistor space calculation of amplitudes beyond the NMHV level that were left open in \cite{Adamo:2011cb}. It would be interesting to settle these point w.r.t.\ the application to form factors. 

Second, we have seen in section~\ref{subsec: non-chiral operators} that the Fourier-type integrations that we needed to perform in order to obtain the form factors of operators with $\dot\alpha$ indices in momentum (twistor) space are tricky and in particular do not commute with the application of the forming operator. It would be very interesting to derive a closed expression for the commutator in order to obtain a
generating object such that the tree-level form factors in momentum space are given by the application of a derivative operator on that object in analogy to what we have achieved at MHV level and for operators without space-time derivatives at NMHV level. Related is the problem of finding explicit expressions for the general Fourier-type integrals that occur for composite operators whose forming operators contain space-time derivatives.

Third, one should connect the results of our article with some new techniques for the evaluation of form factors, such as the connected prescription of \cite{He:2016dol,Brandhuber:2016xue,He:2016jdg} and Gra\ss{}mannian integrals for form factors \cite{Bork:2015fla,Frassek:2015rka,Bork:2016hst, Bork:2016xfn, Bork:2016egt}. This is also a potential pathway towards connections with integrability.

Fourth, it would be important to relate further to the LHC formalism of \cite{Chicherin:2016fac, Chicherin:2016fbj}. The general MHV form factors we  computed \cite{Koster:2016loo} were rederived in \cite{Chicherin:2016qsf} and it would be interesting to see if the LHC methods offer any improvements over the twistor-space ones for NMHV form factors and if they offer any insights on the non-commutativity of integral and derivative encountered in section~\ref{subsec: non-chiral operators}.

Finally, it would be desirable to apply the techniques presented here to higher-point correlation functions, in particular with regard to the recent progress in three-point functions.

\section*{Acknowledgments}

We are grateful to Tim Adamo, Johannes Henn, Simon Caron-Huot,  Arthur Lipstein, Lionel Mason and Erik Panzer for helpful discussions and comments. This research is supported in part by the SFB 647 \emph{``Raum-Zeit-Materie. Analytische und Geometrische Strukturen''}. V.M.\ is also supported by the PRISMA cluster of excellence at the Johannes Gutenberg University in Mainz. V.M.\ would like to thank the Simons Summer Workshop 2015, where part of this work was performed. 
M.W.\ was supported  in  part  by  DFF-FNU  through grant number DFF-4002-00037 and in part by the ERC-Advance grant 291092. L.K.\ would like to thank the CRST at Queen Mary University of London and especially Gabriele Travaglini for hospitality during an early stage of the preparation of this work.

\appendix
\section{Computation of the NMHV form factors}
\label{app:NMHVformfactors}

In this appendix, we elaborate on the notationally more involved aspects in the calculation of the 
 NMHV form factors, both in position and in momentum twistor space.

\subsection{Position twistor space}

We start with the calculation in position twistor space.

In order to compute the form factor, we use the cogwheel-shaped Wilson loop introduced in \cite{Koster:2016loo},%
\footnote{See in particular appendix C therein.}
which we briefly reviewed in section~\ref{subsec: compositeoperatorsfromwilsonloops}.
It has three families of edges $x_i$, $x_i'$ and $x_i''$, cf.\ figure~\ref{fig:CogwheelZoom}.
Each of these edges gives rise to a vertex $\mathbf{W}$.
In order to obtain an $n$-point form factor, we have to sum over all ways to distribute the $n$ external fields on the edges.
Let $m_i$ be the first external field emitted on $x_i$, $m_i'$ the last external field emitted on $x_i$ and $m_i''$ the last external field emitted on $x_i''$.%
\footnote{External fields that are emitted on an interaction line $z$ connected to $x_i$ by a propagator are also considered to be emitted on the edge $x_i$ for this purpose, and analogously for the other edges and the case of several interaction lines.}
We then have to sum over 
\begin{equation}
\label{eq:constraintsforthems}
 1\leq m_1\leq m_1'\leq m_1''< m_2\leq m_2'\leq m_2''<m_3\leq \cdots <m_L\leq m_L'\leq m_L''< m_1+n\leq 2n\eqncom
\end{equation}
where $m_i\leq m_i'$ ensures that there is at least one external field emitted from $x_i$; contributions with no  field on $x_i$ are annihilated by the forming operator \eqref{eq:definitionformingfactoronshellstates}.

\begin{figure}[tbp]
 \centering
  \includegraphics[height=3.2cm]{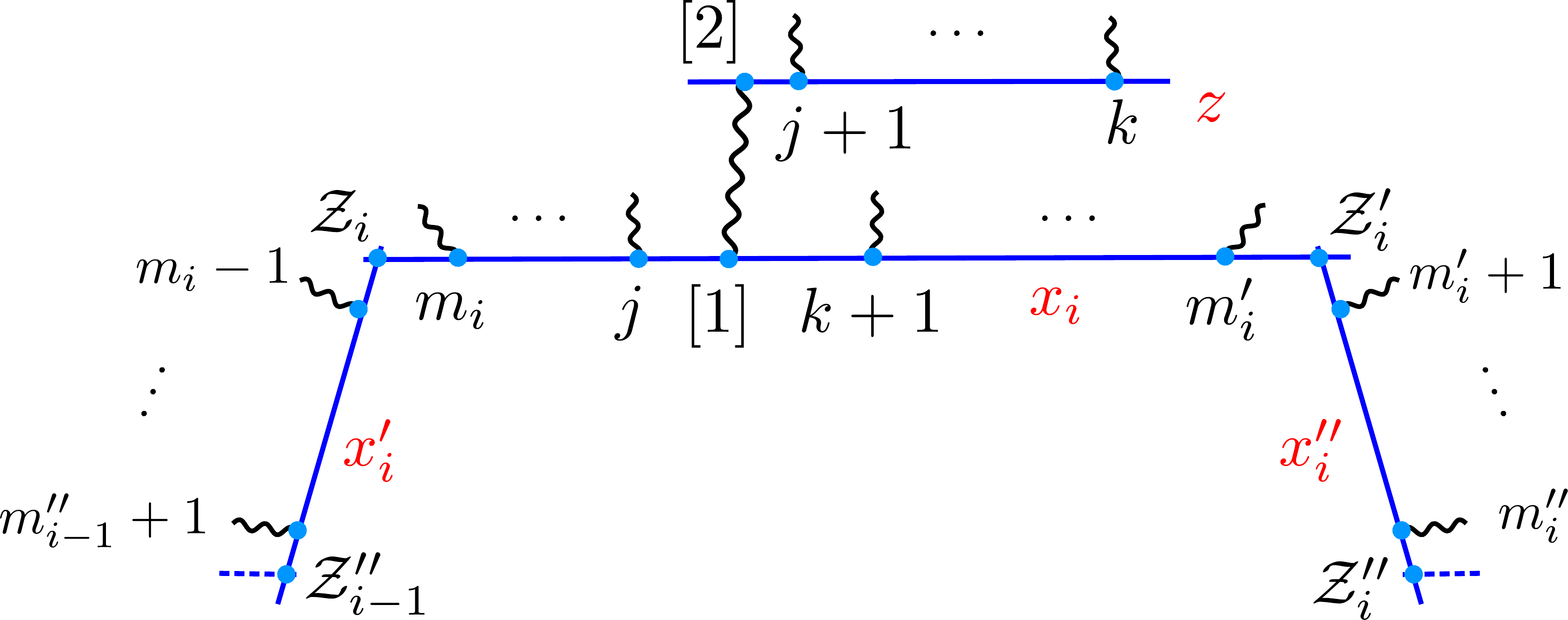}
  \caption{\it An NMHV diagram of type $\bbA$ -- the interaction vertex is connected to the operator-bearing edge $x_i$ of the Wilson loop.}
  \label{fig:AllTermsNMHV1}
\end{figure}

\begin{figure}[tbp]
 \centering
  \includegraphics[height=4.2cm]{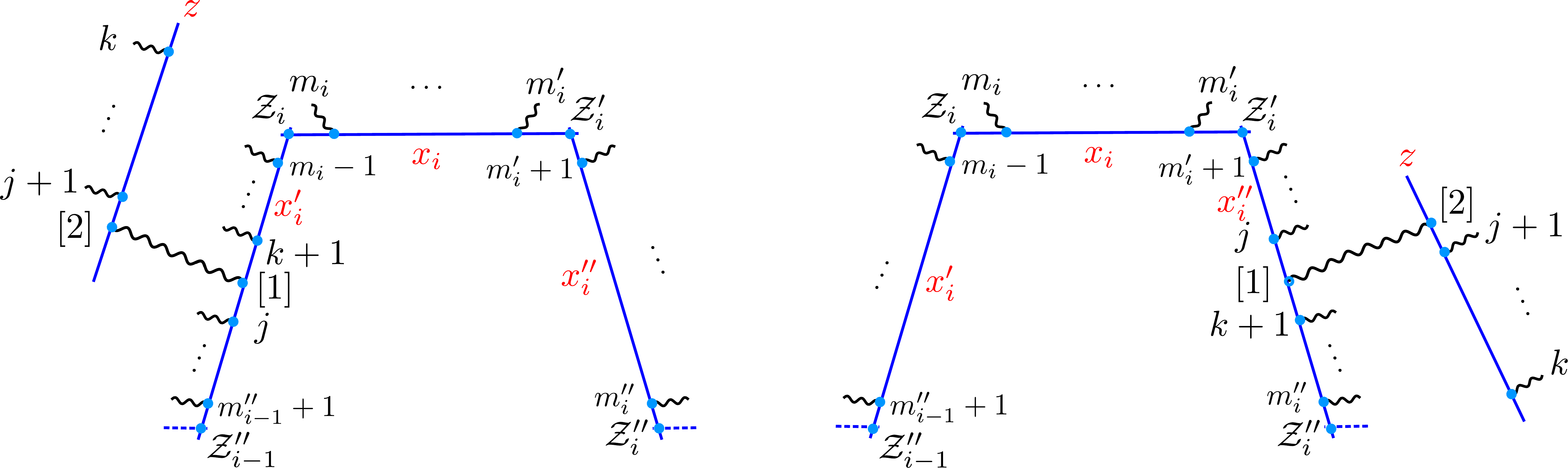}
  \caption{\it The NMHV diagrams of type $\bbB$ and $\bbC$ -- the interaction vertex is connected to the edges $x_i'$ and $x_i''$ of the Wilson loop, which do not carry the operator.}
  \label{fig:AllTermsNMHV2}
\end{figure}

For each cog, we divide the computation of the NMHV form factors into three pieces, illustrated in figure~\ref{fig:AllTermsNMHV1} and figure~\ref{fig:AllTermsNMHV2}. In the first diagram shown in figure~\ref{fig:AllTermsNMHV1}, we need to restrict the range of the parameters as $m_i-1\leq j<k\leq m_i'$, where $j=m_i-1$ implies that the propagator is attached to the leftmost twistor on the line $x_i$, while $k=m_i'$ means that the propagator is attached to the rightmost one. In addition, $k=j+1$ means that we are dealing with a two-point interaction vertex; as in the case of amplitudes, this can be dropped.
Similarly, we need to have $m_{i-1}''\leq j<k\leq m_i-1$ for the left diagrams in figure~\ref{fig:AllTermsNMHV2}, while for the right one we need $m_i'\leq j<k\leq m_i''$. 


In \eqref{eq:NMHVWilsonloopformfactor}, we have given an expression for the interaction vertex connected to a general edge. Specifying \eqref{eq:NMHVWilsonloopformfactor} to the three cases above and suppressing the vertices of the other cogs,
the diagrams of figure~\ref{fig:AllTermsNMHV1} contribute as
\begin{equation}
\label{eq:AcontributiontoNMHVWilsonloopformfactor}
\begin{aligned}
\bbA_i&=\nalpha\sum_{m_i-1\leq j<k\leq m_i'}\verop{x_{i}'}{\calZ_{i-1}'',\calZ_{i}}{\calZt_{m_{i-1}''+1},\ldots, \calZt_{m_i-1}}\\
&\phaneq\times \verop{x_i}{\calZ_i,\calZ_i'}{\calZt_{m_i},\ldots, \calZt_{j},\calZt_{k+1},\ldots,\calZt_{m_i'}}\verop{x_{i}''}{\calZ_{i}',\calZ_{i}''}{\calZt_{m_{i}'+1},\ldots, \calZt_{m_i''}}\\
&\phaneq\times\ver{\calZt_{j+1},\ldots, \calZt_k}[(\calZt_{k+1}\stackrel{k=m_i'}{\longrightarrow}\calZ_i'),(\calZt_j\stackrel{j=m_i-1}{\longrightarrow}\calZ_i),\star,\calZt_{j+1},\calZt_k]\,,
\end{aligned}
\end{equation}
while those on the left- and right-hand side of figure~\ref{fig:AllTermsNMHV2}, called $\bbB_i$ and $\bbC_i$ respectively, contribute 
\begin{equation}
\label{eq:BCcontributiontoNMHVWilsonloopformfactor}
\begin{aligned}
\bbB_i&= \nalpha\sum_{m_{i-1}''\leq j<k\leq m_i-1}\verop{x_{i}'}{\calZ_{i-1}'',\calZ_{i}}{\calZt_{m_{i-1}''+1},\ldots,\calZt_{j},\calZt_{k+1},\ldots, \calZt_{m_i-1}}\\
&\phaneq\times \verop{x_i}{\calZ_i,\calZ_i'}{\calZt_{m_i},\ldots, \calZt_{m_i'}}\verop{x_{i}''}{\calZ_{i}',\calZ_{i}''}{\calZt_{m_{i}'+1},\ldots, \calZt_{m_i''}}\\
&\phaneq\times\ver{\calZt_{j+1},\ldots, \calZt_k}[(\calZt_{k+1}\stackrel{k=m_i-1}{\longrightarrow}\calZ_i),(\calZt_j\stackrel{j=m_{i-1}''}{\longrightarrow}\calZ_{i-1}''),\star,\calZt_{j+1},\calZ_k]\,,\\
\bbC_i &=\nalpha\sum_{m_{i}'\leq j<k\leq m_i''}\verop{x_{i}'}{\calZ_{i-1}'',\calZ_{i}}{\calZt_{m_{i-1}''+1},\ldots, \calZt_{m_i-1}}\\
&\phaneq\times \verop{x_i}{\calZ_i,\calZ_i'}{\calZt_{m_i},\ldots, \calZt_{m_i'}}\verop{x_{i}''}{\calZ_{i}',\calZ_{i}''}{\calZt_{m_{i}'+1},\ldots,\calZt_{j},\calZt_{k+1},\ldots, \calZt_{m_i''}}\\
&\phaneq\times\ver{\calZt_{j+1},\ldots, \calZt_k}[(\calZt_{k+1}\stackrel{k=m_{i}''}{\longrightarrow}\calZ_i),(\calZ_j\stackrel{j=m_{i}'}{\longrightarrow}\calZ_{i-1}''),\star,\calZt_{j+1},\calZt_k]\,.
\end{aligned}
\end{equation}

Finally, the complete tree-level NMHV form factor of the Wilson loop in position twistor space is obtained by adding \eqref{eq:AcontributiontoNMHVWilsonloopformfactor} to the two terms in  \eqref{eq:BCcontributiontoNMHVWilsonloopformfactor}, multiplying with the contributions of the cogs of the Wilson loop that are not connected to the vertex $\mathbf{V}$ and then summing over the cogs $i$ that are as well as over the possible distributions of the integers $m_i,m_i'$ and $m_i''$ subject to \eqref{eq:constraintsforthems}:
\begin{equation}
\label{eq:wilsonloopformfactorinpositiontwistorspace}
\mathbb{F}^{\text{NMHV}}_{\calW}=\sum_{\{m_s,m_s',m_s''\}}\sum_{i=1}^L\Big(\prod_{r=1}^{i-1}\bbD_r\Big)(\bbA_i+\bbB_i+\bbC_i)\Big(\prod_{r=i+1}^{L}\bbD_r\Big)\,,
\end{equation}
where the contribution of a cog that is not connected to the interaction vertex is
\begin{equation}
\begin{aligned}
\bbD_i&= \verop{x_{i}'}{\calZ_{i-1}'',\calZ_{i}}{\calZt_{m_{i-1}''+1},\ldots, \calZt_{m_i-1}} \verop{x_i}{\calZ_i,\calZ_i'}{\calZt_{m_i},\ldots, \calZt_{m_i'}}
\\&\phaneq\times 
\verop{x_{i}''}{\calZ_{i}',\calZ_{i}''}{\calZt_{m_{i}'+1},\ldots, \calZt_{m_i''}}\,.
\end{aligned}
\end{equation}

To obtain the tree-level NMHV form factor of a local operator $\calO$ in position twistor space, we have to act with the forming operator $\mathbf{F}_\calO$ defined in \eqref{formingop} on the \emph{integrand} of \eqref{eq:wilsonloopformfactorinpositiontwistorspace}, i.e.\ \emph{before} doing the integrations over the vertex positions; see the discussion in the main text. 

\subsection{Momentum twistor space}

Let us now describe the transition to momentum (twistor) space.

First, we need to recall some of the notation introduced in \cite{Koster:2016loo}. For the operator $\calO$ of \eqref{eq:composite operator}, we denote by $N_i$ the total number of derivatives (space-time and fermionic) required to generate the field $D^{k_i}A_i$ in \eqref{eq:definitionformingfactoronshellstates}. Specifically, for $A_i=\bar F, \bar \psi $ or $ \phi$, we have $N_i=k_i+2$. Otherwise, $N_i=k_i+3$ for $A_i=\psi$ and $N_i=k_i+4$ for $A_i=F$. 
Moreover, we denote by $n_{\theta_\ri}$ the number of $\frac{\partial}{\partial\theta_\ri}$ derivatives required to generate the field $D^{k_i}A_i$ in \eqref{eq:definitionformingfactoronshellstates}. Concretely, $n_{\theta_\ri}=0,1,2,3,4$ for $A_i=\bar F, \bar \psi, \phi, \psi, F $, respectively. 
For the operators without $\dot\alpha$ indices, we have $k_i=0$ and also $N_i=n_{\theta_i}=2,3,4$.

The computation of the NMHV form factor parallels the calculations done for amplitudes in subsection~\ref{eq:NMHV amplitudes in momentum twistor space}. 
Starting from the position twistor expression \eqref{eq:wilsonloopformfactorinpositiontwistorspace}, we insert on-shell momentum eigenstates.
Inserting on-shell momentum eigenstates in the $\textbf{W}$ vertex gives%
\begin{equation}
\label{eq:definitionmathscrU}
\begin{aligned}
\mathscr{U}_x(\calZ_1,\calZ_2;\fP_1,\ldots, \fP_m)
&\equiv\verop{x}{\calZ_1,\calZ_2}{\calA_{\fP_1},\ldots, \calA_{\fP_m}}
\\&=\frac{\abra{\lambda_1}{\lambda_2}}{\abra{\lambda_1}{p_1}\left(\prod_{k=1}^{m-1}\abra{p_k}{p_{k+1}}\right)\abra{p_m}{\lambda_2}}\e^{i\sum_{k=1}^m(x\fp_k+\theta p_k\eta_k)}\,,
\end{aligned}
\end{equation}
for $m\geq 1$ and $\mathscr{U}_x(\calZ_1,\calZ_2;)=1$ for $m=0$; cf.\ (C.1) in \cite{Koster:2016loo}.
This leads to the general formula \eqref{eq:NMHVWilsonloopformfactor momentumspace}.
Analogously, we obtain for the contribution $\bbA_i$ \eqref{eq:AcontributiontoNMHVWilsonloopformfactor} in which the vertex is connected to the edge $x_i$ of the cog:
\begin{align}
\label{eq:hatAversion1}
\hat{\bbA}_i&=\nalpha \sum_{m_i-1\leq j<k\leq m_i'}\mathscr{U}_{x_{i}'}(\calZ_{i-1}'',\calZ_{i};\fP_{m_{i-1}''+1},\ldots, \fP_{m_i-1})
\mathscr{U}_{x_{i}''}(\calZ_{i}',\calZ_{i}'';\fP_{m_{i}'+1},\ldots, \fP_{m_i''})\nonumber\\
&\phaneq\times \int \frac{\dd^4z\,\dd^8\vartheta}{(2\pi)^4} \frac{\e^{i\sum_{s=j+1}^k(z \fp_s+\vartheta p_s\eta_s)}}{\big(\prod_{s=j+1}^{k-1}\abra{p_s}{p_{s+1}}\big)\abra{p_k}{p_{j+1}}}
 \\&\phaneq\times 
\PPP_{D^{k_i }A_i}\mathscr{U}_{x_i}(\calZ_i,\calZ_i';\fP_{m_i},\ldots, \fP_{j},\fP_{k+1},\ldots,\fP_{m_i'})\nonumber\\&\phaneq\times[\calZ_{(x_i,\theta_i)}(p_{k+1}\stackrel{k=m_i'}{\longrightarrow }\lambda_i'),\calZ_{(x_i,\theta_i)}(p_j\stackrel{j=m_i-1}{\longrightarrow }\lambda_i),\star,\calZ_{(z,\vartheta)}(p_{j+1}),\calZ_{(z,\vartheta)}(p_k)]\,,\nonumber
\end{align}
where we have 
placed a $\hat{\phantom{\cdot}}$ on the contribution $\bbA_i$ from \eqref{eq:AcontributiontoNMHVWilsonloopformfactor} to indicate the fact that it uses the momentum on-shell external states \eqref{eq:definitiononshellmomentumeigenstates} and that we have included the forming operator \eqref{eq:definitionformingfactoronshellstates}.
Expressions similar to \eqref{eq:hatAversion1} can also be obtained for the contributions $\hat{\bbB}_i$ and $\hat{\bbC}_i$ coming from the other edges of the cog, see \eqref{eq:BCcontributiontoNMHVWilsonloopformfactor}. 

In order to proceed as for amplitudes at this point, it is necessary to be able to change the order of the $z$ integration from the vertex $\mathbf{V}$ and the derivatives in the forming operator $\PPP_{D^{k_i }A_i}$. 
For operators without space-time derivatives in the forming operator, i.e.\ $N_i=n_{\theta_i}$, this is possible.
As with amplitudes, we then shift $z\rightarrow z+x_i$ and $\vartheta\rightarrow \vartheta+\theta_i$. After doing the integration, we obtain
\begin{equation}
\label{eq:hatAi}
\begin{aligned}
\hat{\bbA}_i&= \sum_{m_i-1\leq j<k\leq m_i'} \mathscr{U}_{x_{i}'}(\calZ_{i-1}'',\calZ_{i};\fP_{m_{i-1}''+1},\ldots, \fP_{m_i-1})\mathscr{U}_{x_{i}''}(\calZ_{i}',\calZ_{i}'';\fP_{m_{i}'+1},\ldots, \fP_{m_i''})
\\&\phaneq\times \left(\PPP_{D^{k_i }A_i}^{N_i=n_{\theta_i}}\mathscr{U}_{x_i}(\calZ_i,\calZ_i';\fP_{m_i},\ldots,\fP_{m_i'})\right)  \frac{\abra{(p_j\stackrel{j=m_i-1}{\longrightarrow }\lambda_i)}{p_{j+1}}\abra{p_k}{(p_{k+1}\stackrel{k=m_i'}{\longrightarrow }\lambda_i')}}{\abra{(p_j\stackrel{j=m_i-1}{\longrightarrow }\lambda_i)}{(p_{k+1}\stackrel{k=m_i'}{\longrightarrow }\lambda_i')}\abra{p_k}{p_{j+1}}}
\\&\phaneq\times [\calZ_{0}(p_{k+1}\stackrel{k=m_i'}{\longrightarrow }\lambda_i'),\calZ_{0}(p_j\stackrel{j=m_i-1}{\longrightarrow }\lambda_i),\star, \calZ_{\sum_{s=j+1}^{k}\fP_s}(p_{j+1}),\calZ_{\sum_{s=j+1}^{k}\fP_s }(p_{k})]\,,
\end{aligned}
\end{equation}
where we have used \eqref{eq:replacementzvarthetawithmomentumtwistors} and rearranged the second $\mathscr{U}$. 
As in the case of amplitudes, the five-bracket in \eqref{eq:hatAi} is independent of $x_i$ and $\theta_i$. This has the important consequence that the derivatives specifying the operator only act on the exponential functions contained in $\mathscr{U}_{x_i}$. We can now act with the forming operator
\eqref{formingop}, which affects only $\mathscr{U}_{x_i}$ and was computed in appendix C of \cite{Koster:2016loo}. We then take the  operator limit, which sends all $x_i$ to $x$, $\lambda_i,\lambda_i'$ to $\tau_i$ and $\lambda_i''$ to the reference spinor  $\textbf{n}$; see figure~\ref{fig:operatorlimitapp} and also appendix A of \cite{Koster:2016loo}.
The computation is very similar to the one presented in \cite{Koster:2016loo} for the MHV form factor, up to the five-bracket that we need to treat carefully.
We find 
\begin{equation}
\begin{aligned}
\hat{\bbA}_i=& \e^{i\sum_{s=m_{i-1}''+1}^{m_i''}x\fp_s}\sum_{m_i-1\leq j<k\leq m_i'}\tilde{\calI}_i'(m_{i-1}'',m_i)\tilde{\calI}_i(m_i,m_i')\tilde{\calI}_i''(m_i',m_i'')\\
&\times  \Big[\calZ_{0}(p_j\stackrel{j=m_i-1}{\longrightarrow }\tau_i),\calZ_{0}(p_{j+1}),\star, \calZ_{\rmom_k-\rmom_j}(p_{k}),\calZ_{\rmom_k-\rmom_j}(p_{k+1}\stackrel{k=m_i'}{\longrightarrow }\tau_i)\Big]\,,
\end{aligned}
\end{equation}
where we have simplified the five-bracket by absorbing some angular brackets as in \eqref{eq: amplitude simplified}. The factors $\tilde{\calI}_i$, $\tilde{\calI}'_i$ and $\tilde{\calI}''_i$ were defined in appendix C of \cite{Koster:2016loo} and we reproduce them here for the reader's convenience: 
\begin{align}
\label{eq:definitiontildecalI}
\tilde{\calI}_j(m_j,m_j')&=-\frac{\left(\sum_{k=m_j}^{m_j'}\abra{\tau_j}{p_k}\sbra{\bar{p}_k}{\bar{\tau}_j}\right)^{N_j-n_{\theta_j}}\left(\sum_{k=m_j}^{m_j'}\abra{\tau_j}{p_k}\{\xi_j\eta_k\}\right)^{n_{\theta_j}}}{\abra{\tau_j}{p_{m_j}}\prod_{k=m_j}^{m_j'-1}\abra{p_k}{p_{k+1}}\abra{p_{m_j'}}{\tau_j}}\,,\nonumber\\
\tilde{\calI}_j'(m_{j-1}'',m_j)& =\frac{\abra{ \n}{\tau_{j}}}{\abra{\n}{p_{m_{j-1}''+1}}\prod_{k=m_{j-1}''+1}^{m_j-2}\abra{p_k}{p_{k+1}}\abra{p_{m_j-1}}{\tau_j}}\,,\\
\tilde{\calI}_j''(m_j',m_j'')&= \frac{\abra{ \tau_j}{\n}}{\abra{\tau_j}{p_{m_j'+1}}\prod_{k=m_j'+1}^{m_j''-1}\abra{p_k}{p_{k+1}}\abra{p_{m_j''}}{\n}}\,,\nonumber
\end{align}
where we recall that we are currently treating only the case $N_i=n_{\theta_i}$.
These objects are just the operator limit of \eqref{eq:definitionmathscrU} for each of the three edges of the cog with label $i$, stripped of the exponential factors. 
We need to keep in mind the condition \eqref{eq:constraintsforthems} on the integers $\{m_i,m_i',m_i''\}$. Furthermore, when no particle is emitted from a specific edge, the corresponding contribution equals one, i.e.\
\begin{equation}
\label{eq:definitiontildecalI2}
\tilde{\calI}_j'(m_{j-1}'',m_j=m_{j-1}'')=1 
\,,\qquad \tilde{\calI}_j''(m_j',m_j''=m_{j}')=1 
\,.
\end{equation}
In particular, the dependence on the operator $\calO$ is completely contained in the terms $\tilde{\calI}_i(m_i,m_i')$. Using the momentum supertwistors for the form factors and the same technique as in \eqref{eq: amplitude five bracket simplified} for the five-brackets, we write
\begin{equation}
\label{eq:hatA}
\begin{aligned}
\hat{\bbA}_i&= \e^{i\sum_{s=m_{i-1}''+1}^{m_i''}x\fp_s}\tilde{\calI}_i'(m_{i-1}'',m_i)\tilde{\calI}_i(m_i,m_i')\tilde{\calI}_i''(m_i',m_i'')\\
&\phaneq\times  \sum_{m_i-1\leq j<k\leq m_i'} \Big[\calW_j\stackrel{j=m_i-1}{\longrightarrow }\calZ_{y_j}(\tau_i),\calW_{j+1},\star, \calW_{k},\calW_{k+1}\stackrel{k=m_i'}{\longrightarrow }\calZ_{y_k}(\tau_i)\Big]\,.
\end{aligned}
\end{equation}
Similarly, we obtain for $\hat{\bbB}_i$ and $\hat{\bbC}_i$ the formulae
\begin{equation}
\label{eq:hatBhatC}
\begin{aligned}
\hat{\bbB}_i&= \e^{i\sum_{s=m_{i-1}''+1}^{m_i''}x\fp_s}\tilde{\calI}_i'(m_{i-1}'',m_i)\tilde{\calI}_i(m_i,m_i')\tilde{\calI}_i''(m_i',m_i'')\\
&\phaneq\times  \sum_{m_{i-1}''\leq j<k\leq m_i-1}
\Big[\calW_j\stackrel{j=m_{i-1}''}{\longrightarrow }\calZ_{y_j}(\textbf{n}),\calW_{j+1},\star, \calW_{k},\calW_{k+1}\stackrel{k=m_i-1}{\longrightarrow }\calZ_{y_k}(\tau_i)\Big]
\,,\\
\hat{\bbC}_i&= \e^{i\sum_{s=m_{i-1}''+1}^{m_i''}x\fp_s}\tilde{\calI}_i'(m_{i-1}'',m_i)\tilde{\calI}_i(m_i,m_i')\tilde{\calI}_i''(m_i',m_i'')\\
&\phaneq\times  \sum_{m_{i}'\leq j<k\leq m_i''}
\Big[\calW_j\stackrel{j=m_{i}'}{\longrightarrow }\calZ_{y_j}(\tau_i),\calW_{j+1},\star, \calW_{k},\calW_{k+1}\stackrel{k=m_i''}{\longrightarrow }\calZ_{y_k}(\textbf{n})\Big]
\,.
\end{aligned}
\end{equation}
Finally, we also need the contributions from the cogs that are not attached to a vertex
\begin{equation}
\label{eq:hatD}
\hat{\bbD}_i=\e^{i\sum_{s=m_{i-1}''+1}^{m_i''}x\fp_s}\tilde{\calI}_i'(m_{i-1}'',m_i)\tilde{\calI}_i(m_i,m_i')\tilde{\calI}_i''(m_i',m_i'')\,.
\end{equation}

Putting everything together, the NMHV form factor is obtained after summing over all the cogs of the Wilson loop as well as over all the distributions of the integers $\{m_s, m_s',m_s''\}$ ordered as in \eqref{eq:constraintsforthems} and applying the Fourier transformation $\int \frac{\dd^4x}{(2\pi)^4}\e^{-i\fq x}$: 
\begin{equation}
\label{eq:NMHFformfactorMomentumPart1}
\mathscr{F}^{\text{NMHV}}_{\calO}(1,\ldots, n;\fq)=\int \frac{\dd^4x}{(2\pi)^4}\e^{-i\fq x}\sum_{\{m_s,m_s',m_s''\}}\sum_{i=1}^L\Big(\prod_{r=1}^{i-1}\hat{\bbD}_r\Big)(\hat{\bbA}_i+\hat{\bbB}_i+\hat{\bbC}_i)\Big(\prod_{r=i+1}^{L}\hat{\bbD}_r\Big)\,,
\end{equation}
where the quantities $\hat{\bbA}_r,\ldots, \hat{\bbD}_r$ are defined in \eqref{eq:hatA}, \eqref{eq:hatBhatC} and \eqref{eq:hatD}. 
Pulling out the Parke-Taylor denominator and the momentum-conserving delta function, we can bring \eqref{eq:NMHFformfactorMomentumPart1} to the form \eqref{eq:NMHFformfactorMomentumPart2}.

\section{Fourier-type integrals}
\label{app:importantFourier}

In this appendix, we derive several Fourier-type integral identities that are used in section~\ref{sec4}.
Moreover, we demonstrate the non-commutativity of the integral with space-time derivatives coming from the forming operator in a given example.

\subsection{Fourier-type integrals with ratios of scalar products}
\label{app:mainfourier}

The main bosonic Fourier-type integral identity that we want to show reads
\begin{equation}
\label{eq:importantFourieridentity}
\int \frac{\dd^4x }{(2\pi)^4}\frac{1}{x^2}\frac{\langle s_1|x|\zeta]\cdots \langle s_k|x|\zeta]}{\langle t_{1}|x|\zeta]\cdots \langle t_{k}|x|\zeta]}\e^{i\fq x}=\frac{1}{i(4\pi)^2}\frac{1}{\fq^2}\frac{\langle s_1|\fq|\zeta]\cdots \langle s_k|\fq|\zeta]}{\langle t_{1}|\fq|\zeta]\cdots \langle t_{k}|\fq|\zeta]}\eqndot
\end{equation}
In contrast to usual Fourier integrals, this integral, which stems from the bosonic part of the twistor-space propagator, is not only badly behaved for $x^2=0$ but also for $\langle t_{i}|x|\zeta]=0$.
In addition to calculating the integral, we thus also require a prescription that makes it well-defined.
We will first address the latter issue with an explicit calculation of the case $k=1$.
Assuming that a similar prescription also exists for $k\geq2$, we then argue that the integral must be given by \eqref{eq:importantFourieridentity}.

\paragraph{Case $k=1$:}  Regularizing the integral \eqref{eq:importantFourieridentity} is not trivial, due to the fact that the term $\langle t_{1}|x|\zeta]$ introduces an extra pole in the $x_0$ integration. The guiding principle behind the regularization prescription that we choose is based on the following a-posteriori thinking: the expression \eqref{eq:importantFourieridentity} correctly translates the amplitudes from position twistor space to the known amplitudes in momentum space, i.e.\ the additional pole cannot contribute. A way to get rid of the extra pole starts by taking the vector $\ft_1\equiv t_1^{\alpha}\zeta^{\dot{\alpha}}$ and giving it a small mass. Then, we can boost it to be parallel to the $x_0$ direction and replace the $\ft_1 x$ in the denominator by $\ft_1 x+i \varepsilon\fq\ft_1=2\ft_{1,0}(x_0+i\varepsilon \fq_0)$. Since the sign of $\fq_0$ determines how to close the contour due to the exponent $\e^{i\fq x}$, i.e.\ whether we close the contour in the upper or lower $x_0$ plane, the extra pole is always avoided. One issue with this approach is that the intermediate steps of the computation break Lorentz invariance and it is quite non-trivial to see how it is restored in the end. Thus, we prefer to do the computation in Euclidean space and to Wick rotate in the end. 

Let us furthermore replace $s_1^{\alpha}\zeta^{\dot{\alpha}}$ and $t_1^{\alpha}\zeta^{\dot{\alpha}}$ by arbitrary vectors $\fs^{\alpha\dot{\alpha}}$ and $\ft^{\alpha\dot{\alpha}}$. Thus, we consider the integral
\begin{equation}
\label{eq:defIPanzer1}
\tilde{I}
=\int \frac{\dd^4x}{(2\pi)^4}\frac{1}{x^2}\frac{x \fs}{x \ft}\e^{i \fq x}
=\int \frac{\dd^4x}{(2\pi)^4}\frac{1}{x^2}\frac{2x \cdot\fs}{2x \cdot\ft}\e^{2i \fq \cdot x}
=\frac{1}{4}\int \frac{\dd^4x}{(2\pi)^4}\frac{1}{x^2}\frac{x\cdot \fs}{x\cdot \ft}\e^{i \fq\cdot x}\,,
\end{equation}
where we have rescaled $x$ to absorb the factor of 2 in the scalar product; recall that $xy\equiv x^{\alpha\dot\alpha}y_{\alpha\dot\alpha}=2x_\mu y^\mu\equiv2x\cdot y$.
We will be interested in the limit in which the vectors $\fs$ and $\ft$ become complex and obey $\fs^2=\ft^2=\fs \cdot\ft=0$, but for now we take them to be real.
The above integral \eqref{eq:defIPanzer1} is ill-defined due to the $x \cdot\ft$ in the denominator and we propose to define it as%
\footnote{We thank Erik Panzer for a very helpful discussion on this point.}
\begin{equation}
\label{eq:defIPanzer2}
\tilde{I}=\frac{1}{4}\lim_{\varepsilon\rightarrow 0}\int \frac{\dd^4x}{(2\pi)^4}\frac{1}{x^2}\frac{x \cdot \fs}{x \cdot \ft+ i\varepsilon (\fq  \cdot\ft)}\e^{i \fq \cdot  x}\,.
\end{equation}
We decompose $x$ as $x=x_t\hat{\ft}+x_{\perp}$ with $\hat{\ft}=\ft/|\ft|$ and $\ft \cdot x_\perp=0$. We see that we can compute the integration over $x_t$ in \eqref{eq:defIPanzer2} by doing a contour integral that we close in the upper (lower) half plane for $\fq  \cdot \hat{\ft}>0$ ($\fq \cdot  \hat{\ft}<0$). The poles are at $x_t=\pm i |x_\perp|$ and $x_t=-i \varepsilon \fq  \cdot \hat{\ft}$. Due to the way that we close the contour, the last pole never contributes, regardless of the value of $\fq \cdot \hat{\ft}$. Hence, we obtain after computing the residues and taking the limit $\varepsilon\rightarrow 0$
\begin{equation}
\begin{aligned}
\tilde{I}&=\frac{1}{4}\frac{2\pi i}{(2\pi)^4} \int \dd^3x_{\perp}\bigg[\Theta(\fq  \cdot \hat{\ft})\frac{(x_\perp+i |x_\perp|\hat{\ft})\cdot  \fs}{(i  |x_\perp| |\ft|)}\e^{-|x_\perp|\fq\cdot \hat{\ft}}
 \\
 &\hphantom{{}={}\frac{1}{4}\frac{2\pi i}{(2\pi)^4} \int \dd^3x_{\perp}\bigg[}
-\Theta(-\fq  \cdot \hat{\ft})\frac{(x_\perp-i |x_\perp|\hat{\ft}) \cdot \fs}{(i  |x_\perp| |\ft|)}\e^{|x_\perp|\fq \cdot\hat{\ft}}\bigg]\frac{\e^{i \fq \cdot  x_{\perp}}}{2i |x_\perp|}\\
&=\frac{1}{4}\frac{2\pi i}{(2\pi)^4} \int \dd^3x_{\perp}\left[\frac{\fs \cdot  \ft}{\ft^2}+\text{sgn}(\fq  \cdot \hat{\ft})\frac{\fs \cdot x_\perp}{i |\ft||x_\perp|}\right]\frac{\e^{i \fq   \cdot x_{\perp}-|x_\perp||\fq\cdot \hat{\ft}|}}{2i |x_\perp|}\,,
\end{aligned}
\end{equation}
where $\Theta$ denotes the Heaviside step function and $\text{sgn}(\fq  \cdot \hat{\ft})$ the sign of $\fq  \cdot \hat{\ft}$.
Going to spherical coordinates in the space perpendicular to $\ft$ 
via $x_\perp=r \hat{x}_\perp$, we find
\begin{equation}
\begin{aligned}
\tilde{I}&=\frac{1}{4}\frac{2\pi i}{(2\pi)^4}\int_{0}^\infty r^2\dd r\int_{S^2}\dd\hat{x}^2_\perp \frac{1}{2i r}\e^{-r(|\fq  \cdot \hat{\ft}|-i\fq \cdot \hat{x}_\perp )}\left(\frac{\fs \cdot \ft}{\ft^2}+\text{sgn}(\fq \cdot  \hat{\ft})\frac{\fs\cdot  \hat{x}_\perp}{i |\ft|}\right)\\
&=\frac{1}{4}\frac{\pi }{(2\pi)^4}\int_{S^2}\dd\hat{x}^2_\perp \frac{1}{\left(|\fq \cdot \hat{\ft}|-i\fq \cdot \hat{x}_\perp\right)^2}\left(\frac{\fs \cdot \ft}{\ft^2}+\text{sgn}(\fq  \cdot\hat{\ft})\frac{\fs \cdot \hat{x}_\perp}{i |\ft|}\right)\,.
\end{aligned}
\end{equation}
We now define $\fq_\perp=\fq-\ft\frac{\fq \cdot \ft}{\ft^2}$ and similarly for $\fs_\perp$. Setting $\hat{\fq}_\perp=\fq_\perp/|\fq_\perp|$, we decompose $\hat{x}_\perp$ as
\begin{equation}
\hat{x}_\perp=\cos(\theta) \hat{\fq}_\perp+\sin(\theta)\big(\cos(\varphi) \hat{z}+\sin(\varphi) \hat{w}\big)\,,
\end{equation}
where $\hat{z}$ and $\hat{w}$ are two orthonormal vectors spanning the plane orthogonal to $\ft$ and $ \hat{\fq}_\perp$. 
Then, $\dd\hat{x}^2_\perp=\sin(\theta) \dd\theta \dd\varphi$ and the $\varphi$ integral is easily done. It yields
\begin{equation}
\begin{aligned}
\tilde{I}=\frac{1}{4}\frac{2\pi^2}{(2\pi)^4}\int_{0}^\pi \frac{\sin(\theta) \dd\theta}{\left(|\fq  \cdot \hat{\ft}|+i|\fq_\perp|\cos(\theta) \right)^2}\left(\frac{\fs \cdot\ft}{\ft^2}+\text{sgn}(\fq  \cdot\hat{\ft})\cos(\theta)\frac{\fs_{\perp} \cdot\hat{\fq}_\perp}{i |\ft|}\right)\,.
\end{aligned}
\end{equation}
Integrating by parts and using $\log\frac{a+bi}{a-bi}=2i\arctan\frac{b}{a}$, we find
\begin{equation}
\begin{aligned}
\tilde{I}
=\frac{1}{(4\pi)^2\fq^2}\left[\frac{\fs \cdot \ft}{\ft^2}-\frac{(\fq \cdot \hat{\ft})(\fq_\perp \cdot \fs_\perp)}{|\ft|\fq_\perp^2}\right]+\frac{1}{(4\pi)^2}\text{sgn}(\fq\cdot\hat{\ft})\frac{\fq_{\perp}\cdot  \fs_\perp}{|\ft||\fq_\perp|^3}\arctan\frac{|\fq_\perp|}{|\fq \cdot \hat{\ft}|}\,.
\end{aligned}
\end{equation}
We now insert $|\fq_\perp|^2=\fq^2-(\fq\cdot   \ft)^2/\ft^2$ and $\fs_\perp\cdot  \fq_\perp=\fs  \cdot \fq -(\fq \cdot  \ft)(\fs\cdot \ft)/\ft^2$ so that after a couple of trivial manipulations
\begin{equation}
\label{eq:lastversionoftildeI}
\tilde{I}=\frac{1}{(4\pi)^2}\Bigg[\frac{1}{\fq^2}\frac{(\fs \cdot\fq)(\ft \cdot\fq)-(\fs\cdot\ft)\fq^2}{(\fq \cdot \ft)^2-\ft^2\fq^2}
+\frac{(\fs \cdot\ft)(\fq\cdot \ft)-\ft^2(\fs\cdot \fq)}{|(\fq\cdot  \ft)^2-\ft^2\fq^2|^{\frac{3}{2}}}\arctan\frac{\sqrt{|\ft^2\fq^2-(\fq \cdot \ft)^2|}}{|\fq \cdot \hat{\ft}|}\Bigg]\,.
\end{equation}
It is now obvious how to take the limit $\fs^2=\ft^2=\fs\cdot\ft=0$. 
In addition, if we now Wick rotate the integral, we pick up a factor of $-i$ from the measure, so that
\begin{equation}
\label{eq:lastexpressionforI1}
\int \frac{\dd^4x}{(2\pi)^4}\frac{1}{x^2}\frac{\bradot{s_1}{x}{\zeta}}{\bradot{t_1}{x}{\zeta}}\e^{i\fq x}=\frac{1}{\nalpha}\frac{1 }{\fq^2}\frac{\bradot{s_1}{\fq}{\zeta}}{\bradot{t_1}{\fq}{\zeta}}\,.
\end{equation}

\paragraph{Case $k\geq2$:} 
Let us now define $(\fs_i)_{\alpha\dot\alpha}=s_{i\alpha}\zeta_{\dot\alpha}$ and $(\ft_i)_{\alpha\dot\alpha}=t_{i\alpha}\zeta_{\dot\alpha}$ so as to rewrite \eqref{eq:importantFourieridentity} as
\begin{equation}
\label{eq:importantFourieridentity2}
\int \frac{\dd^4x }{(2\pi)^4}\frac{1}{x^2}\frac{(\fs_1 x)\cdots (\fs_k x)}{(\ft_1 x)\cdots (\ft_k x)}\e^{i\fq x}=\frac{1}{i(4\pi)^2}\frac{1}{ \fq^2}\frac{(\fs_1 \fq)\cdots (\fs_k \fq)}{(\ft_1 \fq)\cdots (\ft_k \fq)}\eqncom
\end{equation}
for $\fs_i \fs_j=\fs_i \ft_j=\ft_i \ft_j=0$.
We shall assume that a prescription similar to the case $k=1$ exists that makes the integral well-defined.
We observe first that the left-hand side of \eqref{eq:importantFourieridentity2} is homogeneous of degree $1$ independently in each $\fs_i$ and of degree $-1$ in each $\ft_i$. Furthermore, a simple change of variables shows that it is homogeneous of degree $-2$ in $\fq$. Due to Lorentz invariance and to the conditions that we wish to impose on the vectors $\fs_i$ and $\ft_i$, we can only use the scalar products $\fq^2$, $\fq \fs_i$ and $\fq \ft_i$ to build the answer. Using these ingredients, we cannot build a cross-ration that is invariant under all independent rescalings of the variables. Hence, up to a constant, the answer of the integral can only be equal to the right-hand side of \eqref{eq:importantFourieridentity2}. The constant, however, is fixed by considering the limit where $\fs_i\rightarrow \ft_i$ for $i=2, \dots k$, thus seeing that it is independent of $k$ and consequently given by the one found in the case $k=1$. This concludes the derivation of \eqref{eq:importantFourieridentity}.

\paragraph{Fourier transform of the R-invariants:} Armed with the identity \eqref{eq:importantFourieridentity}, we can prove an important result for the R-invariants, namely the following super Fourier-type identity:
\begin{multline}
\label{eq:superFourierRinvariant}
I_s\equiv \int \frac{\dd^4z\dd^8\vartheta}{(2\pi)^4} \e^{iz\fq+i\vartheta \Gamma}[\calZ_{(x,\theta)}(\lambda_1),\calZ_{(x,\theta)}(\lambda_2),\star,\calZ_{(z,\vartheta)}(\lambda_3),\calZ_{(z,\vartheta)}(\lambda_4)]\\=\frac{1}{\nalpha} \e^{ix\fq+i\theta\Gamma}[\calZ_{(0,0)}(\lambda_1),\calZ_{(0,0)}(\lambda_2),\star,\calZ_{(\fq,\Gamma)}(\lambda_3),\calZ_{(\fq,\Gamma)}(\lambda_4)]\eqncom
\end{multline}
with $\Gamma\equiv \Gamma_{\alpha a}$ being 8 fermionic variables.
\proof 
We start by using the identity 
\begin{equation}
\label{eq:Rinvariantforxandxprime}
[\calZ_{(x,\theta)}(\lambda_1),\calZ_{(x,\theta)}(\lambda_2),\star,\calZ_{(z,\vartheta)}(\lambda_3),\calZ_{(z,\vartheta)}(\lambda_4)]=-\frac{\abra{\lambda_1}{\lambda_2}\abra{\lambda_3}{\lambda_4}}{(x-z)^2}\frac{\prod_{a=1}^4\bradot{(\theta-\vartheta)^a}{x-z}{\zeta}}{\prod_{j=1}^4\bradot{\lambda_j}{x-z}{\zeta}}
\end{equation}
for the R-invariant and by shifting $z\rightarrow z+x$ and $\vartheta\rightarrow \vartheta+\theta$. After expressing $\vartheta$ through $\chi_3$ and $\chi_4$ as $\vartheta^{\alpha a}=\frac{-i}{\abra{\lambda_3}{\lambda_4}}(\lambda_3^{\alpha}\chi_4^a-\lambda_4^{\alpha}\chi_3^a)$, we obtain
\begin{equation}
I_s
=\e^{ix\fq+i\theta \Gamma}\int \frac{\dd^4z\dd^8\vartheta}{(2\pi)^4} \frac{(-1)\abra{\la_1}{\la_2}}{\abra{\lambda_3}{\lambda_4}^3}\frac{1}{z^2}\frac{\prod_{a=1}^4\left(\bradot{\la_3}{z}{\zeta}\chi_4^a-\bradot{\la_4}{z}{\zeta}\chi_3^a\right)}{\prod_{j=1}^4\bradot{\la_j}{z}{\zeta}}\eqndot 
\end{equation}
Using \eqref{eq:importantFourieridentity} to perform the integral over $z$ and expressing the result via $\vartheta$, we find
\begin{equation}
\label{eq:fourierofRinvariantproof}
\begin{aligned}
I_s&=\frac{\e^{ix\fq+i\theta\Gamma}}{\nalpha} \int \dd^8\vartheta \e^{i\vartheta \Gamma} \frac{(-1)\abra{\la_1}{\la_2}\abra{\la_3}{\la_4}}{\fq^2}\frac{\prod_{a=1}^4\bradot{\vartheta^a}{\fq}{\zeta}}{\prod_{j=1}^4\bradot{\la_j}{\fq}{\zeta}}\\
&=\frac{\e^{ix\fq+i\theta\Gamma}}{\nalpha} \int \dd^8\vartheta\frac{(\vartheta \Gamma)^4}{4!} \frac{(-1)\abra{\la_1}{\la_2}\abra{\la_3}{\la_4}}{\fq^2}\frac{\prod_{a=1}^4\bradot{\vartheta^a}{\fq}{\zeta}}{\prod_{j=1}^4\bradot{\la_j}{\fq}{\zeta}}\\
&=\frac{ \e^{ix\fq+i\theta\Gamma}}{\nalpha} \frac{(-1)\abra{\la_1}{\la_2}\abra{\la_3}{\la_4}}{\fq^2}\frac{\prod_{a=1}^4\bradot{\Gamma_a}{\fq}{\zeta}}{\prod_{j=1}^4\bradot{\la_j}{\fq}{\zeta}}\eqncom 
\end{aligned}
\end{equation}
where we used that only the fourth order in the expansion of $\e^{i\vartheta \Gamma}$ contributes as there are exactly four powers of $\vartheta$ in $\prod_{a=1}^4\bradot{\vartheta^a}{\fq}{\zeta}$.
Using an identity analogous to \eqref{eq:Rinvariantforxandxprime}, we see that \eqref{eq:fourierofRinvariantproof} is exactly the desired result \eqref{eq:superFourierRinvariant}. 

\subsection{An integral for a form factor with \texorpdfstring{$\dot\alpha$}{dotted} index}
\label{subapp: integral with derivative}

In section~\ref{subsec: non-chiral operators}, we need to compute the following integral 
\begin{equation}
\int\frac{\dd^4z}{(2\pi)^4}\e^{i z \fq}\frac{\fs (z-x)}{((z-x)^2)^2}
=\e^{ix\fq }\int\frac{\dd^4z}{(2\pi)^4}\e^{i z \fq}\frac{\fs z}{(z^2)^2}
\,.
\end{equation}
We solve it in Euclidean space, i.e.\ we calculate
\begin{equation}
I_x=\int\frac{\dd^4z}{(2\pi)^4}\e^{2i z \cdot\fq}\frac{2\fs\cdot z}{(z^2)^2}
\,,
\end{equation}
where we have also dropped the phase. 
We can rotate $z$ so that $\fs$ is parallel to the $z_0$ direction with component $\fs_0$. Then, we have $\fs\cdot z\equiv\fs_\mu z^\mu =\fs_0 z_0$. We write $\vec\fq$ for the component of $\fq$ that is perpendicular to the $0$-direction. Doing a contour integral for $z_0$, we find
\begin{equation}
I_x=\frac{2\fs_0 }{(2\pi)^4}\int \dd^3\vec{z} \e^{2i \vec{z}\cdot \vec\fq}\int_{-\infty}^{\infty} \dd z_0\frac{z_0\e^{2i z_0\fq_0}}{(z_0^2+|\vec{z}|^2)^2}=\frac{2\fs_0 }{(2\pi)^4}\int \dd^3\vec{z} \e^{2i \vec{z}\cdot\vec\fq}\frac{ \pi i \fq_0\e^{-2|\vec{z}||\fq_0|}}{|\vec{z}|}\,.
\end{equation}
Recognizing $\fs_0\fq_0$ as $\fs\cdot\fq$ and introducing polar coordinates for the remaining $\vec{z}$ integration, we find
\begin{equation}
\label{eq:IxEuclideanFinalResult}
I_x=\frac{\pi i 2\fs\cdot\fq}{(2\pi)^4}2\pi \int_{0}^\infty r^2\dd r\int_{-1}^1\dd u\e^{2i r  u |\vec\fq|}\frac{\e^{-2r|\fq_0|}}{r}
=\frac{\pi i }{(2\pi)^3}\frac{2\fs\cdot \fq}{2\fq^2}=\frac{i}{(4 \pi)^2} \frac{2\fs\cdot\fq}{\fq^2}\,.
\end{equation}
By Wick rotating to Minkowski space as before, which gives  an extra $-i$ factor, we obtain the final result:
\begin{equation}
\label{eq:IxMinkowskiFinalResult}
\int_{}\frac{\dd^4z}{(2\pi)^4}\e^{i z \fq}\frac{\fs (z-x)}{((z-x)^2)^2}=\frac{\fs \fq}{(4\pi)^2} \frac{\e^{i x \fq}}{\fq^2}\,.
\end{equation}

We remark that, unlike in the case that we shall present in the next appendix~\ref{subsec:non-chiralFourierAppendix}, we can compute \eqref{eq:IxMinkowskiFinalResult} by exchanging integral and derivative.\footnote{The reason why we can exchange the derivative and integral here but not in the next section is related to the additional factor $\ft z$ in the denominator.} 
Namely, we can write after taking the derivative out of the integral and shifting $z$ by $x$:
\begin{equation}
\begin{aligned}
\int\frac{\dd^4z}{(2\pi)^4}\e^{i z \fq}\frac{\fs (z-x)}{((z-x)^2)^2}
&=\fs_{\alpha\dot\alpha}\frac{\partial}{\partial x_{\alpha\dot\alpha}} \int\frac{\dd^4z}{(2\pi)^4}\e^{i z \fq}\frac{1}{(z-x)^2}\\
&= \fs_{\alpha\dot\alpha}\frac{\partial}{\partial x_{\alpha\dot\alpha}}\e^{ix\fq}\int \frac{\dd^4z }{(2\pi)^4}\frac{\e^{iz \fq}}{z^2}=\frac{\fs \fq}{(4\pi)^2} \frac{\e^{i x \fq}}{\fq^2}\,,
\end{aligned}
\end{equation}
where we have used \eqref{eq:importantFourieridentity} for $k=0$ and the definition $x^2\equiv\frac{x^{\alpha\dot\alpha}x_{\alpha\dot\alpha}}{2}$ below \eqref{eq:definitionsproducts}.

\subsection{Commutators of integrals and derivatives}
\label{subsec:non-chiralFourierAppendix}

In this subsection, we demonstrate that the derivative and integration in \eqref{eq: start of calculation} do not commute. Therefore, it is not possible to
first perform the Fourier transformations and take the space-time derivative only in the very end. We take as our example the integral in the last line of \eqref{eq: start of calculation}, defining
\begin{equation}
I_{nc}\equiv \int\frac{\dd^{4}z}{(2\pi)^4} \e^{iz(\fp_1+\fp_2)}\left[\tau^{\alpha}\bar{\tau}^{\dot\alpha}\frac{\partial}{\partial x^{\alpha\dot\alpha}}\frac{1}{(x-z)^2}\frac{\bradot{p_2}{x-z}{\zeta}}{\bradot{\tau}{x-z}{\zeta}}\right]\,.
\end{equation}
If we first evaluate the $x$ derivative and then evaluate the integral with the help of appendix~\ref{subapp: integral with derivative}, we obtain the result leading to the last line of \eqref{eq:fpsi1}:
\begin{equation}
\label{eq: start of calculation appendix}
 I_{nc}=\frac{1}{\nalpha}\frac{\e^{ix(\fp_1+\fp_2)}}{(\fp_1+\fp_2)^2} \bradot{p_2}{\fp_1+\fp_2}{\bar\tau}\eqndot
\end{equation}
Alternatively, if we pull out the derivative in front of the integral and then shift $z$ by $x$, we find
\begin{equation}
\begin{aligned}
\tilde{I}_{nc}&\equiv\tau^{\alpha}\bar{\tau}^{\dot\alpha}\frac{\partial}{\partial x^{\alpha\dot\alpha}}\int\frac{\dd^{4}z}{(2\pi)^4} \e^{iz(\fp_1+\fp_2)}\frac{1}{(x-z)^2}\frac{\bradot{p_2}{x-z}{\zeta}}{\bradot{\tau}{x-z}{\zeta}}\\
&=\tau^{\alpha}\bar{\tau}^{\dot\alpha}\frac{\partial}{\partial x^{\alpha\dot\alpha}}\int\frac{\dd^{4}z}{(2\pi)^4} \e^{i(z+x)(\fp_1+\fp_2)}\frac{1}{z^2}\frac{\bradot{p_2}{z}{\zeta}}{\bradot{\tau}{z}{\zeta}}\\&=\tau^{\alpha}\bar{\tau}^{\dot\alpha}\frac{\partial}{\partial x^{\alpha\dot\alpha}}\left\{\e^{i(\fp_1+\fp_2)x}\frac{1}{\nalpha}\frac{1}{(\fp_1+\fp_2)^2}\frac{\bradot{p_2}{\fp_1+\fp_2}{\zeta}}{\bradot{\tau}{\fp_1+\fp_2}{\zeta}}\right\}\,,
\end{aligned}
\end{equation} 
where we have used the integral formula \eqref{eq:lastexpressionforI1}. Evaluating the derivative and comparing the result with \eqref{eq: start of calculation appendix} yields
\begin{equation}
\begin{aligned}
\tilde{I}_{nc}&=\frac{1}{\nalpha}\frac{\e^{ix(\fp_1+\fp_2)} }{(\fp_1+\fp_2)^2}\frac{\bradot{\tau}{\fp_1+\fp_2}{\bar\tau}\bradot{p_2}{\fp_1+\fp_2}{\zeta}}{\bradot{\tau}{\fp_1+\fp_2}{\zeta}}
 \\
 &
=I_{nc}-\frac{1}{\nalpha}\e^{ix(\fp_1+\fp_2)} \frac{\abra{\tau}{p_2}\sbra{\zeta}{\bar\tau}}{\bradot{\tau}{\fp_1+\fp_2}{\zeta}}\,.
\end{aligned}
\end{equation}
where we have used \eqref{eq:cuteidentity} in the last line. Hence, we see that we are in general not allowed to pull the $x$ derivative in front of the integral.

\section{On the one-loop two-point correlation function in the \texorpdfstring{SO$(6)$}{SO(6)} sector}
\label{sec:thethreevertexinthetwopointcorrelator}

In \cite{Koster:2014fva}, some of the present authors calculated the one-loop dilatation operator in the SO$(6)$ sector via the UV-divergent contributions to the one-loop two-point correlation function in twistor space. 
It was stated that the contributions to the dilatation operator stem only from the diagram with the four-point vertex in figure~\ref{fig:oneloopCorrelation4}, see \cite{Koster:2014fva} for details of this calculations. Looking at section~\ref{sec:asketchforloops}, we see that we have in addition a diagram with a three-point vertex and two diagrams with a two-point vertex.
In this appendix, we will discuss their contributions to the dilatation operator and show that they vanish; this complements the computation of \cite{Koster:2014fva}.

Note that, in this appendix, we refer to the spinors $\lambda_i$ in the interaction vertex  \eqref{eq:definitionoftheamplitudevertex 2.1} by $\rho_i$. Furthermore, since we are only interested in showing whether a contribution is divergent or not, we shall often drop irrelevant overall factors.

\paragraph{The two-point vertex connected to both sides of the diagram:}
\begin{figure}[tbp]
 \centering
  \includegraphics[height=3.5cm]{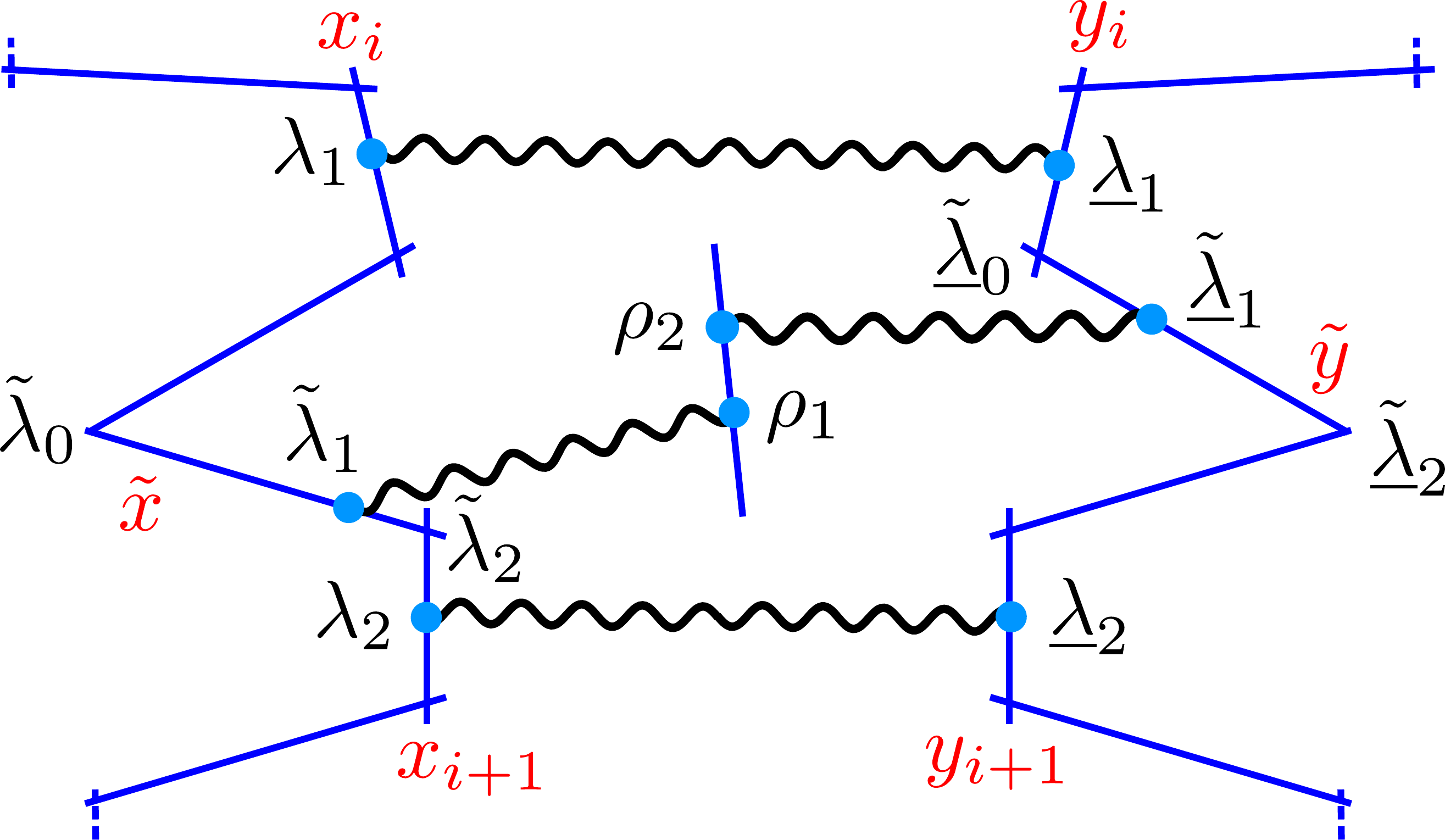}
  \caption{\it A one-loop correlation function diagram with the two-point vertex attached to both operators.}
  \label{fig:appCpar1}
\end{figure}
We are considering diagrams of the type depicted on the left-hand side of figure~\ref{fig:oneloopCorrelation3}, which we reproduce in figure~\ref{fig:appCpar1} including the new notation. The two-point vertex is attached to two gluons $g^+$ emitted from the positions $\tilde{\la}_1$ and $\tilde{\underline{\la}}_1$. We have to sum over the planar ways of attaching the vertex, so that $\tilde{x}$ has to be summed over $x_i$, $x_{i}''$, $x_{i+1}'$ and $x_{i+1}$ and similarly for $\tilde{y}$. Accordingly, the spinors $\tilde{\lambda}_0$, $\tilde{\lambda}_2$, $\tilde{\underline{\lambda}}_0$ and $\tilde{\underline{\lambda}}_2$ take the appropriate values of the vertices of the cogwheel Wilson loops, see figure~\ref{fig:CogwheelZoom}. The result of the particular diagram shown in figure~\ref{fig:appCpar1} is 
\begin{equation}
\label{eq:V2part1}
\begin{aligned}
V_2=&\int\frac{\dd^{4}z\dd^8\vartheta}{(2\pi)^4}\int \DD\la_1\DD\la_{2}\DD \underline{\la}_1\DD \underline{\la}_{2} \DD\tilde{\la}_1\DD\tilde{\underline{\la}}_1\frac{\DD\rho_1\DD\rho_2}{\abra{\rho_1}{\rho_2}\abra{\rho_2}{\rho_1}}\\
&\times \bar{\delta}^{2}\big(\calZ_{x_i}(\la_1),\star,\calZ_{y_i}(\underline{\la}_1)\big)\bar{\delta}^{2}\big(\calZ_{x_{i+1}}(\la_2),\star,\calZ_{y_{i+1}}(\underline{\la}_2)\big) F(\la_1,\la_2,\underline{\la}_1,\underline{\la}_2)
\\
&\times 
\frac{\abra{\tilde{\la}_0}{\tilde{\la}_2}}{\abra{\tilde{\la}_0}{\tilde{\la}_1}\abra{\tilde{\la}_1}{\tilde{\la}_2}}
\frac{\abra{\tilde{\underline{\la}}_0}{\tilde{\underline{\la}}_2}}{\abra{\tilde{\underline{\la}}_0}{\tilde{\underline{\la}}_1}\abra{\tilde{\underline{\la}}_1}{\tilde{\underline{\la}}_2}}
\bar{\delta}^{2|4}\big(\calZ_{\tilde{x}}(\tilde{\la}_1),\star,\calZ_z(\rho_1)\big)\bar{\delta}^{2|4}\big(\calZ_{\tilde{y}}(\tilde{\underline{\la}}_1),\star,\calZ_z(\rho_2)\big)\,,
\end{aligned}
\end{equation}
where $F$ is a function of the appropriate homogeneity that takes into account which operators are inserted at the operator-bearing edges; its precise form does not matter in the present discussion.
The fraction of angular brackets comes from the expansion of the two frames $U$.
We recall that according to our prescription we are to take the integral over $z$ in the interaction vertex last. The fermionic integration over $\vartheta$ can however be taken immediately, which leads to an additional factor of $\abra{\rho_1}{\rho_2}^4$ and removes the fermionic pieces from the $\bar{\delta}^{2|4}$ functions. We can replace 
\begin{equation}
\bar{\delta}^{2}(Z_1,Z_2,Z_3)=\int \frac{\dd s\dd t}{s t}\bar{\delta}^4(Z_1+sZ_2+tZ_3)\,,
\end{equation}
where one has to be careful that the above is homogeneous of degree 0 in $Z_2$ and $Z_3$ but of degree $-4$ in $Z_1$.
The appropriate form of the delta function has to be taken depending on the homogeneity of the integrand in each of the $Z_i$ so that the integral is homogeneous of degree 0. Thus, we obtain
\begin{equation}
\begin{aligned}
\label{eq:V2part2}
V_2\propto \frac{F(\la_1,\la_2,\underline{\la}_1,\underline{\la}_2) \abra{\tilde{\la}_0}{\tilde{\la}_2}\abra{\tilde{\underline{\la}}_0}{\tilde{\underline{\la}}_2}}{|x_i-y_i|^2|x_{i+1}-y_{i+1}|^2}
\int  \frac{\dd^4 z}{|\tilde{x}-z|^2|\tilde{y}-z|^2}\frac{\abra{\rho_1}{\rho_2}^2}{\abra{\tilde{\la}_0}{\tilde{\la}_1}\abra{\tilde{\la}_1}{\tilde{\la}_2}\abra{\tilde{\underline{\la}}_0}{\tilde{\underline{\la}}_1}\abra{\tilde{\underline{\la}}_1}{\tilde{\underline{\la}}_2}}\,,
\end{aligned}
\end{equation}
where the spinors that we integrate over in \eqref{eq:V2part1} have been \emph{fixed} by the delta function to the values
\begin{align}
\label{eq:localizationspinorpart1}
&\la_{1\alpha}=-\underline{\la}_{1\alpha}=\frac{i(x_i-y_i)_{\alpha\dot\alpha}\zeta^{\dot\alpha}}{(x_i-y_i)^2}\,,&
&\la_{2\alpha}=-\underline{\la}_{2\alpha}=\frac{i(x_{i+1}-y_{i+1})_{\alpha\dot\alpha}\zeta^{\dot\alpha}}{(x_{i+1}-y_{i+1})^2}\,,&\nonumber\\
&\tilde{\la}_{1\alpha}=-\rho_{1\alpha}=\frac{i(\tilde{x}-z)_{\alpha\dot\alpha}\zeta^{\dot\alpha}}{(\tilde{x}-z)^2}\,,& &\tilde{\underline{\la}}_{1\alpha}=-\rho_{2\alpha}=\frac{i(\tilde{y}-z)_{\alpha\dot\alpha}\zeta^{\dot\alpha}}{(\tilde{y}-z)^2}\,,&
\end{align}
and the factors of $|\tilde{x}-z|^2|\tilde{y}-z|^2$ and $|x_i-y_i|^2|x_{i+1}-y_{i+1}|^2$ in the denominator arose due to the Jacobians of the integration. Here, as elsewhere,  $\zeta^{\dot\alpha}$ is the lower component of the (bosonic) reference twistor $Z_\star$. 
We now have to take the sum over the insertions points and the operator limit, \emph{before} we perform the integration over $z$. The contributions from the other insertion points are similar and the sum can be done using the Schouten identity in the same way as described in figure 8 of \cite{Koster:2016loo}. It effectively leads to replacing in \eqref{eq:V2part2} $\tilde{\la}_0$ by $\la_1$, $\tilde{\la}_2$ by $\la_2$, $\tilde{\underline{\la}}_0$ by $\underline{\la}_1$ and $\tilde{\underline{\la}}_2$ by $\underline{\la}_2$. Finally, in the operator limit we have $x_i,x_{i+1}\rightarrow x$ and $y_j,y_{j+1}\rightarrow  y$, so that the prefactor $\abra{\la_1}{\la_2}\abra{\underline{\la}_1}{\underline{\la}_2}$ in front of the integral vanishes and hence the contribution is zero.

\begin{figure}[tbp]
 \centering
 \includegraphics[height=4cm]{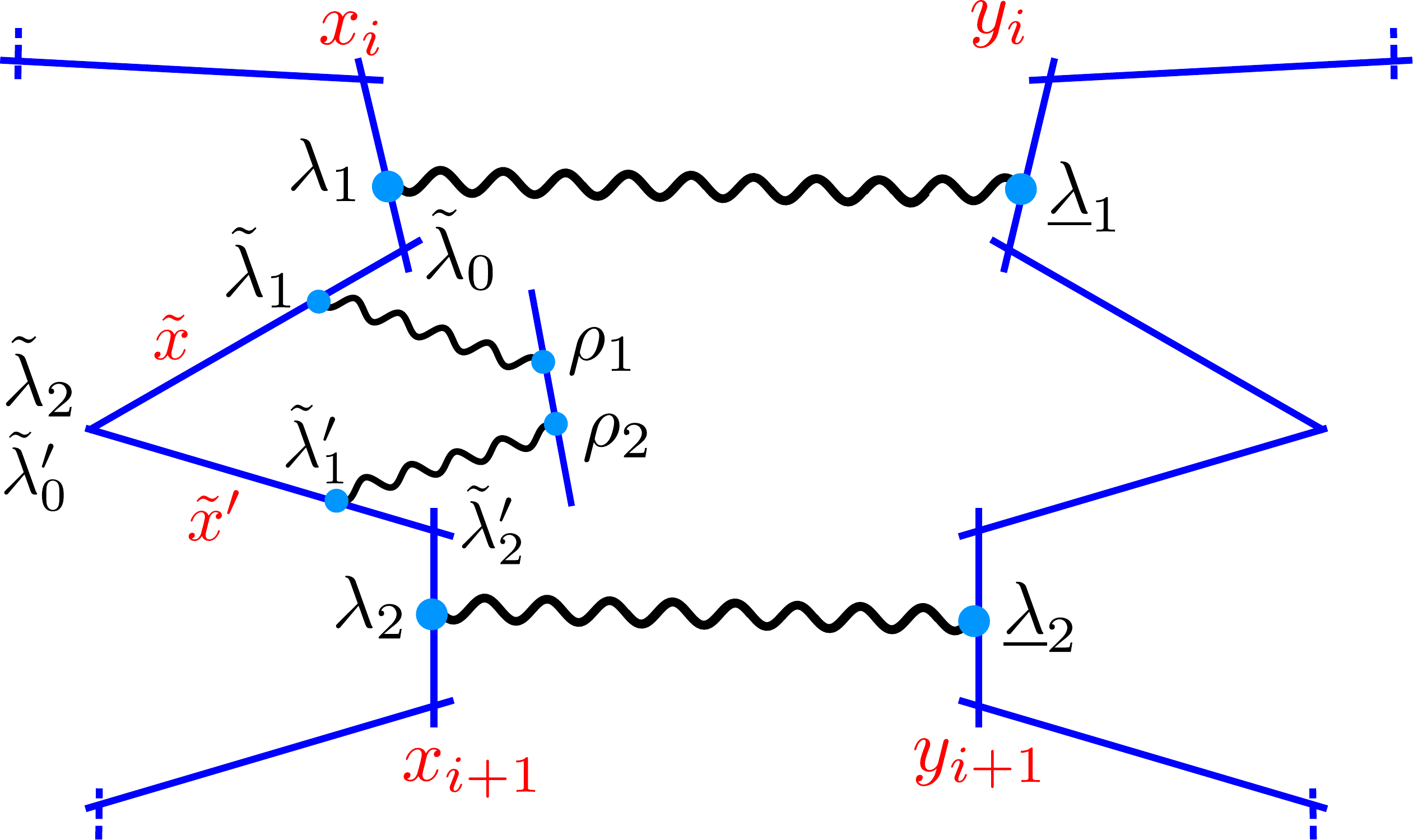}
  \caption{\it A two-point vertex 1-loop correlation function diagram with the two-point vertex attached uniquely to one of the two cogwheel Wilson loops.
  }
  \label{fig:2vertex1LoopDiagram}
\end{figure}

\paragraph{The two-point vertex connected to one side of the diagram:}
For the other diagram containing a two-point vertex, shown in the middle of figure~\ref{fig:oneloopCorrelation3} and in more detail in figure~\ref{fig:2vertex1LoopDiagram}, we have to connect the two propagators coming from the vertex to any combination of the edges between $\la_1$ and $\la_2$. 
Below, we compute just the combination shown in figure~\ref{fig:2vertex1LoopDiagram}.
We find that, 
after the $\vartheta$ integration and the integration over $\lambda_1$, $\lambda_2$, $\underline{\la}_1$  and $\underline{\la}_2$ which localizes these spinors in a similar way as \eqref{eq:localizationspinorpart1},
 it contributes
\begin{align}
\label{twovertex}
V_2'&\propto\frac{F(\la_1,\la_2,\underline{\la}_1,\underline{\la}_2)}{|x_i-y_i|^2|x_{i+1}-y_{i+1}|^2}\int \frac{\dd^4z}{(2\pi)^4}\DD\rho_1\DD\rho_2\DD\tilde\la_1\DD\tilde\la_1'\Bigg[ \frac{\abra{\rho_1}{\rho_2}^4}{\abra{\rho_1}{\rho_2}^2} 
\frac{\abra{\tilde\la_0}{\tilde\la_{2}}}{\abra{\tilde\la_{0}}{\tilde\la_1}\abra{\tilde\la_{1}}{\tilde\la_2}}
\frac{\abra{\tilde\la_0'}{\tilde\la_2'}}{\abra{\tilde\la_0'}{\tilde\la_1'}\abra{\tilde\la_1'}{\tilde\la_2'}} \nonumber\\
&\phaneq\times \bar{\delta}^{2}\big(Z_{\tilde{x}}(\tilde{\la}_1),\star,Z_z(\rho_1)\big)\bar{\delta}^{2}\big(Z_{\tilde{x}'}(\tilde{\la}_1'),\star,Z_z(\rho_2)\big)\Bigg]\,.
\end{align} 
The term $\abra{\rho_1}{\rho_2}^2$ in the denominator comes directly from the interaction vertex in the action, while the factor $\abra{\rho_1}{\rho_2}^4$ in the numerator is the result of the $\dd^8\vartheta$ integration as for the previous two-point vertex.
Doing the remaining integrations and summing over the possible ways to connect the diagram leads in the operator limit (implying $\tilde{x}, \tilde{x}'\rightarrow x$) to
\begin{equation}
\sum_{\text{connections}} V_2' \propto \frac{F(\la_1,\la_2,\underline{\la}_1,\underline{\la}_2)}{|x-y|^4}\int \frac{\dd^4z}{|x-z|^4} \frac{\abra{\la_1}{\la_{2}}\abra{\rho_1}{\rho_2}^{\cancel{2}}}{\abra{\la_{1}}{\tilde\la_1}\cancel{\abra{\tilde\la_{1}}{\tilde\la_1'}}\abra{\tilde\la_1'}{\tilde\la_2}}\eqncom
\end{equation}
where the cancellation happens because we have \textit{localized} the spinors
\begin{equation}
\begin{aligned} 
&\la_{1\alpha}=\underline\la_{1\alpha}\propto(x-y)_{\alpha\dot\alpha}\zeta^{\dot\alpha}\,,&
&\la_{2\alpha}=\underline\la_{2\alpha}\propto(x-y)_{\alpha\dot\alpha}\zeta^{\dot\alpha}\,,&\\
&\tilde\la_{1\alpha}=- \rho_{1\alpha}\propto(x-z)_{\alpha\dot\alpha}\zeta^{\dot\alpha}\,,&
&\tilde\la_{1\alpha}'=-\rho_{2\alpha}\propto(x-z)_{\alpha\dot\alpha}\zeta^{\dot\alpha}\,,&
\end{aligned}
\end{equation}
via the delta functions of the propagators. Hence, we see that the contribution becomes zero due to the vanishing of the product of brackets $\abra{\la_1}{\la_2}\abra{\rho_1}{\rho_2}$ in the operator limit, which we take before performing the integration over $z$.
 
\begin{figure}[tbp]
 \centering
 \includegraphics[height=4cm]{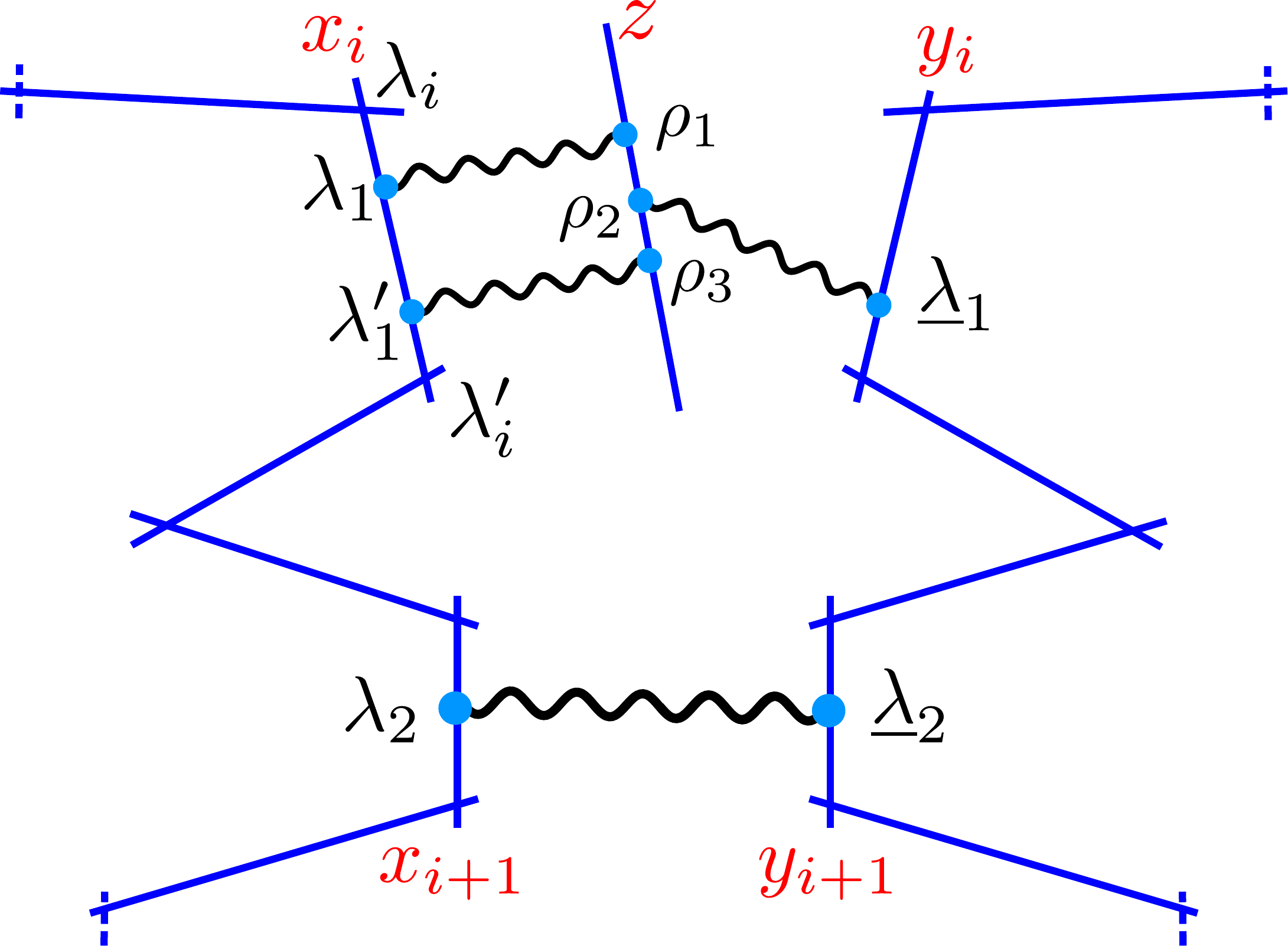}
  \caption{\it Three-point vertex contributions to the one-loop two-point correlation function.
  }
  \label{fig:1loopdiagram}
\end{figure}

\paragraph{Diagrams with a three-point vertex:} 
Finally, we consider the diagrams with a three-point vertex, where we take the operator vertex up to the second term in the expansion of the parallel propagator $U_{(x_i,\theta)}(\calZ_i,\calZ_{i+1})$, see figure~\ref{fig:1loopdiagram}. The particles emitted from the edge $x_i$ of the loop can be either a scalar $\phi$ and a positive-helicity gluon $g^+$ or two anti-fermions $\bar\psi$. 
Therefore, we cannot factor out the forming operator as in the cases with the two-point vertex. 
In the case of two anti-fermions $\bar\psi$, both are attached to the operating-bearing edge $x_i$ of the Wilson loop. However, when there is a scalar $\phi$ and a positive-helicity gluon $g^+$, the gluon can be attached to many different edges and we have to sum over all of them. Here, we will just treat the attachment shown in figure~\ref{fig:1loopdiagram} in detail and then argue that the other terms are similar and can be combined using the Schouten identity after the operator limit is taken. 
We obtain for the vertex on the cog $x_i$ (see appendix B of \cite{Koster:2016loo} for more details)
\begin{equation}
\label{scalar field vertex}
\begin{aligned}
\textbf{W}_{\phi}&=\int \DD{\la}_1\DD{\la}_1'
\Bigg[ \frac{\partial^2\AAA(\calZ_{(x,\theta)}({\la}_1))}{\partial{\chi}_1^a\partial{\chi}_1^b} \frac{\abra{{\la}_1}{ \la_{i}'}}{\abra{{\la}_1}{{\la}_1'}\abra{{\la}_1'}{\la_{i}'}} \AAA(\calZ_{(x,\theta)}({\la}_1'))\\&\qquad\qquad\qquad \qquad + \frac{\partial\AAA(\calZ_{(x,\theta)}({\la}_1))}{\partial{\chi}_1^a}\frac{1}{\abra{{\la}_1}{\la_1'}}\frac{\partial\AAA(\calZ_{(x,\theta)}({\la}_1'))}{\partial{\chi}_1'^b}  \left.\Bigg]\right\vert_{\theta=0}\,\eqndot
\end{aligned}
\end{equation}
The first term corresponds to a scalar $\phi$ and a positive-helicity gluon $g^+$ being emitted from $x_i$ and the second to two anti-fermions $\bar\psi$ being emitted. We treat them separately.%
\footnote{Of course, there is also a term in which the indices $a$ and $b$ in the second line are interchanged, but there is also a term in which the $\AAA(\calZ_{(x,\theta)}(\la_1))$ in the first term is in front of the term with two derivatives acting on $\AAA(\calZ_{(x,\theta)}(\la_1'))$.}
On the opposite side of the correlation function, we have the operator vertex
\begin{equation}
\textbf{W}_{\underline{\phi}}=\left.\int \DD\underline\la_1\frac{\partial^2\AAA(\calZ_{(y,\underline\theta)}(\underline\la_1))}{\partial\underline\chi_1^a\partial\underline\chi_1^b}\right\vert_{\underline\theta=0}\eqndot
\end{equation}
The three $\AAA$ fields, two at line $x$ and one at line $y$, are connected to the three-point vertex
\begin{equation}
\textbf{V}_3=\int \frac{\dd^{4}z\dd^8\vartheta}{(2\pi)^4} \int \frac{\DD\rho_1\DD\rho_2\DD\rho_3}{\abra{\rho_1}{\rho_2}\abra{\rho_2}{\rho_3}\abra{\rho_3}{\rho_1}} \Tr \left(\AAA(\calZ_{(z,\vartheta)}(\rho_1))\AAA(\calZ_{(z,\vartheta)}(\rho_2))\AAA(\calZ_{(z,\vartheta)}(\rho_3))\right) \eqndot
\end{equation}
The expression for the diagram corresponding to the first term in \eqref{scalar field vertex} therefore yields
\begin{align}
&\frac{F(\la_2,\underline\la_2)}{|x_{i+1}-y_{i+1}|}\int \DD{\la}_1\DD{\la}_1' \frac{\dd^{4}z\dd^8\vartheta}{(2\pi)^4}  \frac{\DD\rho_1\DD\rho_2\DD\rho_3}{\abra{\rho_1}{\rho_2}\abra{\rho_2}{\rho_3}\abra{\rho_3}{\rho_1}} 
 \frac{\partial^2}{\partial{\chi}_1^a\partial{\chi}_1^b}\bar\delta^{2|4}(\calZ_{(x_i,\theta)}({\la}_1),\star, Z_{(z,\vartheta)}(\rho_1)) \notag\\
 &\frac{\abra{{\la}_1}{ \la_{i}'}}{\abra{{\la}_1}{{\la}_1'}\abra{{\la}_1'}{\la_{i}'}} 
 \bar\delta^{2|4}(\calZ_{(x_i,\theta)}({\la}_1'),\star, \calZ_{(z,\vartheta)}(\rho_3))\frac{\partial^2}{\partial\underline\chi_1^a\partial\underline\chi_1^b}
  \bar\delta^{2|4}(\calZ_{(z,\vartheta)}({\rho}_2),\star,\calZ_{(y_i,\underline\theta)}(\underline\la_1))\eqncom
\end{align}
where we have localized two spinors as $\la_{2\alpha}\propto \underline{\la}_{2\alpha}\propto (x_{i+1}-y_{i+1})_{\alpha\dot{\alpha}}\zeta^{\dot{\alpha}}$ and where $F(\la_2,\underline\la_2)$ is a function that takes into account what operator are inserted at $\la_2$ and $\underline\la_2$.
Integrating over $\dd^8\vartheta$ gives a factor of $\abra{\rho_3}{\rho_1}^2\abra{\rho_3}{\rho_2}^2$. Evaluating the remaining integrations and summing over the possible ways to connect the diagram leads in the operator limit to
\begin{equation}
\label{eq:3vertexcomputationterm1}
\begin{aligned}
\frac{F(\la_2,\underline\la_2)}{|x-y|^2}\int \frac{\dd^{4}z}{(2\pi)^4} \frac{1}{|x-z|^4|y-z|^2} \Bigg[
 \underbrace{\frac{\abra{\rho_2}{ \la_{i}'}}{\abra{\rho_2}{\rho_1}\abra{\rho_1}{\la_{i}'}} \frac{\abra{\rho_2}{\rho_1}^2\abra{\rho_3}{\rho_1}^2}{\abra{\rho_1}{\rho_2}\abra{\rho_2}{\rho_3}\abra{\rho_3}{\rho_1}}}_{= \frac{\abra{\rho_2}{ \la_{i}'}\abra{\rho_1}{\rho_3}}{\abra{\rho_1}{\la_{i}'}\abra{\rho_2}{\rho_3}}\stackrel{\hexagon\rightarrow \xdot}=1}\Bigg]\eqncom
\end{aligned}
\end{equation}
due to the fact that the spinorial integration leads to
\begin{equation}\label{localizedspinors} \la_1=\rho_2,\quad \la_1'=\rho_1,\quad \underline\la_1=\rho_3\eqndot
\end{equation} 
Similarly, the second term in \eqref{scalar field vertex} yields
\begin{align}
&F(\la_2,\underline\la_2)\int \DD{\la}_1\DD{\la}_1' \frac{\dd^{4}z\dd^8\vartheta}{(2\pi)^4}  \frac{\DD\rho_1\DD\rho_2\DD\rho_3}{\abra{\rho_1}{\rho_2}\abra{\rho_2}{\rho_3}\abra{\rho_3}{\rho_1}} 
 \frac{\partial}{\partial{\chi}_1^a}\bar\delta^{2|4}(\calZ_{(x_i,\theta)}({\la}_1),\star, Z_{(z,\vartheta)}(\rho_1)) \notag\\
 &\frac{1}{\abra{{\la}_1}{{\la}_1'}} 
 \frac{\partial}{\partial{\chi}_1'^b}\ \bar\delta^{2|4}(\calZ_{(x_i,\theta)}({\la}_1'),\star, \calZ_{(z,\vartheta)}(\rho_3))\frac{\partial^2}{\partial\underline\chi_1^a\partial\underline\chi_1^b}
  \bar\delta^{2|4}(\calZ_{(z,\vartheta)}({\rho}_2),\star,\calZ_{(y_i,\underline\theta)}(\underline\la_1))\eqndot
\end{align}
Integrating over $\dd^8\vartheta$ yields a factor of $-\abra{\rho_3}{\rho_2}\abra{\rho_3}{\rho_1}\abra{\rho_2}{\rho_1}^2$ so that in the operator limit we find
\begin{equation}
\label{eq:3vertexcomputation}
\begin{aligned}
&\frac{F(\la_2,\underline\la_2)}{|x-y|^2}\int \frac{\dd^{4}z}{(2\pi)^4} \frac{1}{|x-z|^4|y-z|^2} \Bigg[
- \underbrace{\frac{1}{\abra{\rho_2}{\rho_1}}\frac{\abra{\rho_3}{\rho_2}\abra{\rho_3}{\rho_1}\abra{\rho_1}{\rho_2}^2}{\abra{\rho_1}{\rho_2}\abra{\rho_2}{\rho_3}\abra{\rho_3}{\rho_1}}}_{=1}\Bigg]\eqncom
\end{aligned}
\end{equation}
where also here the spinors are localized according to \eqref{localizedspinors}. We observe that the contributions \eqref{eq:3vertexcomputationterm1} and \eqref{eq:3vertexcomputation} precisely cancel. Hence, the three-point vertex does not contribute to the one-loop UV divergent part of the correlation function. This completes our treatment of the dilatation operator of \cite{Koster:2014fva}.


\bibliographystyle{JHEP}
\bibliography{../biblio}

\providecommand{\href}[2]{#2}\begingroup\raggedright\begin{thebibliography}{10}

\bibitem{Koster:2016ebi}
L.~Koster, V.~Mitev, M.~Staudacher and M.~Wilhelm, \emph{{Composite Operators
  in the Twistor Formulation of $\mathcal{N}=4$ SYM Theory}},
  \href{http://dx.doi.org/10.1103/PhysRevLett.117.011601}{\emph{Phys. Rev.
  Lett.} {\bf 117} (2016) 011601},
  [\href{https://arxiv.org/abs/1603.04471}{{\tt 1603.04471}}].

\bibitem{Koster:2016loo}
L.~Koster, V.~Mitev, M.~Staudacher and M.~Wilhelm, \emph{{All Tree-Level MHV
  Form Factors in $\mathcal{N}=4$ SYM from Twistor Space}},
  \href{http://dx.doi.org/10.1007/JHEP06(2016)162}{\emph{JHEP} {\bf 06} (2016)
  162}, [\href{https://arxiv.org/abs/1604.00012}{{\tt 1604.00012}}].

\bibitem{Koster:2014fva}
L.~Koster, V.~Mitev and M.~Staudacher, \emph{{A Twistorial Approach to
  Integrability in $\mathcal N=$ 4 SYM}},
  \href{http://dx.doi.org/10.1002/prop.201400085}{\emph{Fortsch. Phys.} {\bf
  63} (2015) 142--147}, [\href{https://arxiv.org/abs/1410.6310}{{\tt
  1410.6310}}].

\bibitem{Witten:2003nn}
E.~Witten, \emph{{Perturbative gauge theory as a string theory in twistor
  space}},
  \href{http://dx.doi.org/10.1007/s00220-004-1187-3}{\emph{Commun.Math.Phys.}
  {\bf 252} (2004) 189--258}, [\href{https://arxiv.org/abs/hep-th/0312171}{{\tt
  hep-th/0312171}}].

\bibitem{Boels:2006ir}
R.~Boels, L.~Mason and D.~Skinner, \emph{{Supersymmetric Gauge Theories in
  Twistor Space}},
  \href{http://dx.doi.org/10.1088/1126-6708/2007/02/014}{\emph{JHEP} {\bf 0702}
  (2007) 014}, [\href{https://arxiv.org/abs/hep-th/0604040}{{\tt
  hep-th/0604040}}].

\bibitem{Boels:2007qn}
R.~Boels, L.~J. Mason and D.~Skinner, \emph{{From twistor actions to MHV
  diagrams}},
  \href{http://dx.doi.org/10.1016/j.physletb.2007.02.058}{\emph{Phys. Lett.}
  {\bf B648} (2007) 90--96}, [\href{https://arxiv.org/abs/hep-th/0702035}{{\tt
  hep-th/0702035}}].

\bibitem{ArkaniHamed:2009dn}
N.~Arkani-Hamed, F.~Cachazo, C.~Cheung and J.~Kaplan, \emph{{A Duality For The
  S Matrix}}, \href{http://dx.doi.org/10.1007/JHEP03(2010)020}{\emph{JHEP} {\bf
  03} (2010) 020}, [\href{https://arxiv.org/abs/0907.5418}{{\tt 0907.5418}}].

\bibitem{Mason:2009qx}
L.~J. Mason and D.~Skinner, \emph{{Dual Superconformal Invariance, Momentum
  Twistors and Grassmannians}},
  \href{http://dx.doi.org/10.1088/1126-6708/2009/11/045}{\emph{JHEP} {\bf 11}
  (2009) 045}, [\href{https://arxiv.org/abs/0909.0250}{{\tt 0909.0250}}].

\bibitem{Bullimore:2010pj}
M.~Bullimore, L.~J. Mason and D.~Skinner, \emph{{MHV Diagrams in Momentum
  Twistor Space}}, \href{http://dx.doi.org/10.1007/JHEP12(2010)032}{\emph{JHEP}
  {\bf 12} (2010) 032}, [\href{https://arxiv.org/abs/1009.1854}{{\tt
  1009.1854}}].

\bibitem{Adamo:2011cb}
T.~Adamo and L.~Mason, \emph{{MHV diagrams in twistor space and the twistor
  action}},
  \href{http://dx.doi.org/10.1103/PhysRevD.86.065019}{\emph{Phys.Rev.} {\bf
  D86} (2012) 065019}, [\href{https://arxiv.org/abs/1103.1352}{{\tt
  1103.1352}}].

\bibitem{Adamo:2011pv}
T.~Adamo, M.~Bullimore, L.~Mason and D.~Skinner, \emph{{Scattering Amplitudes
  and Wilson Loops in Twistor Space}},
  \href{http://dx.doi.org/10.1088/1751-8113/44/45/454008}{\emph{J.Phys.} {\bf
  A44} (2011) 454008}, [\href{https://arxiv.org/abs/1104.2890}{{\tt
  1104.2890}}].

\bibitem{Mason:2010yk}
L.~J. Mason and D.~Skinner, \emph{{The Complete Planar S-matrix of
  $\mathcal{N}=4$ SYM as a Wilson Loop in Twistor Space}},
  \href{http://dx.doi.org/10.1007/JHEP12(2010)018}{\emph{JHEP} {\bf 12} (2010)
  018}, [\href{https://arxiv.org/abs/1009.2225}{{\tt 1009.2225}}].

\bibitem{Bullimore:2011ni}
M.~Bullimore and D.~Skinner, \emph{{Holomorphic Linking, Loop Equations and
  Scattering Amplitudes in Twistor Space}},
  \href{https://arxiv.org/abs/1101.1329}{{\tt 1101.1329}}.

\bibitem{Chicherin:2014uca}
D.~Chicherin, R.~Doobary, B.~Eden, P.~Heslop, G.~P. Korchemsky, L.~Mason and E.~Sokatchev, \emph{{Correlation functions of the chiral stress-tensor multiplet in
  $ \mathcal{N}=4 $ SYM}},
  \href{http://dx.doi.org/10.1007/JHEP06(2015)198}{\emph{JHEP} {\bf 06} (2015)
  198}, [\href{https://arxiv.org/abs/1412.8718}{{\tt 1412.8718}}].

\bibitem{Chicherin:2016fac}
D.~Chicherin and E.~Sokatchev, \emph{{$\mathcal{N}=4$ super-Yang-Mills in LHC
  superspace. Part I: Classical and quantum theory}},
  \href{https://arxiv.org/abs/1601.06803}{{\tt 1601.06803}}.

\bibitem{Chicherin:2016fbj}
D.~Chicherin and E.~Sokatchev, \emph{{$\mathcal{N}=4$ super-Yang-Mills in LHC
  superspace. Part II: Non-chiral correlation functions of the stress-tensor
  multiplet}},  \href{https://arxiv.org/abs/1601.06804}{{\tt 1601.06804}}.

\bibitem{Chicherin:2016qsf}
D.~Chicherin and E.~Sokatchev, \emph{{Composite operators and form factors in
  $\mathcal{N}=4$ SYM}},  \href{https://arxiv.org/abs/1605.01386}{{\tt
  1605.01386}}.

\bibitem{Chicherin:2016soh}
D.~Chicherin and E.~Sokatchev, \emph{{Demystifying the twistor construction of
  composite operators in $\mathcal{N}=4$ super-Yang-Mills theory}},
  \href{https://arxiv.org/abs/1603.08478}{{\tt 1603.08478}}.

\bibitem{Mueller:1979ih}
A.~H. Mueller, \emph{{On the Asymptotic Behavior of the Sudakov Form-factor}},
  \href{http://dx.doi.org/10.1103/PhysRevD.20.2037}{\emph{Phys. Rev.} {\bf D20}
  (1979) 2037}.

\bibitem{Collins:1980ih}
J.~C. Collins, \emph{{Algorithm to Compute Corrections to the Sudakov
  Form-factor}}, \href{http://dx.doi.org/10.1103/PhysRevD.22.1478}{\emph{Phys.
  Rev.} {\bf D22} (1980) 1478}.

\bibitem{Sen:1981sd}
A.~Sen, \emph{{Asymptotic Behavior of the Sudakov Form-Factor in QCD}},
  \href{http://dx.doi.org/10.1103/PhysRevD.24.3281}{\emph{Phys. Rev.} {\bf D24}
  (1981) 3281}.

\bibitem{Magnea:1990zb}
L.~Magnea and G.~F. Sterman, \emph{{Analytic continuation of the Sudakov
  form-factor in QCD}},
  \href{http://dx.doi.org/10.1103/PhysRevD.42.4222}{\emph{Phys. Rev.} {\bf D42}
  (1990) 4222--4227}.

\bibitem{vanNeerven:1985ja}
W.~L. van Neerven, \emph{{Infrared Behavior of On-shell Form-factors in a
  $\mathcal{N}=4$ Supersymmetric {Yang-Mills} Field Theory}},
  \href{http://dx.doi.org/10.1007/BF01571808}{\emph{Z. Phys.} {\bf C30} (1986)
  595}.

\bibitem{Brandhuber:2010ad}
A.~Brandhuber, B.~Spence, G.~Travaglini and G.~Yang, \emph{{Form Factors in
  $\mathcal{N}=4$ Super Yang-Mills and Periodic Wilson Loops}},
  \href{http://dx.doi.org/10.1007/JHEP01(2011)134}{\emph{JHEP} {\bf 1101}
  (2011) 134}, [\href{https://arxiv.org/abs/1011.1899}{{\tt 1011.1899}}].

\bibitem{Bork:2010wf}
L.~V. Bork, D.~I. Kazakov and G.~S. Vartanov, \emph{{On form factors in
  $\mathcal{N}=4$ sym}},
  \href{http://dx.doi.org/10.1007/JHEP02(2011)063}{\emph{JHEP} {\bf 02} (2011)
  063}, [\href{https://arxiv.org/abs/1011.2440}{{\tt 1011.2440}}].

\bibitem{Brandhuber:2011tv}
A.~Brandhuber, O.~Gurdogan, R.~Mooney, G.~Travaglini and G.~Yang,
  \emph{{Harmony of Super Form Factors}},
  \href{http://dx.doi.org/10.1007/JHEP10(2011)046}{\emph{JHEP} {\bf 10} (2011)
  046}, [\href{https://arxiv.org/abs/1107.5067}{{\tt 1107.5067}}].

\bibitem{Bork:2011cj}
L.~V. Bork, D.~I. Kazakov and G.~S. Vartanov, \emph{{On MHV Form Factors in
  Superspace for $\mathcal{N}=4$ SYM Theory}},
  \href{http://dx.doi.org/10.1007/JHEP10(2011)133}{\emph{JHEP} {\bf 10} (2011)
  133}, [\href{https://arxiv.org/abs/1107.5551}{{\tt 1107.5551}}].

\bibitem{Henn:2011by}
J.~M. Henn, S.~Moch and S.~G. Naculich, \emph{{Form factors and scattering
  amplitudes in $\mathcal{N}=4$ SYM in dimensional and massive
  regularizations}},
  \href{http://dx.doi.org/10.1007/JHEP12(2011)024}{\emph{JHEP} {\bf 12} (2011)
  024}, [\href{https://arxiv.org/abs/1109.5057}{{\tt 1109.5057}}].

\bibitem{Gehrmann:2011xn}
T.~Gehrmann, J.~M. Henn and T.~Huber, \emph{{The three-loop form factor in
  $\mathcal{N}=4$ super Yang-Mills}},
  \href{http://dx.doi.org/10.1007/JHEP03(2012)101}{\emph{JHEP} {\bf 03} (2012)
  101}, [\href{https://arxiv.org/abs/1112.4524}{{\tt 1112.4524}}].

\bibitem{Brandhuber:2012vm}
A.~Brandhuber, G.~Travaglini and G.~Yang, \emph{{Analytic two-loop form factors
  in $\mathcal{N}=4$ SYM}},
  \href{http://dx.doi.org/10.1007/JHEP05(2012)082}{\emph{JHEP} {\bf 05} (2012)
  082}, [\href{https://arxiv.org/abs/1201.4170}{{\tt 1201.4170}}].

\bibitem{Bork:2012tt}
L.~V. Bork, \emph{{On NMHV form factors in $\mathcal{N}=4$ SYM theory from
  generalized unitarity}},
  \href{http://dx.doi.org/10.1007/JHEP01(2013)049}{\emph{JHEP} {\bf 01} (2013)
  049}, [\href{https://arxiv.org/abs/1203.2596}{{\tt 1203.2596}}].

\bibitem{Engelund:2012re}
O.~T. Engelund and R.~Roiban, \emph{{Correlation functions of local composite
  operators from generalized unitarity}},
  \href{http://dx.doi.org/10.1007/JHEP03(2013)172}{\emph{JHEP} {\bf 1303}
  (2013) 172}, [\href{https://arxiv.org/abs/1209.0227}{{\tt 1209.0227}}].

\bibitem{Johansson:2012zv}
H.~Johansson, D.~A. Kosower and K.~J. Larsen, \emph{{Two-Loop Maximal Unitarity
  with External Masses}},
  \href{http://dx.doi.org/10.1103/PhysRevD.87.025030}{\emph{Phys. Rev.} {\bf
  D87} (2013) 025030}, [\href{https://arxiv.org/abs/1208.1754}{{\tt
  1208.1754}}].

\bibitem{Boels:2012ew}
R.~H. Boels, B.~A. Kniehl, O.~V. Tarasov and G.~Yang, \emph{{Color-kinematic
  Duality for Form Factors}},
  \href{http://dx.doi.org/10.1007/JHEP02(2013)063}{\emph{JHEP} {\bf 02} (2013)
  063}, [\href{https://arxiv.org/abs/1211.7028}{{\tt 1211.7028}}].

\bibitem{Penante:2014sza}
B.~Penante, B.~Spence, G.~Travaglini and C.~Wen, \emph{{On super form factors
  of half-BPS operators in $\mathcal{N}=4$ super Yang-Mills}},
  \href{http://dx.doi.org/10.1007/JHEP04(2014)083}{\emph{JHEP} {\bf 04} (2014)
  083}, [\href{https://arxiv.org/abs/1402.1300}{{\tt 1402.1300}}].

\bibitem{Brandhuber:2014ica}
A.~Brandhuber, B.~Penante, G.~Travaglini and C.~Wen, \emph{{The last of the
  simple remainders}},
  \href{http://dx.doi.org/10.1007/JHEP08(2014)100}{\emph{JHEP} {\bf 08} (2014)
  100}, [\href{https://arxiv.org/abs/1406.1443}{{\tt 1406.1443}}].

\bibitem{Bork:2014eqa}
L.~V. Bork, \emph{{On form factors in $ \mathcal{N}=4 $ SYM theory and
  polytopes}}, \href{http://dx.doi.org/10.1007/JHEP12(2014)111}{\emph{JHEP}
  {\bf 12} (2014) 111}, [\href{https://arxiv.org/abs/1407.5568}{{\tt
  1407.5568}}].

\bibitem{Wilhelm:2014qua}
M.~Wilhelm, \emph{{Amplitudes, Form Factors and the Dilatation Operator in
  $\mathcal{N}=4$ SYM Theory}},
  \href{http://dx.doi.org/10.1007/JHEP02(2015)149}{\emph{JHEP} {\bf 02} (2015)
  149}, [\href{https://arxiv.org/abs/1410.6309}{{\tt 1410.6309}}].

\bibitem{Nandan:2014oga}
D.~Nandan, C.~Sieg, M.~Wilhelm and G.~Yang, \emph{{Cutting through form factors
  and cross sections of non-protected operators in $ \mathcal{N}=4 $ SYM}},
  \href{http://dx.doi.org/10.1007/JHEP06(2015)156}{\emph{JHEP} {\bf 06} (2015)
  156}, [\href{https://arxiv.org/abs/1410.8485}{{\tt 1410.8485}}].

\bibitem{Loebbert:2015ova}
F.~Loebbert, D.~Nandan, C.~Sieg, M.~Wilhelm and G.~Yang, \emph{{On-Shell
  Methods for the Two-Loop Dilatation Operator and Finite Remainders}},
  \href{http://dx.doi.org/10.1007/JHEP10(2015)012}{\emph{JHEP} {\bf 10} (2015)
  012}, [\href{https://arxiv.org/abs/1504.06323}{{\tt 1504.06323}}].

\bibitem{Bork:2015fla}
L.~V. Bork and A.~I. Onishchenko, \emph{{On soft theorems and form factors in $
  \mathcal{N}=4 $ SYM theory}},
  \href{http://dx.doi.org/10.1007/JHEP12(2015)030}{\emph{JHEP} {\bf 12} (2015)
  030}, [\href{https://arxiv.org/abs/1506.07551}{{\tt 1506.07551}}].

\bibitem{Frassek:2015rka}
R.~Frassek, D.~Meidinger, D.~Nandan and M.~Wilhelm, \emph{{On-shell diagrams,
  Graßmannians and integrability for form factors}},
  \href{http://dx.doi.org/10.1007/JHEP01(2016)182}{\emph{JHEP} {\bf 01} (2016)
  182}, [\href{https://arxiv.org/abs/1506.08192}{{\tt 1506.08192}}].

\bibitem{Boels:2015yna}
R.~Boels, B.~A. Kniehl and G.~Yang, \emph{{Master integrals for the four-loop
  Sudakov form factor}},
  \href{http://dx.doi.org/10.1016/j.nuclphysb.2015.11.016}{\emph{Nucl. Phys.}
  {\bf B902} (2016) 387--414}, [\href{https://arxiv.org/abs/1508.03717}{{\tt
  1508.03717}}].

\bibitem{Huang:2016bmv}
R.~Huang, Q.~Jin and B.~Feng, \emph{{Form Factor and Boundary Contribution of
  Amplitude}}, \href{http://dx.doi.org/10.1007/JHEP06(2016)072}{\emph{JHEP}
  {\bf 06} (2016) 072}, [\href{https://arxiv.org/abs/1601.06612}{{\tt
  1601.06612}}].

\bibitem{Brandhuber:2016fni}
A.~Brandhuber, M.~Kostacinska, B.~Penante, G.~Travaglini and D.~Young,
  \emph{{The SU(2$|$3) dynamic two-loop form factors}},
  \href{http://dx.doi.org/10.1007/JHEP08(2016)134}{\emph{JHEP} {\bf 08} (2016)
  134}, [\href{https://arxiv.org/abs/1606.08682}{{\tt 1606.08682}}].

\bibitem{Bork:2016hst}
L.~V. Bork and A.~I. Onishchenko, \emph{{Grassmannians and form factors with
  $q^2=0$ in $\mathcal{N}=4$ SYM theory}},
  \href{https://arxiv.org/abs/1607.00503}{{\tt 1607.00503}}.

\bibitem{Bork:2016xfn}
L.~V. Bork and A.~I. Onishchenko, \emph{{Wilson lines, Grassmannians and Gauge
  Invariant Off-shell Amplitudes in $\mathcal{N}=4$ SYM}},
  \href{https://arxiv.org/abs/1607.02320}{{\tt 1607.02320}}.

\bibitem{He:2016dol}
S.~He and Y.~Zhang, \emph{{Connected formulas for amplitudes in standard
  model}},  \href{https://arxiv.org/abs/1607.02843}{{\tt 1607.02843}}.

\bibitem{Caron-Huot:2016cwu}
S.~Caron-Huot and M.~Wilhelm, \emph{{Renormalization group coefficients and the
  S-matrix}},  \href{https://arxiv.org/abs/1607.06448}{{\tt 1607.06448}}.

\bibitem{Brandhuber:2016xue}
A.~Brandhuber, E.~Hughes, R.~Panerai, B.~Spence and G.~Travaglini, \emph{{The
  connected prescription for form factors in twistor space}},
  \href{https://arxiv.org/abs/1608.03277}{{\tt 1608.03277}}.

\bibitem{He:2016jdg}
S.~He and Z.~Liu, \emph{{A note on connected formula for form factors}},
  \href{https://arxiv.org/abs/1608.04306}{{\tt 1608.04306}}.

\bibitem{Yang:2016ear}
G.~Yang, \emph{{Color-Kinematics Duality and Sudakov Form Factor at Five
  Loops}},  \href{https://arxiv.org/abs/1610.02394}{{\tt 1610.02394}}.

\bibitem{Ahmed:2016vgl}
T.~Ahmed, P.~Banerjee, P.~K. Dhani, N.~Rana, V.~Ravindran and S.~Seth,
  \emph{{Konishi Form Factor at Three Loop in ${\cal N}=4$ SYM}},
  \href{https://arxiv.org/abs/1610.05317}{{\tt 1610.05317}}.

\bibitem{Loebbert:2016xkw}
F.~Loebbert, C.~Sieg, M.~Wilhelm and G.~Yang, \emph{{Two-Loop SL(2) Form
  Factors and Maximal Transcendentality}},
  \href{https://arxiv.org/abs/1610.06567}{{\tt 1610.06567}}.

\bibitem{Bork:2016egt}
L.~V. Bork and A.~I. Onishchenko, \emph{{Grassmannian Integral for General
  Gauge Invariant Off-shell Amplitudes in $\mathcal{N}=4$ SYM}},
  \href{https://arxiv.org/abs/1610.09693}{{\tt 1610.09693}}.

\bibitem{Alday:2007he}
L.~F. Alday and J.~Maldacena, \emph{{Comments on gluon scattering amplitudes
  via AdS/CFT}},
  \href{http://dx.doi.org/10.1088/1126-6708/2007/11/068}{\emph{JHEP} {\bf 11}
  (2007) 068}, [\href{https://arxiv.org/abs/0710.1060}{{\tt 0710.1060}}].

\bibitem{Maldacena:2010kp}
J.~Maldacena and A.~Zhiboedov, \emph{{Form factors at strong coupling via a
  Y-system}}, \href{http://dx.doi.org/10.1007/JHEP11(2010)104}{\emph{JHEP} {\bf
  11} (2010) 104}, [\href{https://arxiv.org/abs/1009.1139}{{\tt 1009.1139}}].

\bibitem{Gao:2013dza}
Z.~Gao and G.~Yang, \emph{{Y-system for form factors at strong coupling in
  $AdS_5$ and with multi-operator insertions in $AdS_3$}},
  \href{http://dx.doi.org/10.1007/JHEP06(2013)105}{\emph{JHEP} {\bf 06} (2013)
  105}, [\href{https://arxiv.org/abs/1303.2668}{{\tt 1303.2668}}].

\bibitem{Wilhelm:2016izi}
M.~Wilhelm, \emph{{Form factors and the dilatation operator in $\mathcal{N}=4$
  super Yang-Mills theory and its deformations}}.
\newblock PhD thesis, 2016.
\newblock \href{https://arxiv.org/abs/1603.01145}{{\tt 1603.01145}}.

\bibitem{Penante:2016ycx}
B.~Penante, \emph{{On-shell methods for off-shell quantities in $\mathcal{N}=4$
  Super Yang-Mills: from scattering amplitudes to form factors and the
  dilatation operator}}.
\newblock PhD thesis, 2016.
\newblock \href{https://arxiv.org/abs/1608.01634}{{\tt 1608.01634}}.

\bibitem{Adamo:2013cra}
T.~Adamo, \emph{{Twistor actions for gauge theory and gravity}},
  \href{https://arxiv.org/abs/1308.2820}{{\tt 1308.2820}}.

\bibitem{Nair:1988bq}
V.~Nair, \emph{{A Current Algebra for Some Gauge Theory Amplitudes}},
  \href{http://dx.doi.org/10.1016/0370-2693(88)91471-2}{\emph{Phys.Lett.} {\bf
  B214} (1988) 215}.

\bibitem{Hodges:2009hk}
A.~Hodges, \emph{{Eliminating spurious poles from gauge-theoretic amplitudes}},
  \href{http://dx.doi.org/10.1007/JHEP05(2013)135}{\emph{JHEP} {\bf 1305}
  (2013) 135}, [\href{https://arxiv.org/abs/0905.1473}{{\tt 0905.1473}}].

\bibitem{Adamo:2011dq}
T.~Adamo, M.~Bullimore, L.~Mason and D.~Skinner, \emph{{A Proof of the
  Supersymmetric Correlation Function / Wilson Loop Correspondence}},
  \href{http://dx.doi.org/10.1007/JHEP08(2011)076}{\emph{JHEP} {\bf 1108}
  (2011) 076}, [\href{https://arxiv.org/abs/1103.4119}{{\tt 1103.4119}}].

\bibitem{Beisert:2004ry}
N.~Beisert, \emph{{The Dilatation operator of $\mathcal{N}=4$ super Yang-Mills
  theory and integrability}},
  \href{http://dx.doi.org/10.1016/j.physrep.2004.09.007}{\emph{Phys. Rept.}
  {\bf 405} (2004) 1--202}, [\href{https://arxiv.org/abs/hep-th/0407277}{{\tt
  hep-th/0407277}}].

\bibitem{Minahan:2010js}
J.~A. Minahan, \emph{{Review of AdS/CFT Integrability, Chapter I.1: Spin Chains
  in $\mathcal{N}=4$ Super Yang-Mills}},
  \href{http://dx.doi.org/10.1007/s11005-011-0522-9}{\emph{Lett.Math.Phys.}
  {\bf 99} (2012) 33--58}, [\href{https://arxiv.org/abs/1012.3983}{{\tt
  1012.3983}}].

\bibitem{Nandan:2012rk}
D.~Nandan and C.~Wen, \emph{{Generating All Tree Amplitudes in $\mathcal{N}=4$
  SYM by Inverse Soft Limit}},
  \href{http://dx.doi.org/10.1007/JHEP08(2012)040}{\emph{JHEP} {\bf 08} (2012)
  040}, [\href{https://arxiv.org/abs/1204.4841}{{\tt 1204.4841}}].

\end{thebibliography}\endgroup

\end{document}